\documentclass[aps,prd,nofootinbib,twocolumn,superscriptaddress,preprintnumbers,balancelastpage,longbibliography]{revtex4-1}

\usepackage{placeins}
\usepackage{amsmath,amssymb,mathtools,bm}
\usepackage{graphicx, color, hepunits}
\usepackage[dvipsnames]{xcolor}
\usepackage{float}
\usepackage{filecontents}
\usepackage{multirow}
 \usepackage{hyperref} 
\hypersetup{
    colorlinks=true,       
    linkcolor=blue,        
    citecolor=blue,        
    filecolor=magenta,     
    urlcolor=blue          
}

\usepackage[utf8]{inputenc}
\usepackage[english]{babel}

\newcommand{\es}[2] {\begin{equation} \label{#1} \begin{split} #2 \end{split} \end{equation}}
\usepackage{placeins}

\begin{document}

\title{Higgsino Dark Matter Confronts 14 years of Fermi Gamma Ray Data}

\author{Christopher Dessert}
\affiliation{Berkeley Center for Theoretical Physics, University of California, Berkeley, CA 94720, U.S.A.}
\affiliation{Theoretical Physics Group, Lawrence Berkeley National Laboratory, Berkeley, CA 94720, U.S.A.}
\affiliation{Leinweber Center for Theoretical Physics, Department of Physics, University of Michigan, Ann Arbor, MI 48109 U.S.A.}

\author{Joshua W. Foster}
\affiliation{Center for Theoretical Physics, Massachusetts Institute of Technology, Cambridge, Massachusetts 02139, U.S.A}

\author{Yujin Park}
\affiliation{Berkeley Center for Theoretical Physics, University of California, Berkeley, CA 94720, U.S.A.}
\affiliation{Theoretical Physics Group, Lawrence Berkeley National Laboratory, Berkeley, CA 94720, U.S.A.}

\author{Benjamin R. Safdi}
\affiliation{Berkeley Center for Theoretical Physics, University of California, Berkeley, CA 94720, U.S.A.}
\affiliation{Theoretical Physics Group, Lawrence Berkeley National Laboratory, Berkeley, CA 94720, U.S.A.}

\author{Weishuang Linda Xu}
\affiliation{Berkeley Center for Theoretical Physics, University of California, Berkeley, CA 94720, U.S.A.}
\affiliation{Theoretical Physics Group, Lawrence Berkeley National Laboratory, Berkeley, CA 94720, U.S.A.}

\date{\today}
\preprint{MIT-CTP/5454}

\begin{abstract}
Thermal higgsino dark matter (DM), with mass around 1 TeV, is a well-motivated, minimal DM scenario that arises in supersymmetric extensions of the Standard Model.  Higgsinos may naturally be the lightest superpartners in Split-supersymmetry models that decouple the scalar superpartners while keeping higgsinos and gauginos close to the TeV scale.  Higgsino DM may annihilate today to give continuum gamma-ray emission at energies less than a TeV in addition to a line-like signature at energies equal to the mass. Previous searches for higgsino DM, for example with the H.E.S.S. gamma-ray telescope, have 
not reached the necessary sensitivity to probe the higgsino annihilation cross-section.  In this work we make use of 14 years of {\it Fermi} gamma-ray data at energies above $\sim$10 GeV to search for the continuum emission near the Galactic Center from higgsino annihilation.  
We interpret our results using DM profiles from Milky Way analogue galaxies in the FIRE-2 hydrodynamic cosmological simulations.
We set the strongest constraints to-date on higgsino-like DM.
Our results show a mild, $\sim$2$\sigma$ preference for higgsino DM with a mass near the thermal higgsino mass and, depending on the DM density profile, the expected  cross-section.
\end{abstract} 

\maketitle

Dark matter (DM) makes up $\sim$27\% of the energy in our Universe today~\cite{Planck:2018vyg}, with only $\sim$5\% of the energy density in ordinary matter, yet its microscopic nature remains unknown.  One tantalizing possibility, which has driven decades of experimental and theoretical effort, is that DM arises as the lightest superpartner (LSP) in supersymmetric (SUSY) extensions of the Standard Model that address the hierarchy problem related to the unnaturally low Higgs mass parameter (see~\cite{Jungman:1995df} for a review).  LSP DM at the $\sim$TeV scale may naturally acquire the correct DM abundance through thermal freeze-out.
On the other hand, electroweak scale SUSY and LSP DM have come under increasing tension in recent years from null searches for new physics at the Large Hadron Collider (LHC)~\cite{CMS:2013wdg,ATLAS:2015wrn}, direct detection experiments~\cite{XENON:2018voc}, and indirect searches~\cite{Cohen:2013ama,Fermi-LAT:2016uux}.  

Natural LSP candidates 
are the neutral gauginos -- namely, bino and wino LSPs -- and the higgsino.  Pure bino DM is excluded by direct searches at the LHC~\cite{ATLAS:2014jxt,CMS:2015flg}.  Wino DM is in strong tension with null results from DM annihilation searches with the H.E.S.S. gamma-ray telescope~\cite{Fan:2013faa,Cohen:2013ama}.  Nearly-pure higgsino LSPs, on the other hand, remain one of the better motivated and sought after, yet unprobed, DM scenarios (see, {\it e.g.},~\cite{Co:2021ion} for a recent summary).  Higgsinos, which are the superpartners of the two Higgs doublets in the minimal supersymmetric standard model (MSSM), are especially motivated in light of (i) null results for wino DM, and (ii) the fact that null searches for superpartners at {\it e.g.} the LHC suggest that nature may implement a split-spectrum version of SUSY such as Split-SUSY~\cite{Wells:2003tf,Giudice:2004tc,Arkani-Hamed:2004ymt,Arvanitaki:2012ps,Arkani-Hamed:2012fhg}, which naturally leads to higgsino or wino LSP DM.  Split-SUSY, mini-Split, and similar constructions~\cite{Hall:2011jd} aim to preserve LSP DM and high-scale gauge unification but give up on trying to fully solve the hierarchy problem; in such models the scalar superpartners are taken to have large masses, with the gauginos and higgsinos remaining near the TeV scale. Such split-spectrum models may accommodate the observed Higgs mass and solve a number of troublesome problems with the MSSM, such as the lack of flavor changing neutral currents~\cite{Gabbiani:1996hi} and the non-observation of new CP violation in electric dipole moment searches~\cite{Altmannshofer:2013lfa,Cesarotti:2018huy}.

In this work we search for annihilation signatures of higgsino DM with {\it Fermi} gamma-ray data.  The higgsino interactions with the Standard Model are specified by its representation under the electroweak force, and thus the requirement that freeze-out of annihilation produces the correct DM abundance fully determines the higgsino mass under thermal cosmologies.
The higgsino mass arises from the Lagrangian term \mbox{${\mathcal L} \supset - \mu \tilde H_u \cdot \epsilon \cdot \tilde H_d + {\rm h.c.}$}, where $\mu$ is the MSSM $\mu$-parameter, $\tilde H_u$ ($\tilde H_d$) is the up-type (down-type) higgsino electroweak doublet, and $\epsilon$ is the totally anti-symmetric symbol in $SU(2)_L$ space.  There are two neutral higgsino fermions, which are generically split into two non-degenerate Majorana mass eigenstates by dimension-five operators that have the effect of inducing a slight mixing between the neutral gauginos and  higgsinos (see, {\it e.g.},~\cite{Nagata:2014wma}).  The charged higgsino states are heavier than the neutral states by at least $\sim$350 MeV because of radiative contributions to the charged higgsino masses below electroweak symmetry breaking~\cite{Thomas:1998wy}.  The mass splitting between the two neutral Majorana states, which we call $\Delta m$, must be greater than around 200 keV to avoid direct detection constraints from inelastic $Z$-exchange, where the lower Majorana state scatters into the heavier mass eigenstate (see, {\it e.g.},~\cite{Nagata:2014wma}).  When $\Delta m \gtrsim 200$ keV, direct detection of higgsino DM proceeds through elastic scattering with higher-dimensional operators and is thought to be below the neutrino floor~\cite{Hisano:2011cs,Hill:2013hoa,Hill:2014yka,Hill:2014yxa}.    

The relic abundance of higgsinos from thermal freezeout matches the observed DM abundance~\cite{Planck:2018vyg} for mass $m_\chi = 1.08 \pm 0.02 $ TeV, accounting for uncertainties on the DM abundance~\cite{Bottaro:2022one}. We refer to the higgsino with such a mass as the thermal higgsino. 
Apart from the SUSY motivations, higgsino DM may be viewed through the lens of minimal DM~\cite{Cirelli:2005uq}.   If $m_\chi$ is less than the TeV scale, then higgsinos are a computable but subdominant component of the DM, unless there is a non-standard cosmological history that increases their abundance.  The same annihilation processes that set the higgsino DM abundance in the early universe also lead to annihilation signatures of higgsinos today.
Sommerfeld enhancement introduces dependencies of the present-day annihilation rate on the mass splittings between the neutral and charged higgsino states, though this effect is relatively minor at the thermal higgsino mass~\cite{Hisano:2004ds}. 

The strongest existing indirect detection constraints on the thermal higgsino arise from Galactic Center (GC) searches for the line emission expected due to the $\gamma\gamma$ and $\gamma Z$ final states with H.E.S.S.~\cite{HESS:2018cbt}, which constrain $\langle\sigma v\rangle_{\gamma\gamma} \lesssim 4\times 10^{-28}$ cm$^{-3}$/s assuming an Einasto DM profile, whereas the thermal cross section is a factor of 4 smaller, and for H.E.S.S. searches for continuum emission from annihilation to $W^+W^-$~\cite{Montanari:2021yic}. The forthcoming Cherenkov Telescope Array (CTA), on the other hand, with 500 hours of exposure, is expected to have sensitivity to higgsino DM at the thermal mass~\cite{Rinchiuso:2020skh}.
Future lepton or hadron colliders may also be able to discover thermal higgsinos~\cite{Capdevilla:2021fmj,Arkani-Hamed:2015vfh,FCC:2018byv}.

In this Letter we use the existing 14 years of {\it Fermi} data to achieve world-leading sensitivity to higgsino DM by searching for annihilation to continuum gamma-rays through the $W^+W^-$ and $ZZ$ final states.  Our upper limits on the annihilation cross-section surpass those from H.E.S.S. for higgsino-like DM with mass $m_\chi \sim$ TeV, though our upper limits are weaker than expected due to the presence of a modest ($\sim$2$\sigma$) preference for the signal model over the null hypothesis. We interpret our results in the context of DM profiles from the FIRE-2 hydrodynamic cosmological simulations~\cite{Hopkins:2017ycn,2022MNRAS.513...55M} to show that the best-fit annihilation cross-section we recover may be consistent with the expected higgsino cross-section, potentially providing the first hint of thermal higgsino DM.

\begin{figure}[!t]
	\begin{center}
		\includegraphics[width=0.49\textwidth]{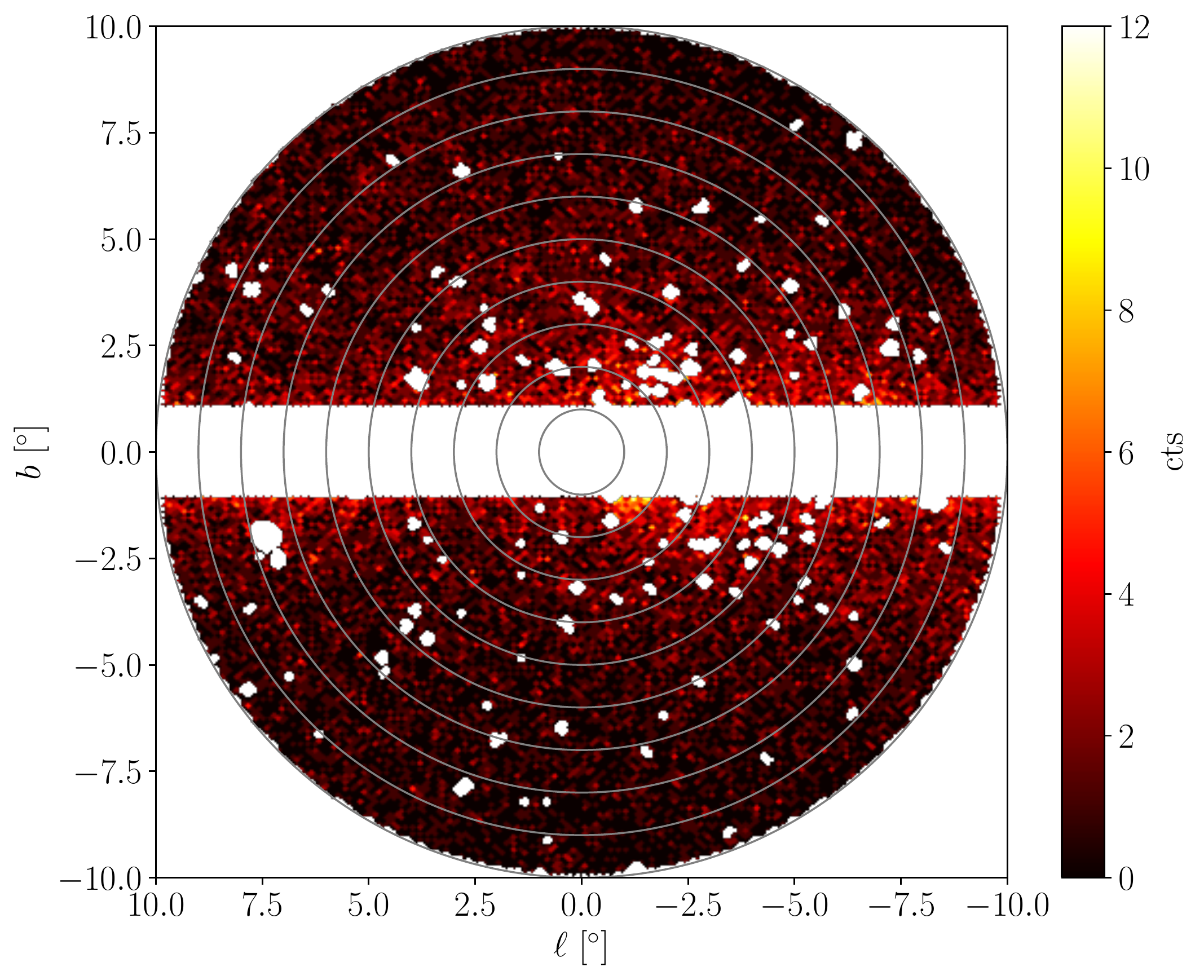} \\
	\end{center}
	\caption{The photon count data used in this work in our ROI, which has the Galactic plane masked at $|b| \geq 1^\circ$ along with 4FGL PSs and pixels more than 10$^\circ$ from the GC.  For illustration the data are summed above 10 GeV. We analyze the data in 9 concentric annuli, as indicated. 
	}
	\label{fig:data}
\end{figure}

\begin{figure}[!t]
	\begin{center}
		\includegraphics[width=0.49\textwidth]{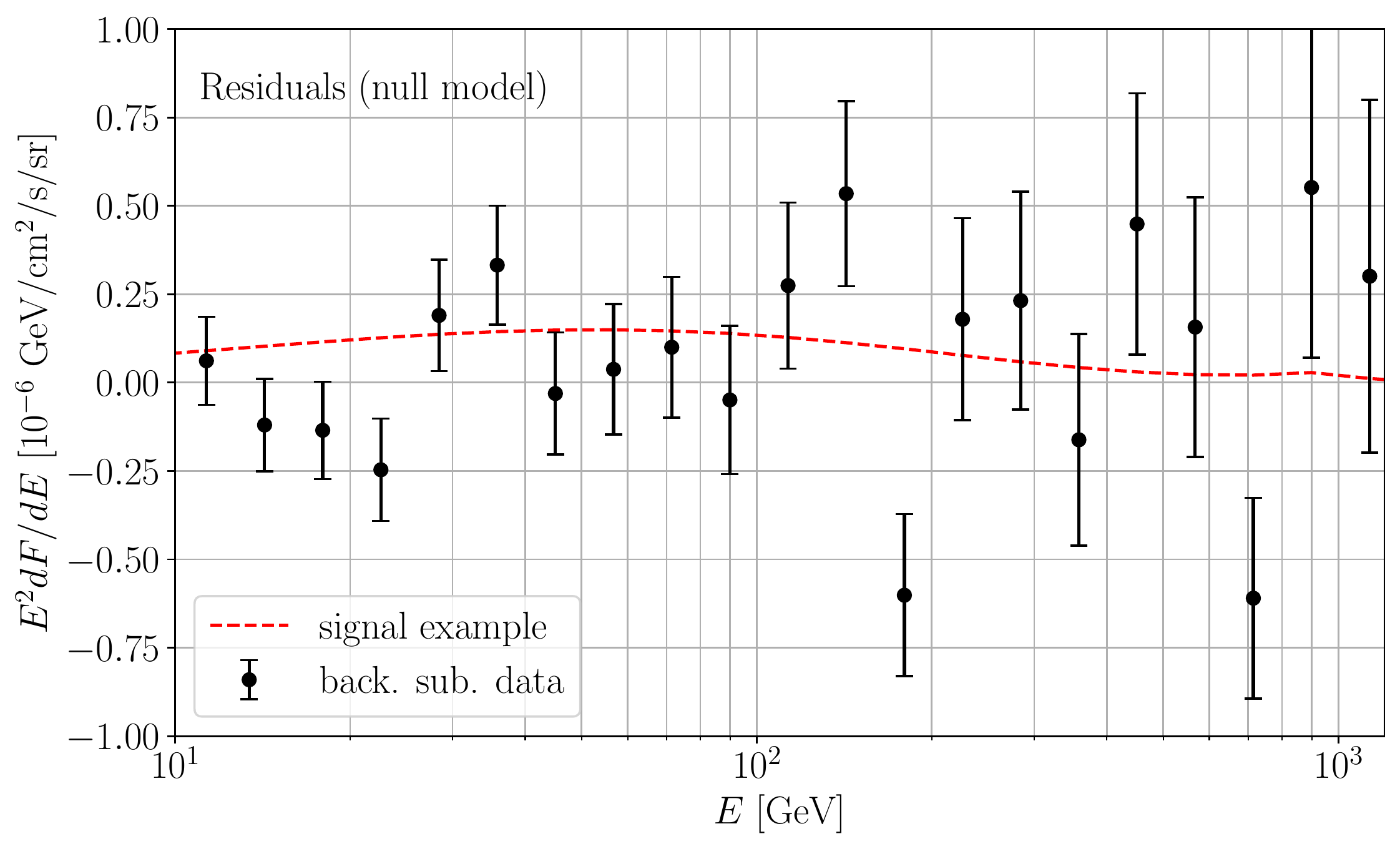}
	\end{center}
	\caption{ The data with the best-fit null hypothesis model subtracted and then summed over all annuli.  For reference we illustrate  a higgsino-like signal with $m_\chi = 1.1$ TeV, $\langle \sigma v \rangle = 5 \times 10^{-26}$ cm$^3/$s, and an NFW DM profile. 
	}
	\label{fig:spectrum}
\end{figure}

\noindent
{\bf Data reduction and analysis.---}
We reduce 722 weeks of Pass 8 {\it Fermi} gamma-ray data with SOURCE selection criterion taken between August 4, 2008 and June 10, 2022 with the recommended quality cuts \mbox{\texttt{DATAQUAL>0}} and {\mbox{\texttt{LATCONFIG==1}}} along with \texttt{zenithangle} less than 90$^\circ$.  We include the top 3 of 4 quartiles of the data as ranked by the point spread function (PSF). As in~\cite{Cohen:2016uyg}, we initially bin the data into 40 logarithmically-spaced energy bins between 200 MeV and 2 TeV, and we bin spatially using \texttt{HEALPIX}~\cite{Gorski:2004by} with \texttt{nside=512}.  However, in our analysis we only analyze data starting in the 18$^{\rm th}$ energy bin, with minimum energy $\sim 10.02$ GeV, since our signal peaks at higher energies and since the lower energies are more contaminated by Galactic diffuse emission.  Starting at 10 GeV we also mostly avoid the {\it Fermi} Galactic Center Excess, which is an excess of $\sim$GeV gamma-rays observed near the GC~\cite{Goodenough:2009gk,Hooper:2010mq,Hooper:2011ti,Abazajian:2012pn,Daylan:2014rsa,Fermi-LAT:2017opo};  the excess has not been found to extend above 10 GeV with our Galactic emission model~\cite{Linden:2016rcf}.   We include energies up to the DM mass $m_\chi$, as the signal spectrum has no support beyond that. 

Our region of interest (ROI) for the analysis is that within 10$^\circ$ of the GC, with the Galactic plane masked ($|b| \geq 1^\circ$) in addition to a 4FGL point source (PS) mask~\cite{Fermi-LAT:2019yla}.  The PS mask is constructed through the following procedure. First, we use~\cite{Fermi-LAT:2019yla} to construct a PS model in our first analysis energy bin in terms of predicted photon counts accounting for the detector response and, in particular, the PSF. Then, we find the brightest pixel, as ranked by predicted PS emission, within 30$^\circ$ of the GC. We mask all pixels that have a predicted photon flux from PSs as low as $0.5\%$ of that in the brightest pixel. This masking approach preferentially masks pixels from bright sources, as opposed to more conventional masks based off of PSF containment that may under- (over-) mask bright (dim) sources.  We then further divide our ROI into 9 concentric annuli, going out to $10^\circ$ from the GC starting at $1^\circ$, with angular spacings of $1^\circ$. We stack and analyze the spectral data in each of these annuli independently.  
The photon counts in our analysis ROI above 10 GeV are illustrated in the top panel of Fig.~\ref{fig:data}.  Even after the Galactic plane mask, more photons are observed near the GC and the Galactic plane from diffuse emission within the Milky Way.

We model the spectral data in each annulus 
under the null hypothesis using a linear combination of: (i) the spectral template derived from the {\it Fermi} Galactic emission model \texttt{gll\_iem\_v07} (\texttt{p8r3}), reprocessed for our data set and selection criterion; (ii) the 4FGL PS spectral template appropriate for our ROI; and (iii) the isotropic diffuse emission appropriate for our Galactic emission model and data set.
PS and isotropic emission, however, are sub-dominant compared to Galactic diffuse emission, as illustrated in Supplementary Material (SM) Fig.~\ref{fig:model}. As shown in the SM, not including the PS and isotropic templates leads to nearly identical results.  We define the ensemble of null-hypothesis spectral models (Galactic emission, PS, and isotropic) as the background templates. 
In each radial bin we construct a likelihood to constrain the spectral model, which consists of the background templates along with the signal template that is discussed shortly, by taking the product of the Poisson probabilities to observe the data counts in each energy bin given the model prediction. The background templates are given individual nuisance parameters that rescale the overall normalization of that template; we require the nuisance parameters to be positive. At a given DM mass $m_\chi$ and in a given radial annulus, we construct the profile likelihood for the annihilation cross-section $\langle \sigma v \rangle$ profiling over the background nuisance parameters. We then construct the joint profile likelihood for $\langle \sigma v \rangle$, at fixed $m_\chi$, by taking the product of the profile likelihoods over all radial bins.  We use the joint profile likelihood to constrain the signal model. (See the SM for details, along with an alternative analysis that incorporates spatial information into the likelihood.)  

In Fig.~\ref{fig:spectrum} we show the background-subtracted counts data, with the best-fit null hypothesis model, summed over all annuli up through 1.1 TeV for an analysis looking for a higgsino with $m_\chi = 1.1$ TeV.  
The data are largely consistent with the null hypothesis. Note that this figure is for illustrative purposes only and is not used in the analysis, which treats the radial bins separately.  Our sensitivity is dominated by the energy bins less than around 100 GeV, as we further illustrate in the SM; this should be contrasted with the sensitivity of upcoming experiments like CTA that will probe thermal higgsino DM at energies near a TeV but lose sensitivity below $\sim$100 GeV.  Our inclusion of photons with energies between $\sim$10--100 GeV is what makes us competitive with CTA, even though CTA will have a much larger effective area than {\it Fermi}.

\begin{figure*}[!htb]
	\begin{center}
		\includegraphics[width=0.49\textwidth]{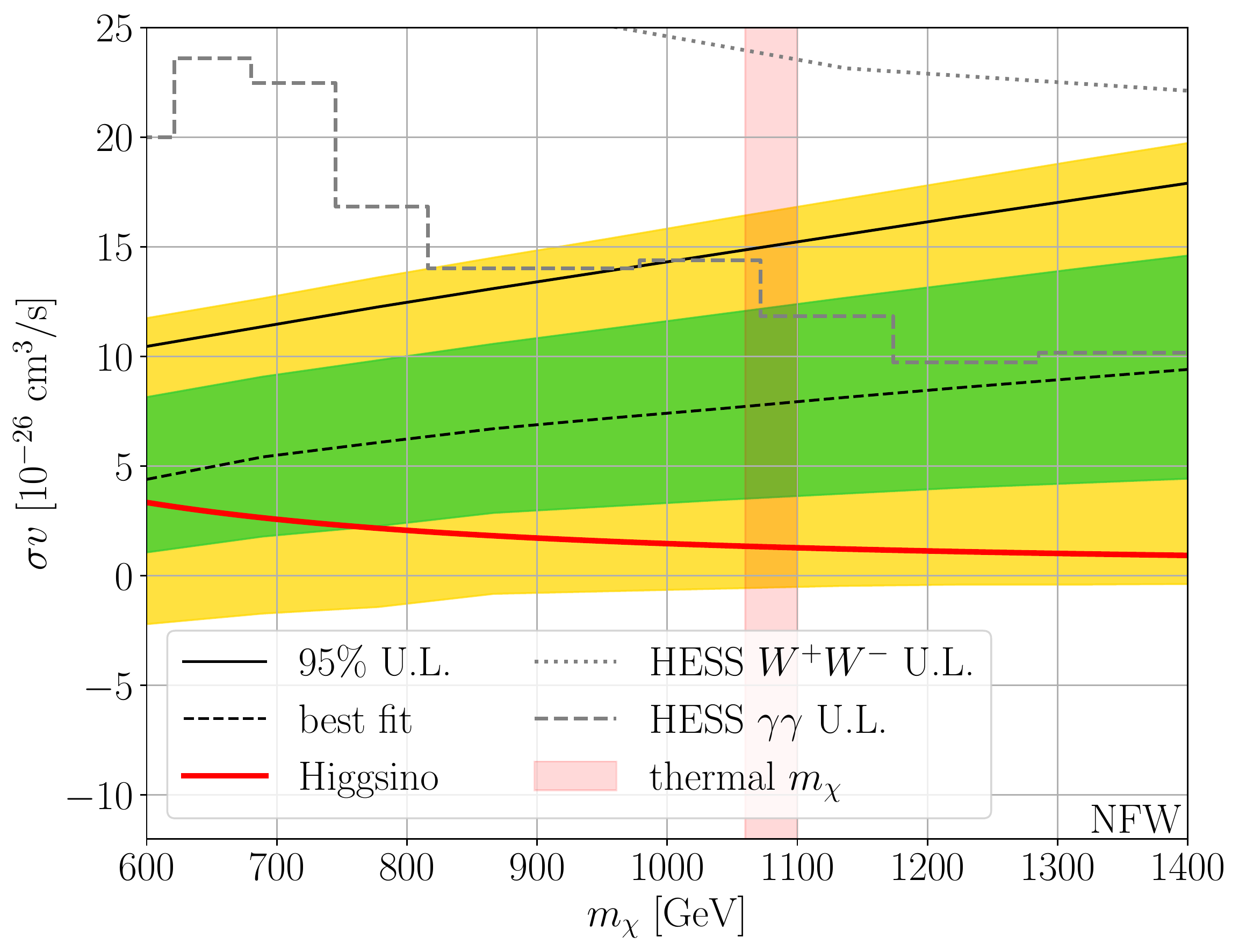}
		\includegraphics[width=0.49\textwidth]{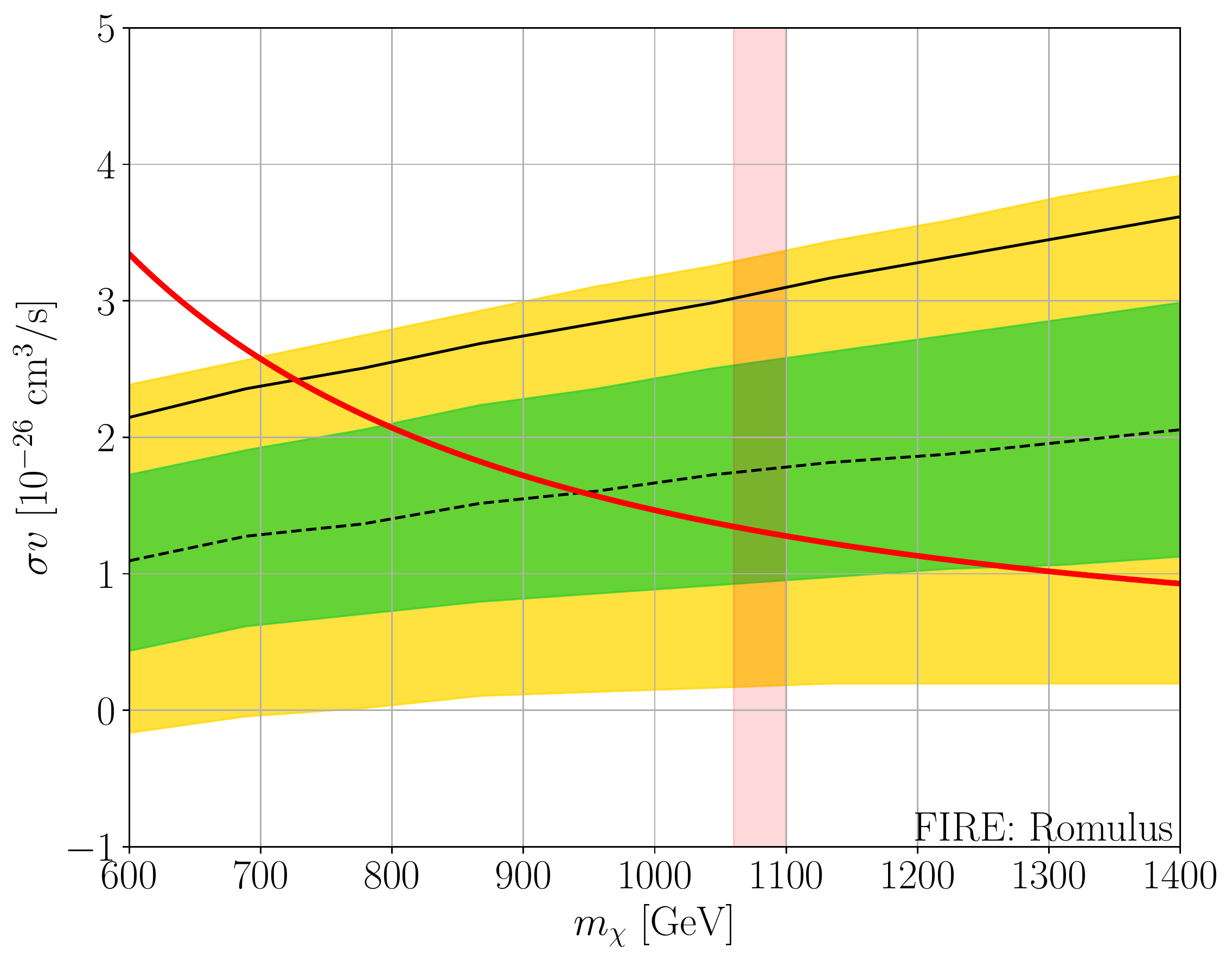}
	\end{center}
	\caption{The best-fit annihilation cross-section for higgsino-like DM as a function of the DM mass for our fiducial analysis assuming an NFW DM profile (left) and the FIRE-2 \texttt{Romulus} profile (right), which gives the best-fit to the data of all profiles considered. We illustrate the best-fit, 1$\sigma$ (green) and 2$\sigma$ (gold) confidence intervals for the recovered cross-section, in addition to the 95\% one-sided upper limit.  We compare our results to the expected higgsino annihilation cross-section (red) and to the 95\% upper limits from the H.E.S.S. searches for annihilation to $W^+W^-$ and gamma-ray lines.  The $m_\chi$ range, accounting for uncertainties, where higgsinos make up the correct DM abundance in the standard thermal cosmology is shaded (thermal $m_\chi$).  For the \texttt{Romulus} profile the recovered cross-section is consistent with the expected cross-section for the thermal higgsino within 1$\sigma$ (green band) and inconsistent with the null hypothesis of no higgsino at $\sim$2$\sigma$, as illustrated by the gold band.  Note that $\langle \sigma v \rangle$ is allowed to be both positive and negative in the analysis, even though negative cross-sections are unphysical.  For all $m_\chi$ we assume that higgsinos make up all of the observed DM.  }
	\label{fig:results}
\end{figure*}

\noindent
{\bf Results.---} To interpret the data in the context of the higgsino model we need to compute the gamma-ray spectrum $dN/dE$ per annihilation from the decays of the unstable particles produced during higgsino annihilation.  The dominant annihilation channels are $\chi \chi \to W^+ W^-$ and $\chi \chi \to Z Z$.  At $m_\chi = 1.1$ TeV the branching ratio to $W$ ($Z$) pairs is $\sim$60\% ($\sim$40\%).  Since the higgsino annihilates through its electroweak interactions, for a given $m_\chi$ there is a fixed and calculable $\langle \sigma v\rangle$.  This annihilation cross-section is illustrated in Fig.~\ref{fig:results} as a function of $m_\chi$. Note that there is minor dependence on the mass splittings between higgsino states, though as we show in the SM these differences do not qualitatively affect our results.  The $dN/dE$ for annihilations to $W$ and $Z$ pairs are calculated using \texttt{PPPC 4 DM ID}~\cite{Cirelli:2010xx}.   When constraining the higgsino model we treat the overall cross-section $\langle \sigma v\rangle$ as a free parameter, which could be negative, but with the branching ratio to $W$ and $Z$ pairs fixed.

In addition to the spectrum per annihilation we also need to know the astrophysical $J$-factor, $J \equiv \int ds \rho^2_{\rm DM}(s)$, in our ROI in order to compute the expected signal in our radial annuli.  Here, $\rho_{\rm DM}(s)$ is the DM density along the line-of-sight, parameterized by the distance $s$ from Earth.  Our benchmark DM profile is the spherically-symmetric Navarro–Frenk–White (NFW)~\cite{Navarro:1995iw,Navarro:1996gj} profile, normalized to produce a local DM density $\rho_{\rm DM}(s=0)  = 0.4$ GeV$/$cm$^3$, with a scale radius $r_s = 20$ kpc, and with the distance from the GC to the Sun of $r_\odot = 8.23$ kpc~\cite{2022arXiv220412551L}.  Our local DM density choice is motivated by the recent review~\cite{deSalas:2020hbh}, which concludes that local DM density measurements, from analyses of stellar motions perpendicular to the disk within a few kpc, tend to favor the range $0.4-0.6$ GeV$/$cm$^3$, broadly consistent with rotation curve data, though one should keep in mind that the true local DM density may be slightly larger or smaller than our choice. Furthermore, the scale radius is currently poorly constrained, such that $r_s = 20$ kpc represents a reasonable choice as opposed to a value strongly preferred by data.  

Given the ad-hoc nature of the NFW profile, and that it is motivated by DM-only $N$-body simulations, a potentially more promising approach to computing the $J$-factor profiles is to use the results for Milky Way analogue galaxies in hydrodynamic cosmological simulations that include baryonic effects.  Towards that end, we also compute the $J$-factor profiles in 12 FIRE-2 zoom-in Milky Way analogue galaxies~\cite{Hopkins:2017ycn}, using simulation outputs provided in~\cite{2022MNRAS.513...55M}. The FIRE-2 simulations are state-of-the-art hydrodynamic simulations, which provide the highest angular resolution to-date for $J$-factor profiles, that incorporate {\it e.g.} stellar feedback and radiative transfer amongst baryons, which dominate the inner potential wells of Milky Way sized galaxies, in addition to gravitational dynamics. Six of these twelve galaxies, including \texttt{Romulus} and \texttt{Romeo} which we discuss more below, were evolved in pair configurations to mimic the interactions between the Milky Way and Andromeda. The Milky Way analogues are chosen to have stellar masses in the range $(3,11)\times 10^{10}$ $M_\odot$ with virial masses in $(0.9,1.8) \times 10^{12}$ $M_\odot$~\cite{2022MNRAS.513...55M}.  The particle masses and positions were then adjusted in~\cite{2022MNRAS.513...55M} such that the local DM density is $0.38$ GeV$/$cm$^3$ at the distance to the Sun $r_\odot = 8.3$ kpc, which are similar to our fiducial values for the NFW profile.  We then compute the azimuthally-averaged $J$-factors in our ROI annuli; these $J$-factors are compared to those from the NFW profile in the SM Fig.~\ref{fig:J-factor}. 

In the left panel of Fig.~\ref{fig:results} we show the results of our analysis of the {\it Fermi} data interpreted for higgsino DM using the NFW DM profile. We illustrate the best-fit cross-section, along with $1$ and $2\sigma$ significance containment intervals, as functions of the higgsino mass, assuming that at each mass the higgsino makes up all of the DM (see SM Fig.~\ref{fig:sf} for our results assuming a subfraction of the DM). Our one-sided 95\% upper limit is also illustrated. For the fiducial NFW profile we are unable to exclude the higgsino cross-sections over the mass range shown. However, our upper limit is weaker than expected due to a slight statistical preference for the signal model over the null model.  At the thermal mass the local significance in favor of the signal model is $\sim$2$\sigma$ (see SM Fig.~\ref{fig:TS} for the discovery test statistic as a function of mass). Note that in Fig.~\ref{fig:spectrum} we illustrate the higgsino model prediction relative to the background-subtracted and fully-stacked data for a reference cross-section.  

The FIRE-2 $J$-factor profiles are typically enhanced relative to that of the NFW model due to adiabatic contraction, as illustrated in SM Fig.~\ref{fig:J-factor}, though there is significant spread over the 12 realizations. In the right panel of Fig.~\ref{fig:results} we show our results interpreted in the context of the FIRE-2 halo, \texttt{Romulus}, with the largest $J$-factor (the other halos are illustrated in the SM).  Note that we use the FIRE-2 naming conventions for the Milky Way analogue galaxies~\cite{Hopkins:2017ycn}.  The FIRE-2 halo profiles lead to comparable discovery significances compared to the NFW analysis, as illustrated in SM Fig.~\ref{fig:TS}, with \texttt{Romulus} providing the best fit. Intriguingly, with the \texttt{Romulus} profile and multiple other FIRE-2 profiles the excess in favor of the signal model has a best-fit cross-section consistent with the higgsino model at the thermal mass.
Over the ensemble of 12 FIRE-2 $J$-factor profiles that we consider, the best-fit $\langle \sigma v\rangle$ for $m_\chi \approx 1.08$ TeV ranges from $1.7 \times 10^{-26}$ cm$^3/$s to $7.3 \times 10^{-26}$ cm$^3/$s, with the median value of $3.4 \times 10^{-26}$ cm$^3/$s; the higgsino cross-section at this mass is $\langle \sigma v\rangle \approx 1.3 \times 10^{-26}$ cm$^3/$s.  The \texttt{Romulus} profile leads to the best-fit cross-section at $m_\chi \approx 1.08$ TeV of $\langle \sigma v\rangle = 1.7 \pm 0.8 \times 10^{-26}$ cm$^3/$s.  The \texttt{Romeo} halo may be the most Milky Way-like, due to the similarities of its thick disk, circular velocity, and stellar mass to the Milky Way; using this halo we recover a cross-section  $\langle \sigma v\rangle = 1.9 \pm 1.0 \times 10^{-26}$ cm$^3/$s at the thermal higgsino mass. 

In Fig.~\ref{fig:results} (left) we show the 95\% upper limit from an analysis of H.E.S.S. data looking for continuum emission above $\sim$200 GeV associated with $\chi \chi \to W^+ W^-$~\cite{Montanari:2021yic}.  H.E.S.S. is less sensitive to $\chi \chi \to ZZ$, since annihilation to $Z$ pairs produces significantly fewer photons above $\sim$200 GeV than annihilation to $W$ pairs, so to convert the results presented in~\cite{Montanari:2021yic} to higgsino-like DM limits we use only the $W^+W^-$ result (additionally, H.E.S.S. does not present results for annihilation to $Z$ pairs).  Furthermore,~\cite{Montanari:2021yic} uses an ROI ranging from 0.5$^\circ$ to 2.9$^\circ$ from the GC and assumes an Einasto profile; we rescale their results to those appropriate for an NFW profile in the left panel of Fig.~\ref{fig:results}.  The H.E.S.S. upper limits are less constraining than our upper limits across the mass range shown. (The FIRE-2 simulations do not have resolution down to $\sim$0.5$^\circ$, so we do not show the results from ~\cite{Montanari:2021yic} in the right panel of Fig.~\ref{fig:results}.)

Constraints on higgsino DM using H.E.S.S. searches for gamma-ray lines, which are from the loop-suppressed processes $\chi \chi \to \gamma\gamma$ and $\chi \chi \to Z \gamma$, are also relevant, though the hard-photon spectrum is affected by electroweak radiative effects~\cite{Beneke:2019gtg}.  We translate the H.E.S.S. gamma-ray line limits in~\cite{HESS:2018cbt}, which were computed using the same ROI as in their continuum $W^+W^-$ search described above, to limits on the total annihilation cross-section using the NLL' calculation for the energy spectrum near the gamma-ray endpoint in~\cite{Beneke:2019gtg}; the recasted limit is illustrated in the left panel of Fig.~\ref{fig:results}.  The H.E.S.S. upper limit surpasses our upper limit at large masses, though it should be kept in mind that (i) the H.E.S.S. analysis is significantly closer to the GC than ours and thus the comparison relies on the possibly incorrect shape of the NFW profile, and (ii) the result in~\cite{Beneke:2019gtg} may be subject to ${\mathcal O}(1)$ uncertainties when applied to the H.E.S.S. analysis because the energy binning used in the analysis does not directly match the assumptions in the calculations in~\cite{Beneke:2019gtg}. (See also~\cite{Baumgart:2018yed}.) 

\noindent
{\bf Discussion.---}In this work we set the strongest constraints to-date on higgsino-like DM that annihilates to $W^+W^-$ and $ZZ$ using nearly the entire {\it Fermi} data set collected since the mission's launch in 2008. We search for the continuum gamma-ray emission above 10 GeV associated with the decays of these massive vector bosons. Interpreting our results in the context of the NFW profile our upper limits on the annihilation cross-section are world-leading but do not constrain the predicted higgsino cross-section.  On the other hand, hydrodynamic simulations show that the DM profile may be contracted in the inner Galaxy, enhancing the annihilation rate.

Our upper limits are weaker than expected because of a modest ($\sim$2$\sigma$) preference for the higgsino model over the null hypothesis of background-only emission. The best-fit cross-section is consistent with the expected higgsino cross-section for a thermal ($m_\chi \approx 1.1$ TeV) higgsino making up all of the DM for multiple FIRE-2 DM density profiles. Given that higgsino DM is well motivated from supersymmetry and, in particular, Split-SUSY type models that seem to be favored by the lack of observed superpartners to-date at the LHC in addition to the observed Higgs mass and precision Grand Unification, the possibility that the data present the first hint for higgsino DM is promising. This possibility will be tested with the upcoming CTA~\cite{Rinchiuso:2020skh}, which should be sensitive to higgsino DM annihilation.

\noindent
{\bf Acknowledgements.---}
{\it 
We thank Dan Hooper, Simon Knapen, Matthew McCullough, Lina Necib, Nick Rodd, Tracy Slatyer, and Tim Cohen for useful discussions, and we thank Daniel McKeown and the FIRE-2 project for providing us with the simulated halo profiles used in our analysis.  J.W.F was supported by a Pappalardo Fellowship. C.D., Y.P., and  B.R.S. were supported  in  part  by  the  DOE  Early Career  Grant  DESC0019225. W.L.X. thanks the Mainz Institute of Theoretical Physics of the Cluster of Excellence PRISMA+ (Project ID 39083149) for its hospitality during completion of part of this work, and is supported by the U.S. Department of Energy under Contract DE-AC02-05CH11231. This research used resources from the Lawrencium computational cluster provided by the IT Division at the Lawrence Berkeley National Laboratory, supported by the Director, Office of Science, and Office of Basic Energy Sciences, of the U.S. Department of Energy under Contract No.  DE-AC02-05CH11231.

}

\bibliography{refs}

\clearpage

\onecolumngrid
\begin{center}
  \textbf{\large Supplementary Material for Higgsino Dark Matter Confronts 14 years of Fermi Gamma Ray Data}\\[.2cm]
  \vspace{0.05in}
  {Christopher Dessert, Joshua W. Foster, Yujin Park, Benjamin R. Safdi, and Weishuang Linda Xu}
\end{center}

\twocolumngrid

\setcounter{equation}{0}
\setcounter{figure}{0}
\setcounter{table}{0}
\setcounter{section}{0}
\setcounter{page}{1}
\makeatletter
\renewcommand{\theequation}{S\arabic{equation}}
\renewcommand{\thefigure}{S\arabic{figure}}
\renewcommand{\theHfigure}{S\arabic{figure}}%
\renewcommand{\thetable}{S\arabic{table}}

\onecolumngrid

This Supplementary Material contains supporting material for the main Letter.  Sec.~\ref{sec:methods} provides additional details of the methods used in our analyses, Sec.~\ref{sec:extended} provides additional results for our fiducial analysis presented in the main Letter, while Sec.~\ref{sec:analysis} gives the results of systematic analysis variations that provide additional support to our main conclusions.

\section{Methods}
\label{sec:methods}

In this section we describe the methods used in our fiducial analysis and in our analysis variations.

\subsection{Data analysis}

As described in the main Letter, the {\it Fermi} data is initially reduced into 40 log-spaced energy bins between 200 MeV and 2 TeV and into spatial maps using the \texttt{HEALPIX} pixelation~\cite{Gorski:2004by} scheme with \texttt{nside=512}.  While reducing the data we also reprocess the \texttt{p8r3} 
Galactic emission model, along with the 4FGL PS model, extended source emission model, and the associated isotropic flux model.  Note that the \texttt{p8r3} model is not recommended for use in searches for extended emission because detected, extended emission regions (specifically, unassociated sources with extension beyond 2$^\circ$) were added in as residuals to the \texttt{p8r3} model. Our signal, corresponding spatially to the $J$-factor profile, is extended beyond the instrument PSF.  However, our fiducial analysis only uses spectral information in the diffuse model and not spatial information.  Note also that there are no significant extended sources in the extended source catalog within our ROI, and in addition the GC Excess below 10 GeV has been successfully observed with the \texttt{p8r3} model~\cite{Linden:2016rcf}.
We also use the \texttt{p8r3} Galactic emission model in our spatial template analysis described below, though ideally it should be verified if {\it Fermi} added any residuals into \texttt{p8r3} that are spatially coincident with our signal template.  In future work it would also be useful to perform such spatial template analyses with Galactic emission models that do not include data-driven residuals.  Note that later in the SM, in Sec.~\ref{sec:dd}, we also present spectral analysis results using a purely data-driven background model that is constructed from data further away from the GC than our signal ROI. 

We perform two types of analyses in this work.  First, our fiducial analysis, which is described in the main Letter, uses spectral information only within 9 annuli centered at the GC (see Fig.~\ref{fig:data} for the geometry).  The second analysis is used as a systematic check in the SM, and it consists of spatial template fits performed independently in each energy bin. We describe both analyses below, starting with our fiducial (spectral) analysis.  Note that in all of our analyses we focus on the GC because it provides the most sensitive probe of DM annihilation; searches using dwarf galaxies, for example, would likely be less sensitive~\cite{Fermi-LAT:2015att}.

Consider the data ${\bm d} = \{n_{i}\}$ stacked over pixels within a given annulus, numbered one through nine with one being the inner-most annulus from 1$^\circ$ to 2$^\circ$.  Here, $n_i$ stands for the number of counts observed in energy bin $i$, with $i$ in principle going from 1 to 40 though in practice we restrict the analysis energy range to energies above 10 GeV (for our fiducial analysis) as well as energies below the DM mass.  In our fiducial analysis we describe the data using a spectral model that predicts the expected number of counts $\mu_i({\bm \theta})$ in energy bin $i$ as a function of the model parameters ${\bm \theta}$.  In our fiducial analysis ${\bm \theta} = \{ A_{\rm sig}, {\bf A_{\rm bkg}}\}$, with $A_{\rm sig}$ the signal model parameter that controls the strength of the signal ({\it i.e.}, the cross-section), at a fixed DM mass, while ${\bf A_{\rm bkg}} = \{ A_{\rm gal}, A_{\rm PS}, A_{\rm iso} \}$ changes the normalization of the associated background model components.  In particular, $\mu_i({\bm \theta}) = A_{\rm sig} \mu^{\rm sig}_i +   A_{\rm gal} \mu^{\rm gal}_i + A_{\rm PS} \mu^{\rm PS}_i + A_{\rm iso} \mu^{\rm iso}_i $, with $\mu^{\rm gal}$ the Galactic diffuse spectrum, $\mu^{\rm PS}$ the residual PS spectrum (which extends beyond our PS mask), and $\mu^{\rm iso}$ the isotropic spectrum as predicted by {\it Fermi} for our event class and diffuse model.  The background parameters are not allowed to take on negative values, though our signal parameter is, and the background parameters are also constrained to be less than five times their expected normalizations to aid with convergence of the global optimizer used to maximize the likelihood.

\begin{figure*}[!htb]
	\begin{center}
	\includegraphics[width=0.49\textwidth]{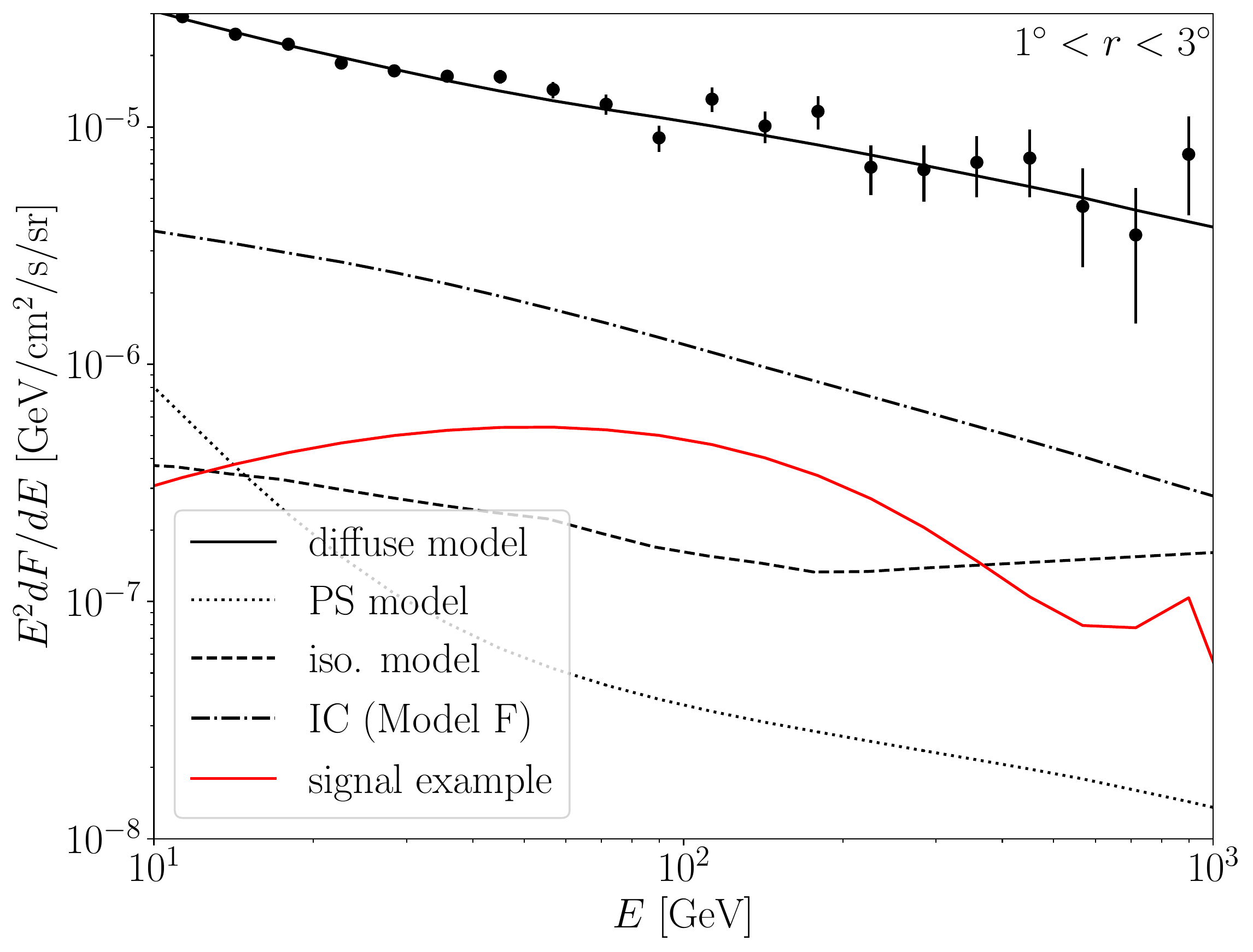}
		\end{center}
	\caption{The spectrum within our first two annuli compared to the spectral templates used in our fiducial analysis, including the \texttt{p8r3} Galactic emission model, the residual PS model, and the isotropic model. The signal model is illustrated for a DM mass of $m_\chi = 1.1$ TeV and a cross-section of $\langle \sigma v \rangle = 5 \times 10^{-26}$ cm$^3/$s.  Note that we do not include line emission in the signal spectrum, with the peak near a TeV arising from electroweak radiative effects in the $W^+W^-$ spectrum.  We also show the IC spectral template used in Sec.~\ref{sec:IC}.}
	\label{fig:model}
\end{figure*}

In Fig.~\ref{fig:model} we show the spectral templates summed over our first two annuli ($1^\circ < r < 3^\circ$) with default normalizations (${\bf A_{\rm bkg}} = {\bf 1}$) and a signal normalization corresponding to $\langle \sigma v \rangle = 5 \times 10^{-26}$ cm$^3/$s for an NFW DM profile and $m_\chi = 1.1$ TeV.
The Galactic diffuse template dominates over PS and isotropic emission.  Indeed, as we show later removing those latter two templates completely does not qualitatively affect our results.  Note that we also illustrate an Inverse Compton (IC) spectral template in Fig.~\ref{fig:model}; this component is not used in our fiducial analysis, but it is used in a systematic test described in Sec.~\ref{sec:IC}.

We constrain the model parameters using the data through a Poisson likelihood
\es{ }{
p({\bm d}| {\bm \theta}) = \prod_i {\mu_i({\bm \theta})^{n_i} e^{-\mu_i({\bm \theta})} \over n_i !} \,,
}
with the product being over energy bins.  In particular, we construct the test statistic (TS) for upper limits
\es{}{
t(A_{\rm sig}) \equiv -2 \left[ \log p({\bm d}| {A_{\rm sig},   {\bf \hat A_{\rm bkg}}}) - \log p({\bm d}| \hat {\bm \theta})\right] \,,
}
where in the second term $\hat {\bm \theta}$ refers to the model parameters that maximize the likelihood while in the first term ${\bf \hat A_{\rm bkg}}$ refers to the background nuisance parameter vector that maximizes the log likelihood at fixed $A_{\rm sig}$.  We assume Wilks' theorem such that the 95\% one-sided upper limit is given by the value of $A_{\rm sig}$ greater than the best-fit value $\hat A_{\rm sig}$ where $t(A_{\rm sig}) \approx 2.71$ (see, {\it e.g.},~\cite{Cowan:2010js}).  The discovery TS for a one-sided test, which determines the significance of the signal model over the null model, is given by $q = t(A_{\rm sig} = 0)$ if ${ \hat A}_{\rm sig} > 0$ and $q = 0$ otherwise.  Note that we allow $A_{\rm sig} <0$, even though such values are unphysical, to ensure that we find the likelihood maximum. 

In the SM we perform an alternate analysis using a spatial template fit instead of a spectral template fit.  In each energy bin we assign independent nuisance parameters for the PS template and the Galactic emission template. The spatial PS template is constructed from the 4FGL catalog along with the instrument PSF, though we treat the overall normalization of the template as a nuisance parameter. In each energy bin we assign independent nuisance parameters to the PS model and to the Galactic emission model. Thus, the likelihood may be written as  
\es{eq:template}{
p({\bm d}| {\bm \theta}) = \prod_{ {\rm energies} \,\, i} \prod_{{\rm pixels}\,\, j} {\mu_{ij}({\bm \theta})^{n_{ij}} e^{-\mu_{ij}({\bm \theta})} \over n_{ij} !} \,,
}
where $n_{ij}$ is the observed number of counts in energy bin $i$ and spatial pixel $j$, while $\mu_{ij}$ is the model prediction in that pixel.  There is still a single signal parameter $A_{\rm sig}$, which -- given a DM mass -- rescales the signal amplitude in each energy bin according to the DM spectrum. The spatial profile of the signal is given by the $J$-factor. The number of nuisance parameters is twice the number of energy bins.

\FloatBarrier
\subsection{Higgsino annihilation signal}

For a nearly-pure higgsino,  the freeze-out abundance under a thermal cosmology is essentially determined by its mass, and thus the requirement that it saturates the total amount of observed DM fixes a precise mass prediction.  Given the cold DM abundance measurement  $\Omega_{\rm CDM} h^2 = 0.12 \pm 0.0012$~\cite{Planck:2018vyg}, the thermal higgsino mass is determined to be $m_{\rm thermal} = 1.08 \pm 0.02 $ TeV~\cite{Bottaro:2022one}.

In the present day, when the higgsino is much colder and Sommerfeld effects become important, the cross-section of its annihilation and branching ratios of its decay products acquire a weak dependence on the two mass splitting parameters: that between its two neutral components and that between the LSP and charged component. In this Letter and associated SM, we present results for the scenario where the charged splitting is fixed to its radiative value 350 MeV and the neutral splitting saturates the up-scattering bound from nuclear recoil experiments, 200 keV~\cite{Baumgart:2014vma,Rinchiuso:2020skh}.     

The differential flux of $\gamma$-rays expected from higgsino annihilation is given as 
\begin{equation}
\frac{{\rm d} F }{{\rm d} E} = \frac{J}{8\pi m_{\chi}^2} \langle \sigma v \rangle \frac{{\rm d}N_\gamma}{{\rm d} E}, \qquad \qquad J(\theta) \equiv \int {\rm d} s \, \rho_{\rm DM}^2 (s, \theta).
\end{equation}

The astrophysical $J$-factor encodes information on the DM distribution within the ROI. 
Note that the higgsino need not constitute the entirety of the observed DM abundance, and indeed the thermal abundance of a higgsino LSP with $m_\chi < 1.1 \,\rm{TeV}$ is a fixed fraction of the observed total, $\Omega_\chi/\Omega_{\rm cdm} \sim (m_\chi/ 1.1 {\rm TeV})^2$. However, this fractional abundance is relatively easily enhanced with non-thermal cosmologies.

In this paper we search for only the continuum part of the higgsino annihilation spectra; while this annihilation does produce a monochromatic line signal that is generally easier to target in searches, the expected strength of this signal is comparatively suppressed even after accounting for Sommerfeld enhancement. The continuum spectra consists of contributions from  $W^+W^-$ and $ZZ$ final states (and a small contribution from the $\gamma Z$ channel weighed by a factor of $1/2$), and the cross sections are computed at tree-level. These cross sections are modified by the mixing of the various higgsino states due to Sommerfeld enhancement, which is parametrized by \begin{equation}
    \langle \sigma v \rangle_{\chi \chi \to XY} =  2 v_{i} \Gamma^{ij}_{XY} v_{j},
\end{equation}
where the mixed-state vector $v_{i}$ is determined by solving the Schr\"odinger equation, whose potential is set by the mass gaps between higgsino states and the electroweak interaction $\alpha_W {\rm e}^{-M r} / r$~\cite{Rinchiuso:2020skh,Cirelli:2007xd,Hisano:2004ds}.

\begin{figure*}[!htb]
	\begin{center}
		\includegraphics[width=0.7\textwidth]{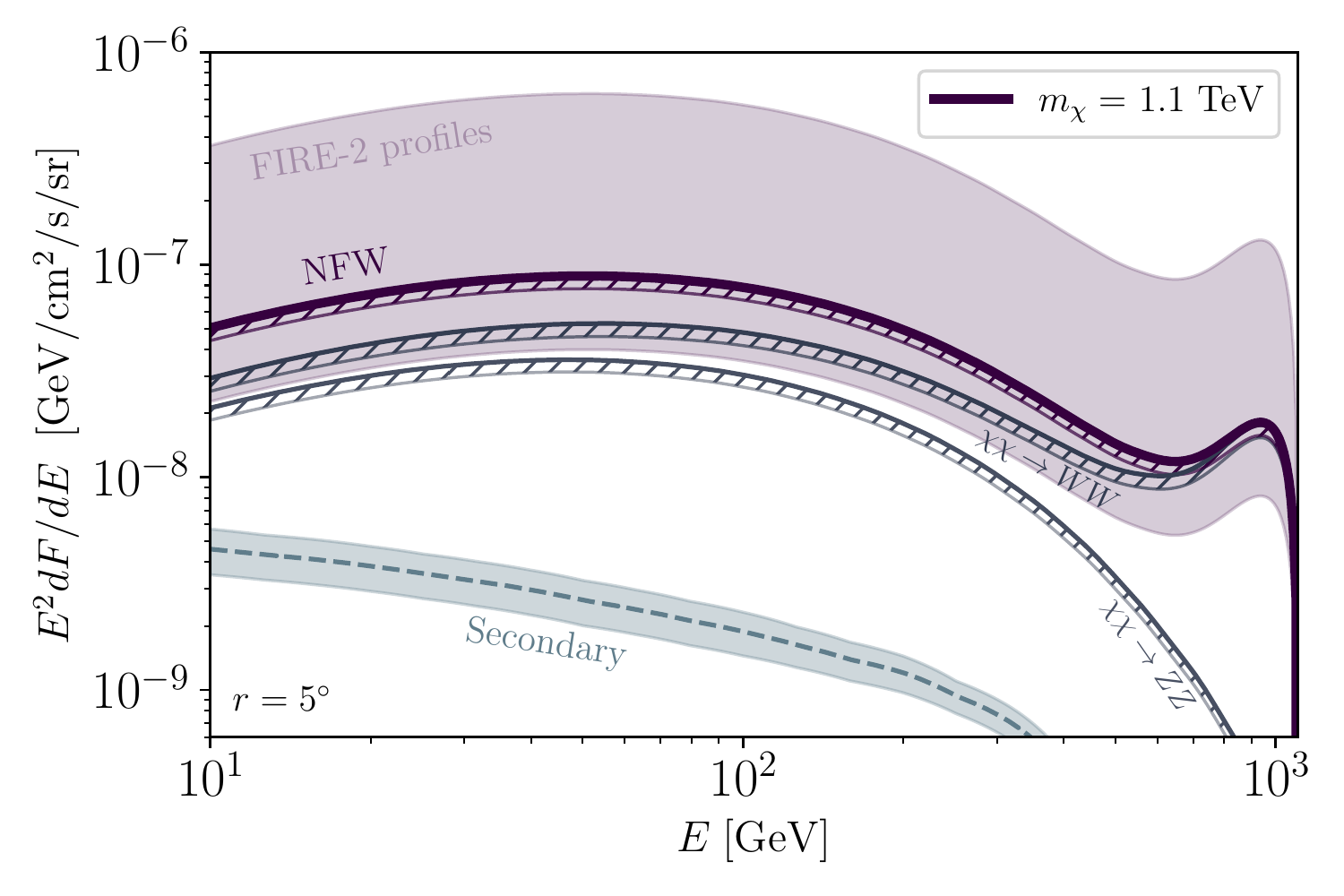}
		\end{center}
	\caption{The expected gamma-ray flux at $r = 5^\circ$ from the GC from prompt annihilation of thermal higgsinos and its breakdown into the contributing $WW$ and $ZZ$ channels~\cite{Cirelli:2010xx}. The fiducial spectra, shown as a solid line, assumes a higgsino with mass gaps $\Delta m = 200$ keV, $\Delta m_+ = 350$ MeV, and an NFW halo profile with $r_s = 20$ kpc. The purple shaded region represents the expected variance from DM halo profile across 12 FIRE-2 hydro simulations~\cite{2022MNRAS.513...55M}, the top edge of which corresponds to the profile of {\tt Romulus}, which has the most enhanced $J$-factor of the FIRE-2 suite. Much subdominant to this is the expected variation in spectra due to mass gaps of the various higgsino states, represented by the hatched shaded regions; here, we consider scenarios where $\Delta m \in [200{\rm keV}, 2{\rm GeV}]$ and $\Delta m_+ \in [350{\rm MeV}, 2{\rm GeV}]$. Finally, an additional subdominant contribution is given by the secondary emission from inverse Compton, Bremsstrahlung, and synchotron processes, with an associated shaded region indicating uncertainties from propagation and magnetic field models~\cite{Buch:2015iya}.  We do not include this contribution in our analysis.}
	\label{fig:theory_spectra}
\end{figure*}

The injection spectra of these annihilation modes into gamma-rays, ${\rm d}N_\gamma/{\rm d} E$, are obtained via the prescription of {\tt PPPC 4 DM ID}~\cite{Cirelli:2010xx}, where radiative electroweak corrections have been accounted for and final $W/Z$ polarizations averaged over. Our main analysis assumes a signal that consists of only the prompt emission in order to minimize dependency on propagation functions, and we do not expect secondary contributions to have a significant impact on our analysis. The expected primary flux from a 1.1 TeV thermal higgsino for a fiducial halo profile is shown in Figure~\ref{fig:theory_spectra}.

\section{Extended Results for the Fiducial Analysis}
\label{sec:extended}

\begin{figure*}[!htb]
	\begin{center}
		\includegraphics[width=0.49\textwidth]{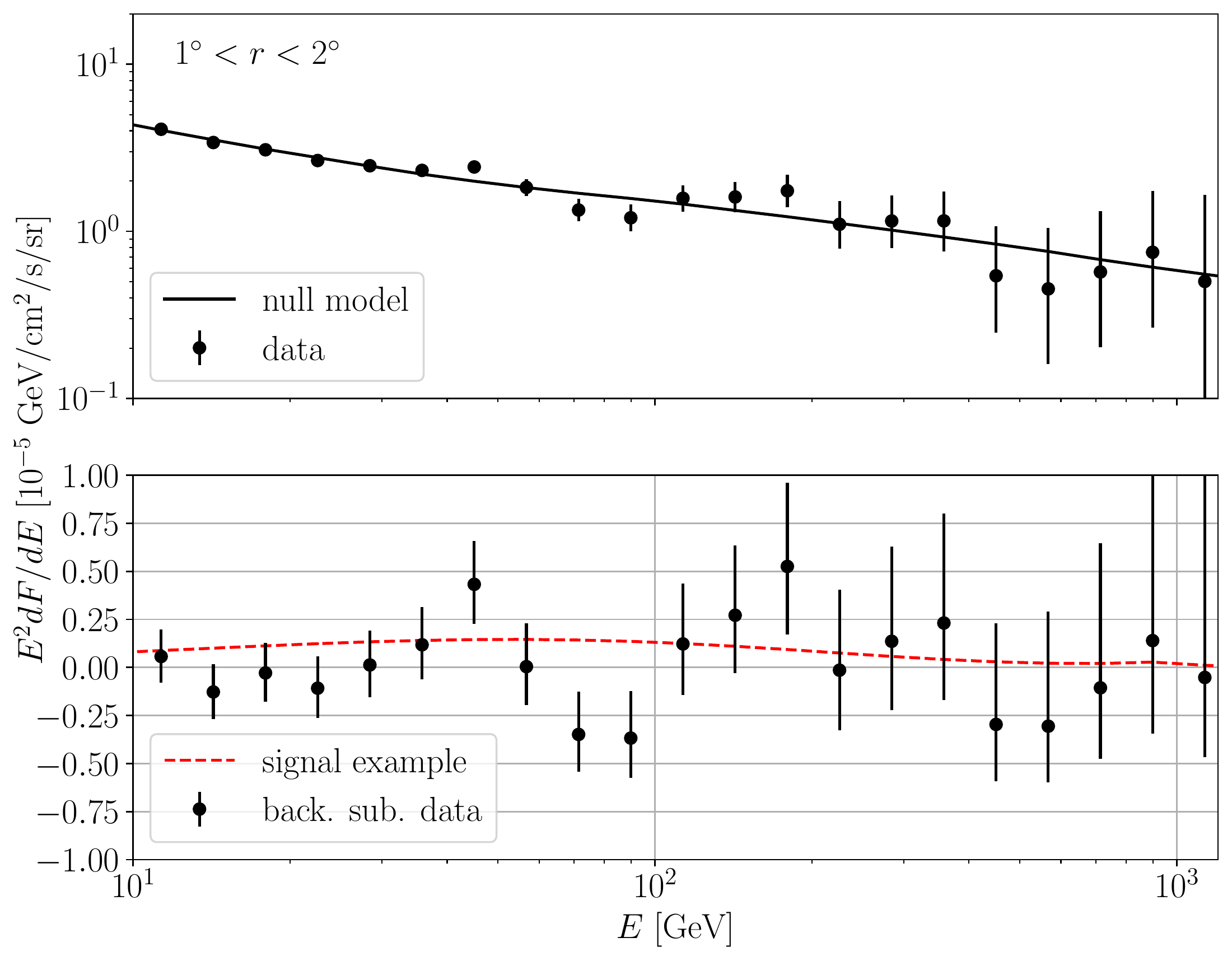}
		\includegraphics[width=0.49\textwidth]{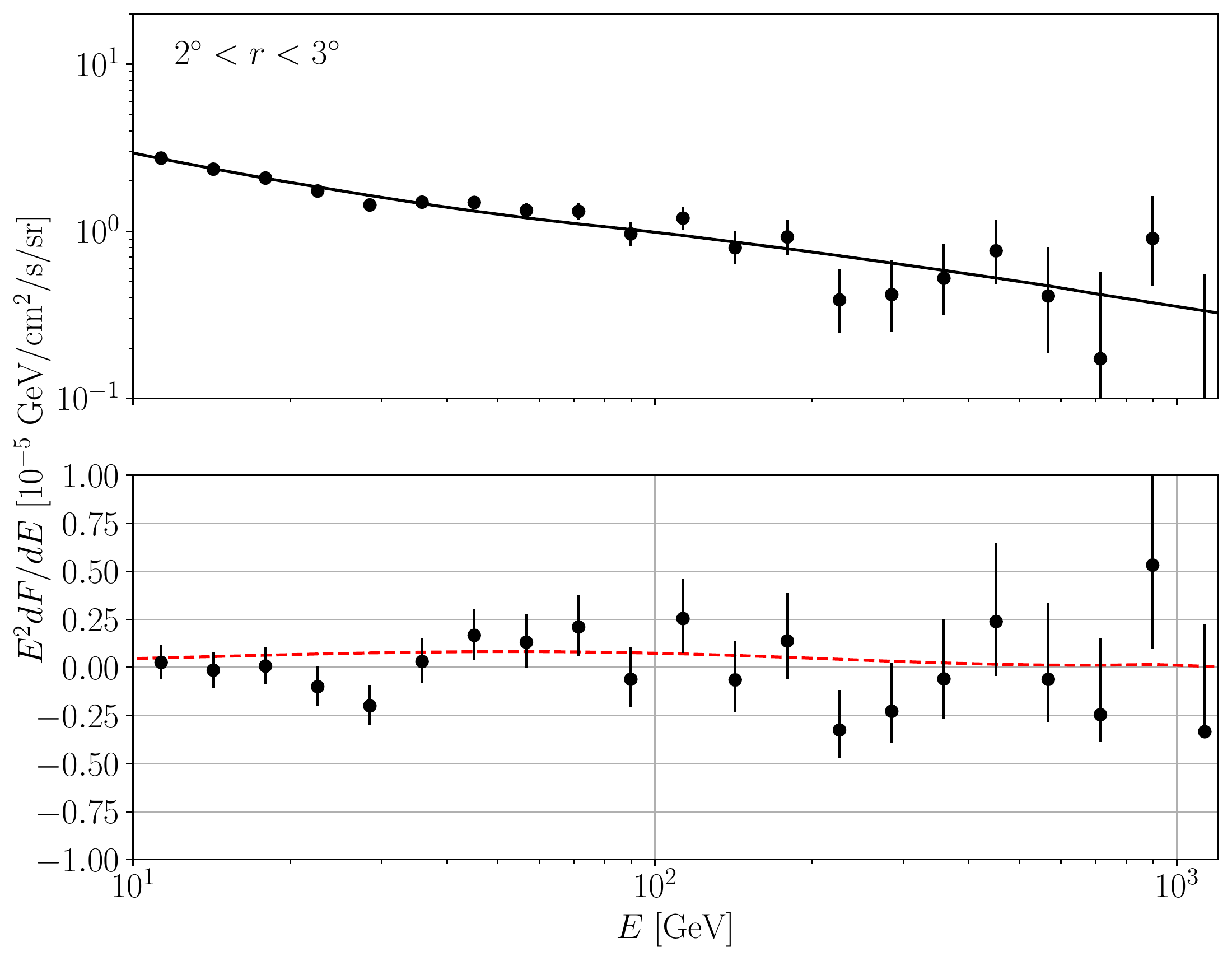}
		\includegraphics[width=0.49\textwidth]{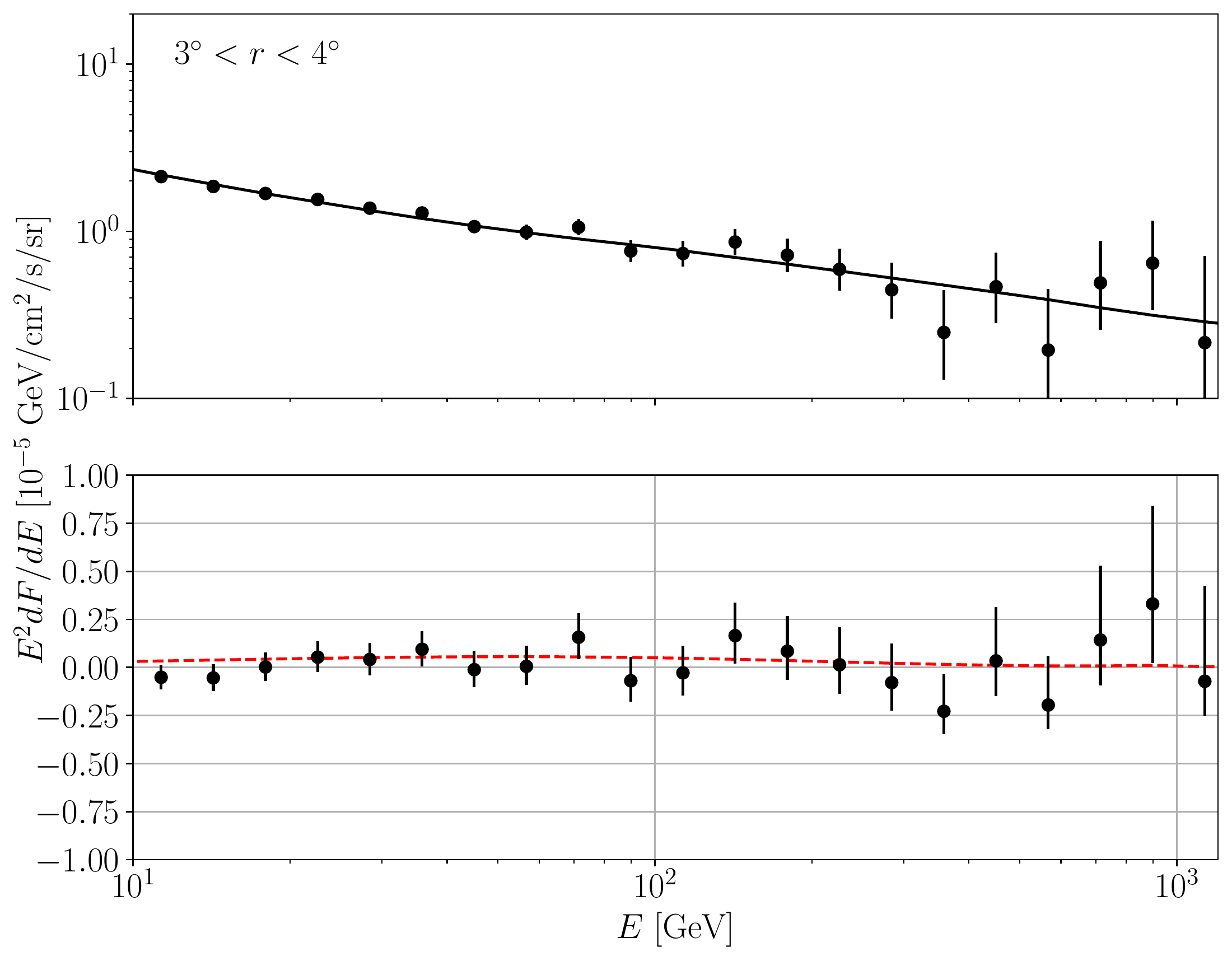}
		\includegraphics[width=0.49\textwidth]{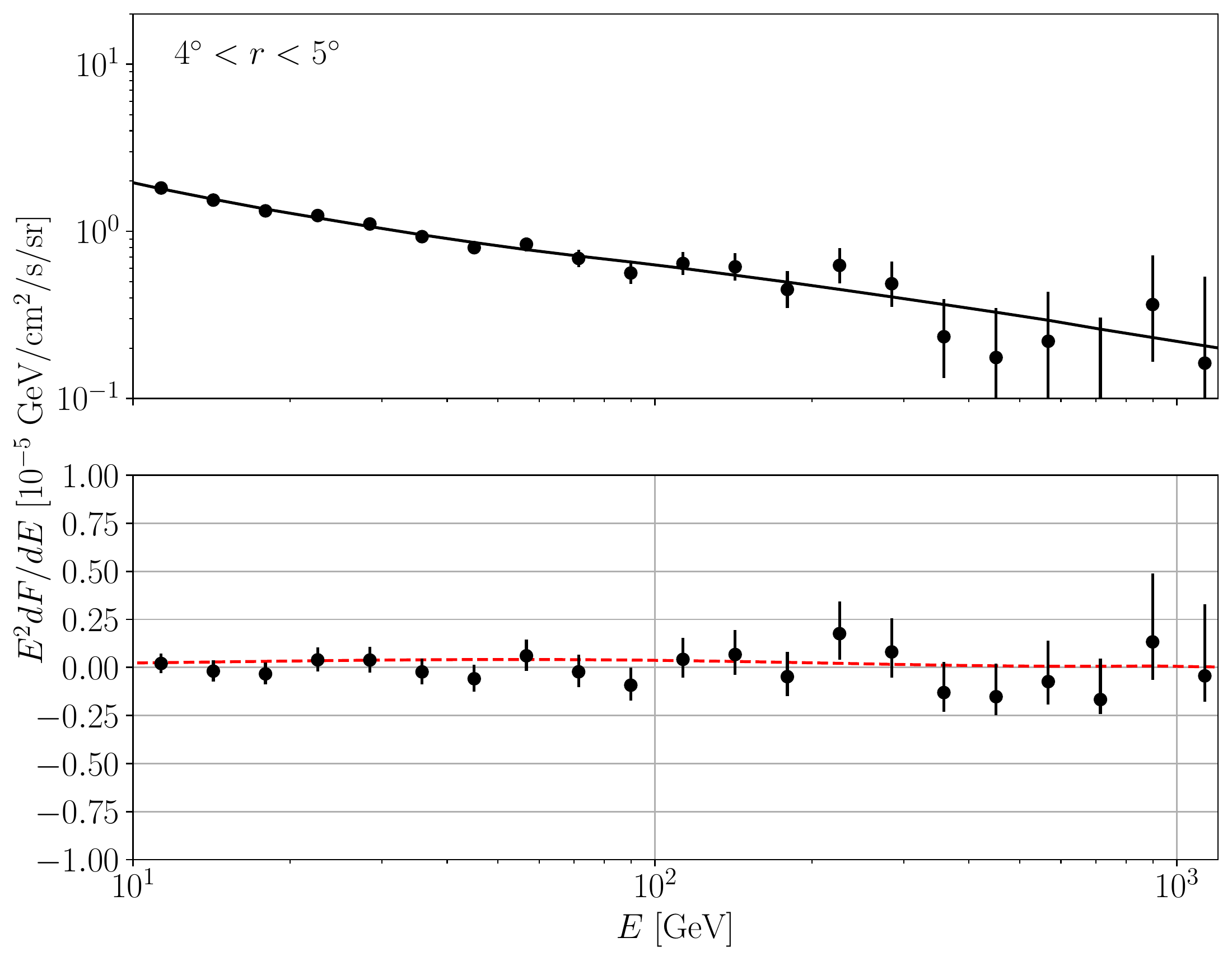}
		\end{center}
	\caption{(top panels) The observed flux in the first four annuli, as indicated, along with the best-fit null model that is the \texttt{p8r3} Galactic emission model plus isotropic and PS emission.  (bottom panels) The residuals -- data minus best-fit background -- constructed from the information in the top panels.  We illustrate an example signal for a higgsino with $m_\chi = 1.1$ TeV and $\langle \sigma v \rangle = 5 \times 10^{26}$ cm$^3/$s for the NFW DM profile.  }
	\label{fig:residuals_1}
\end{figure*}

\begin{figure*}[!htb]
	\begin{center}
		\includegraphics[width=0.49\textwidth]{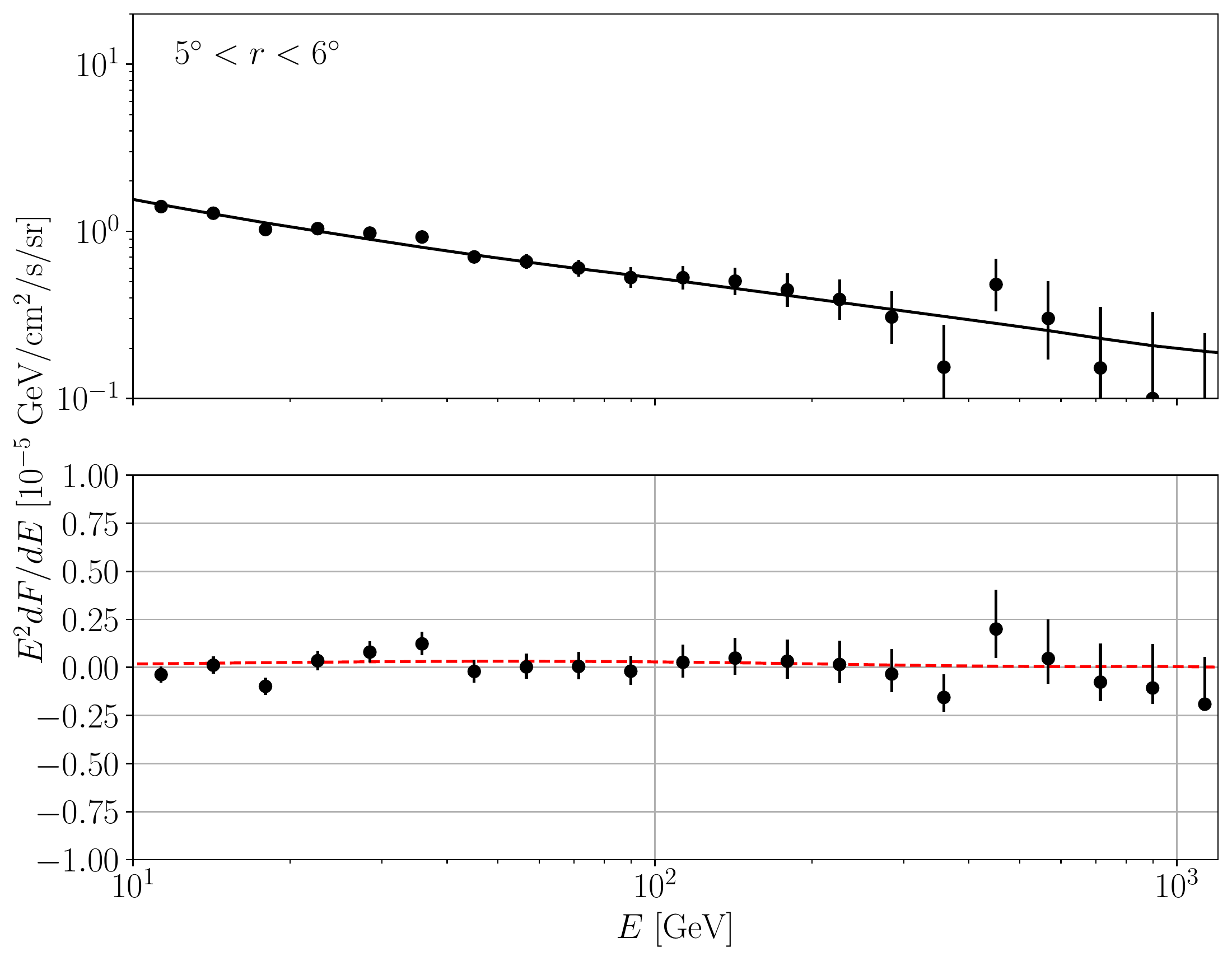}
		\includegraphics[width=0.49\textwidth]{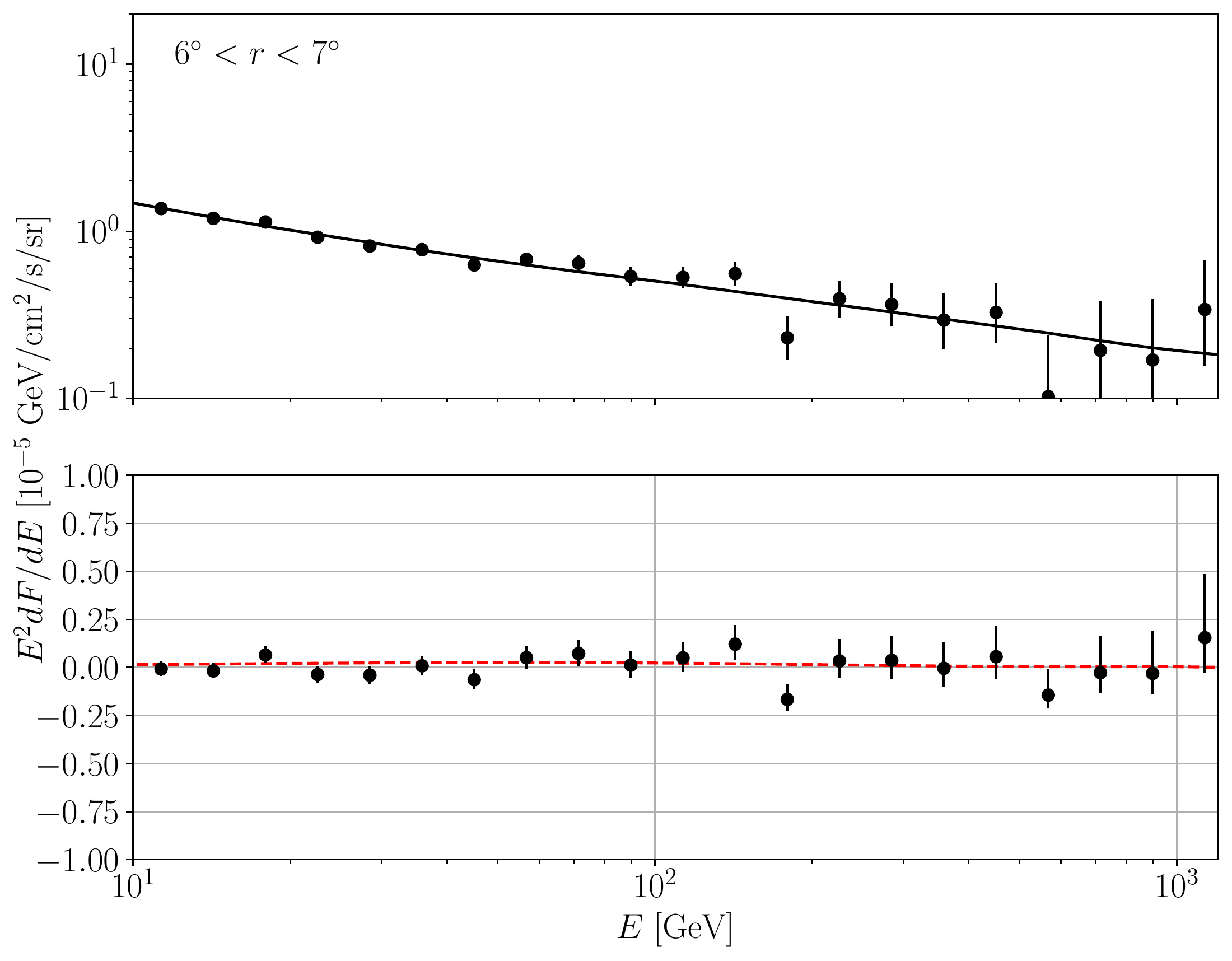}
		\includegraphics[width=0.49\textwidth]{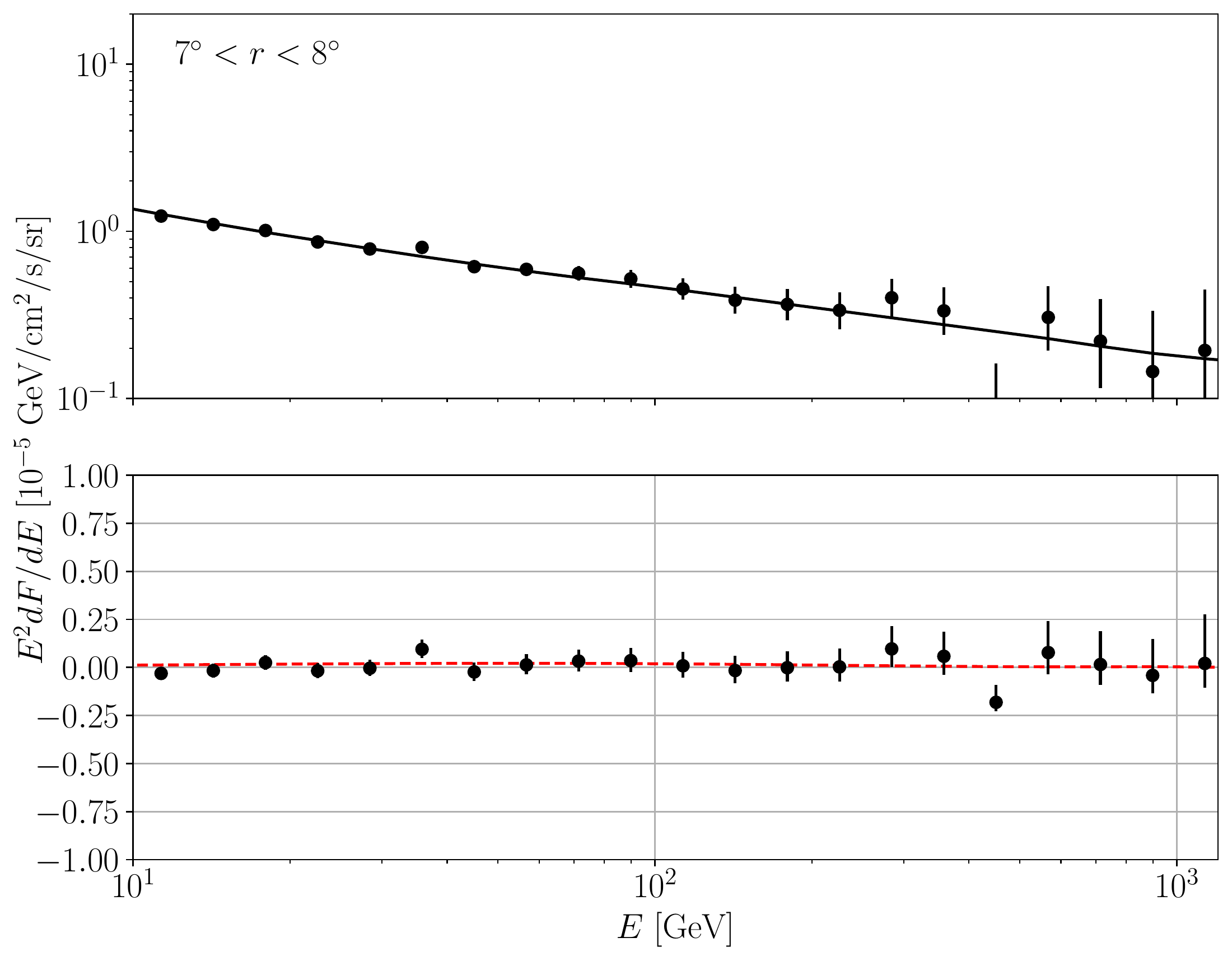}
		\includegraphics[width=0.49\textwidth]{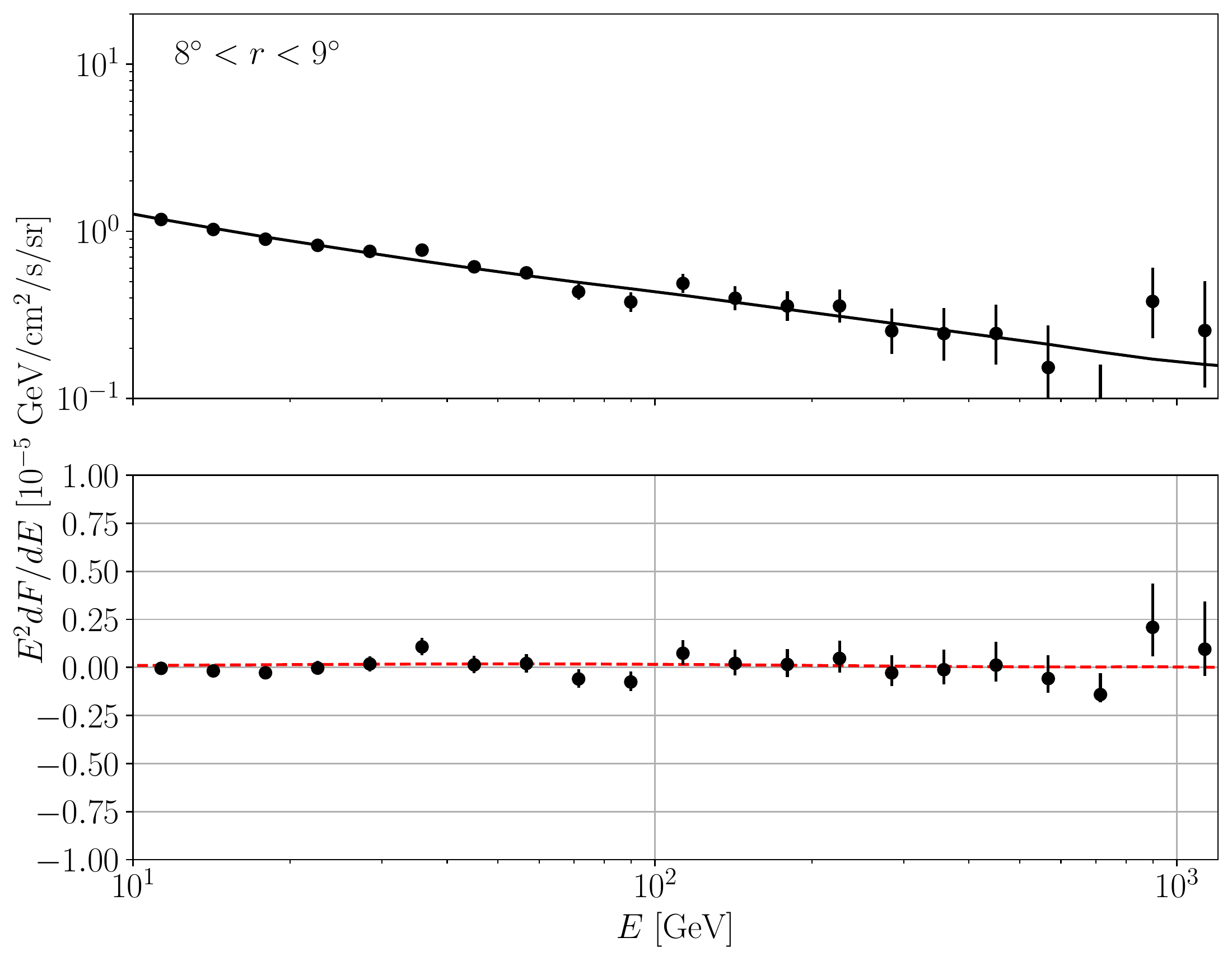}
		\includegraphics[width=0.49\textwidth]{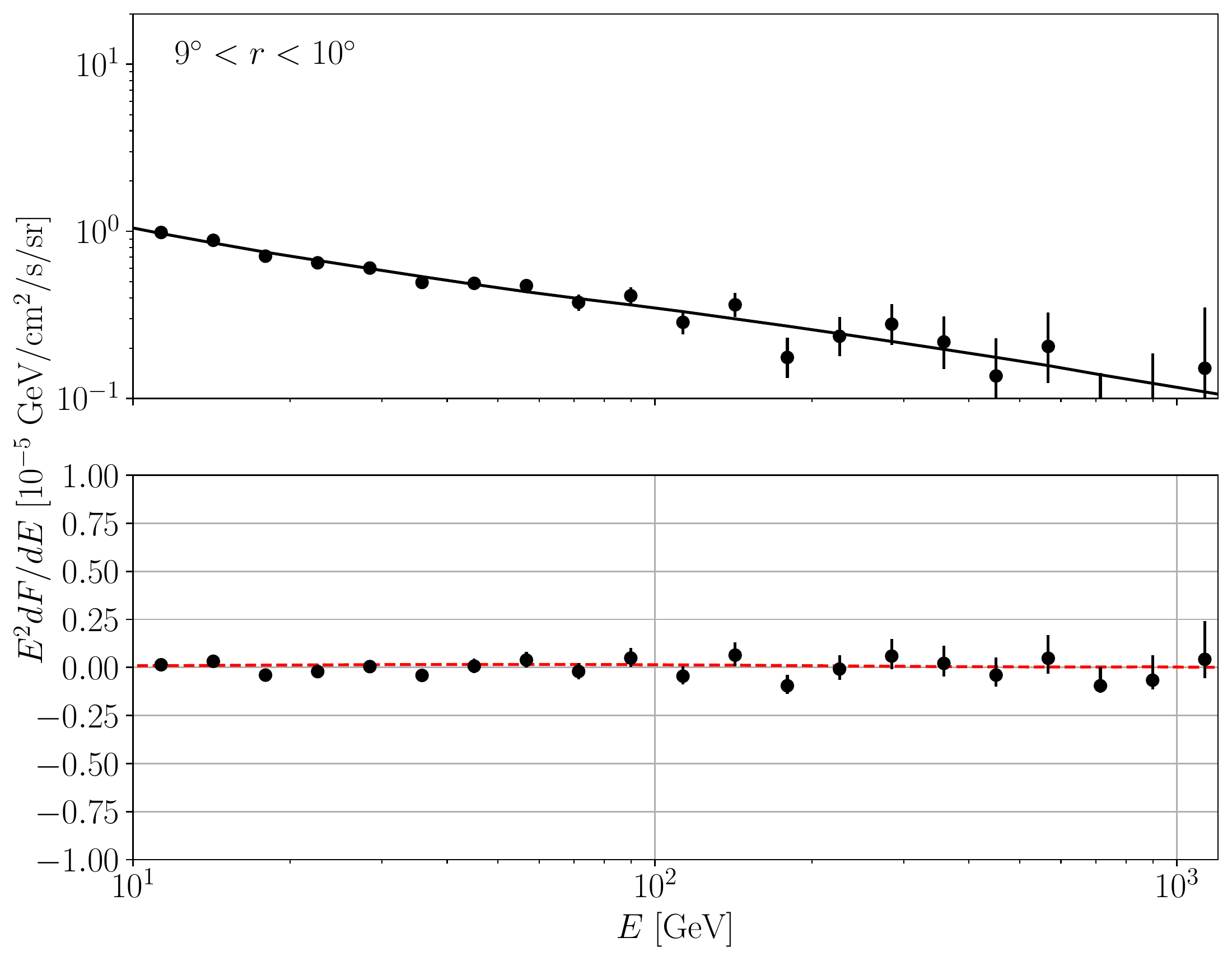}
	\end{center}
	\caption{As in Fig.~\ref{fig:residuals_1} but for the latter 5 annuli. }
	\label{fig:residuals_2}
\end{figure*}

In this section we present additional results from the fiducial analysis described in the main Letter. In Figs.~\ref{fig:residuals_1} and~\ref{fig:residuals_2} we show the best-fit null spectra and residuals in each of the 9 concentric annuli. We illustrate the higgsino signal on top of the background-subtracted data for an NFW DM profile with a higgsino mass $m_\chi = 1.1$ TeV and a cross-section $\langle \sigma v \rangle = 5 \times 10^{-26}$ cm$^3/$s. Note that these figures are for presentation purposes only and are not used in our analysis, as our analysis profiles over the background nuisance parameter when constructing the profile likelihood for the cross section.  However, visual inspection of these figures shows that there are no obvious sources of systematic mismodeling present in our fiducial analysis.

\begin{figure*}[!htb]
	\begin{center}
		\includegraphics[width=0.55\textwidth]{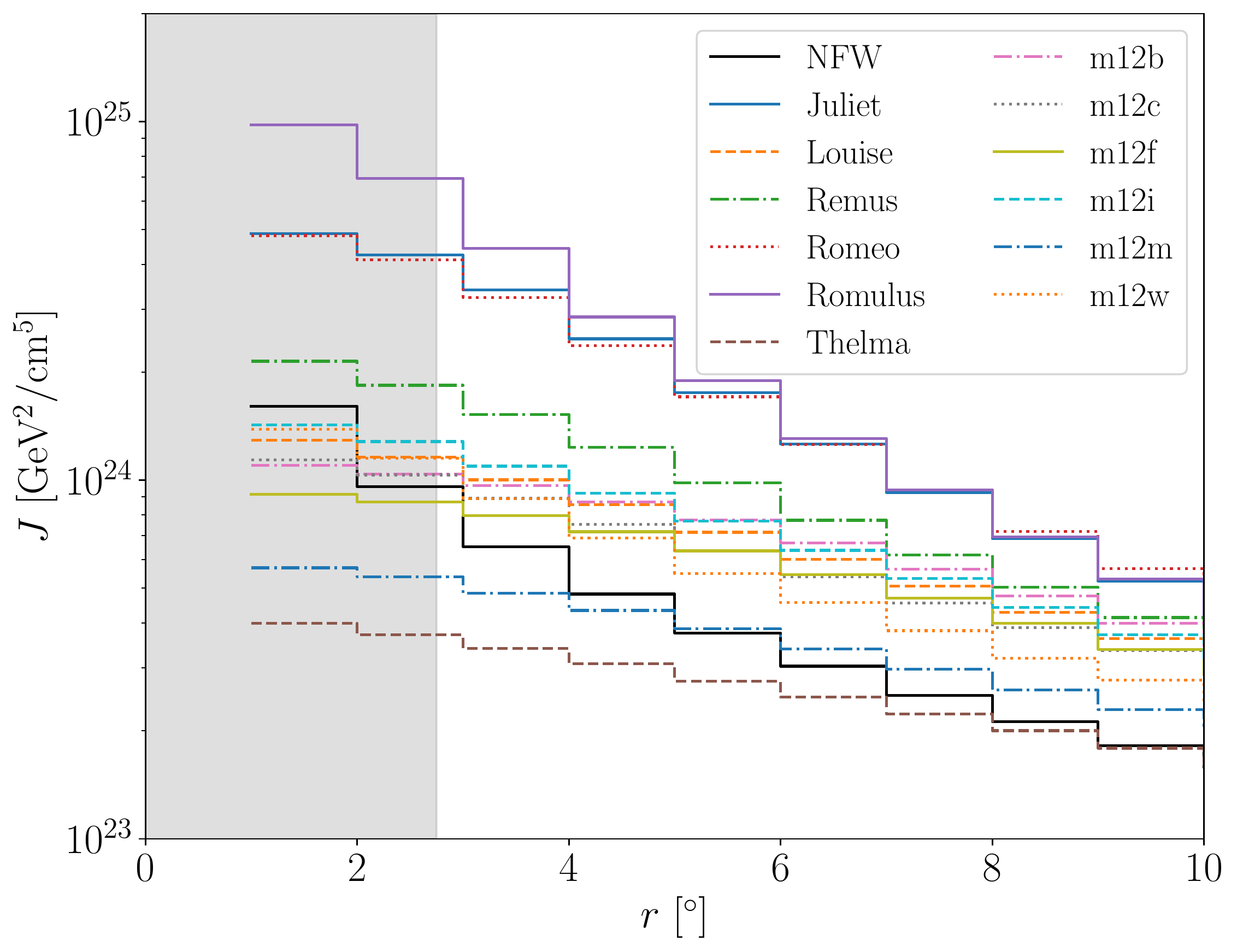}
		\end{center}
	\caption{The $J$-factor profiles averaged over our analysis annuli for our fiducial NFW profile and for the 12 FIRE-2 $J$-factor profiles from~\cite{2022MNRAS.513...55M} for Milky Way analogue galaxies.  Note that the resolution of the FIRE-2 zoom-in simulations is indicated in shaded grey, which implies that the first 2 annuli may misestimate  the true $J$-factors in those galaxies.  The \texttt{Romeo} galaxy may be the most Milky Way like, while we illustrate our results in the main Letter for the NFW and \texttt{Romulus} profiles. The named galaxy simulations, including \texttt{Remus} and \texttt{Romulus}, were evolved in pairs in order to mimic the Milky Way's co-evolution with M31, and are in this sense more realistic representations compared to the six unnamed ones.}
	\label{fig:J-factor}
\end{figure*}

In the main Letter we presented results for the NFW DM profile and for the \texttt{Romulus} FIRE-2 profile, which provided the best-fit to the {\it Fermi} data.  Here, we present results for all 12 of the FIRE-2 $J$-factor profiles for Milky Way analogue galaxies.  The $J$-factors for the FIRE-2 Milky Way analogue galaxies are illustrated in Fig.~\ref{fig:J-factor} (reproduced from~\cite{2022MNRAS.513...55M}), compared to the $J$-factor profile for our fiducial NFW profile. Interestingly, almost all of the FIRE-2 $J$-factors are larger in the inner few degrees relative to those from the NFW profile.  On the other hand, the angular resolution in the FIRE-2 simulations, which have the best resolution of any current hydrodynamic simulation suites, is estimated at 2.75$^\circ$~\cite{2022MNRAS.513...55M}, meaning that the $J$-factors in the inner $\sim$3 degrees may be unreliable.

\begin{figure*}[!htb]
	\begin{center}
		\includegraphics[width=0.49\textwidth]{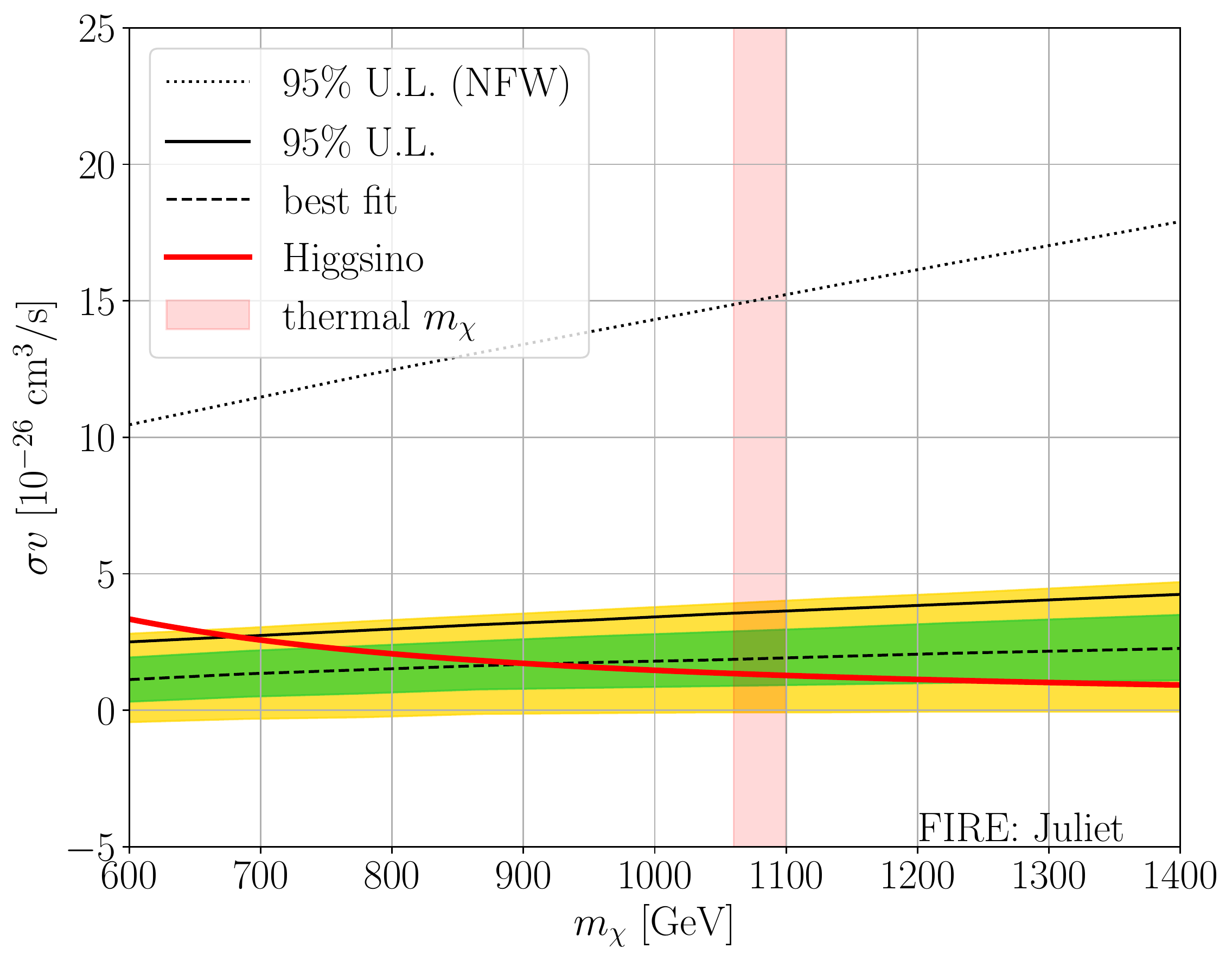}
		\includegraphics[width=0.49\textwidth]{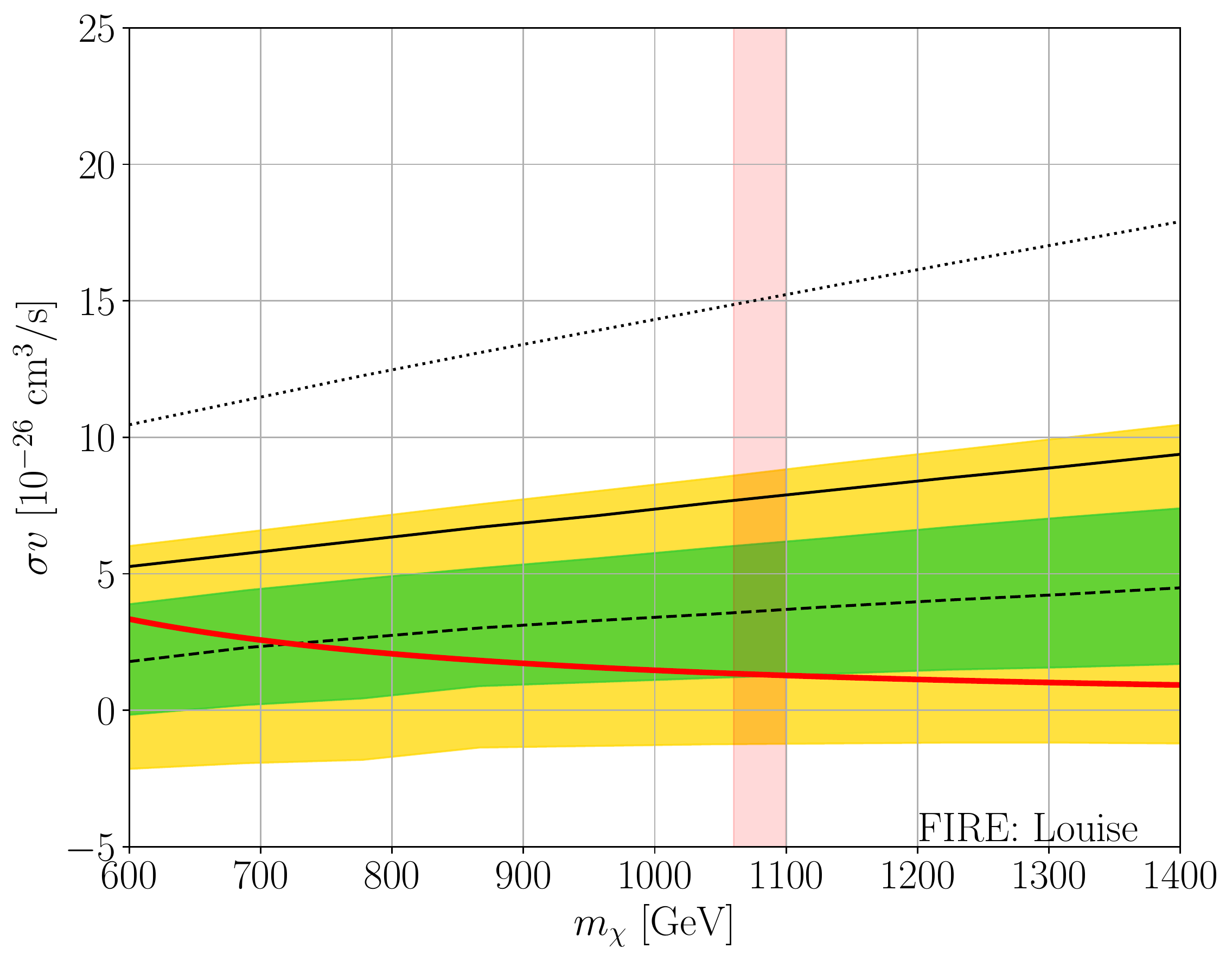}
		\includegraphics[width=0.49\textwidth]{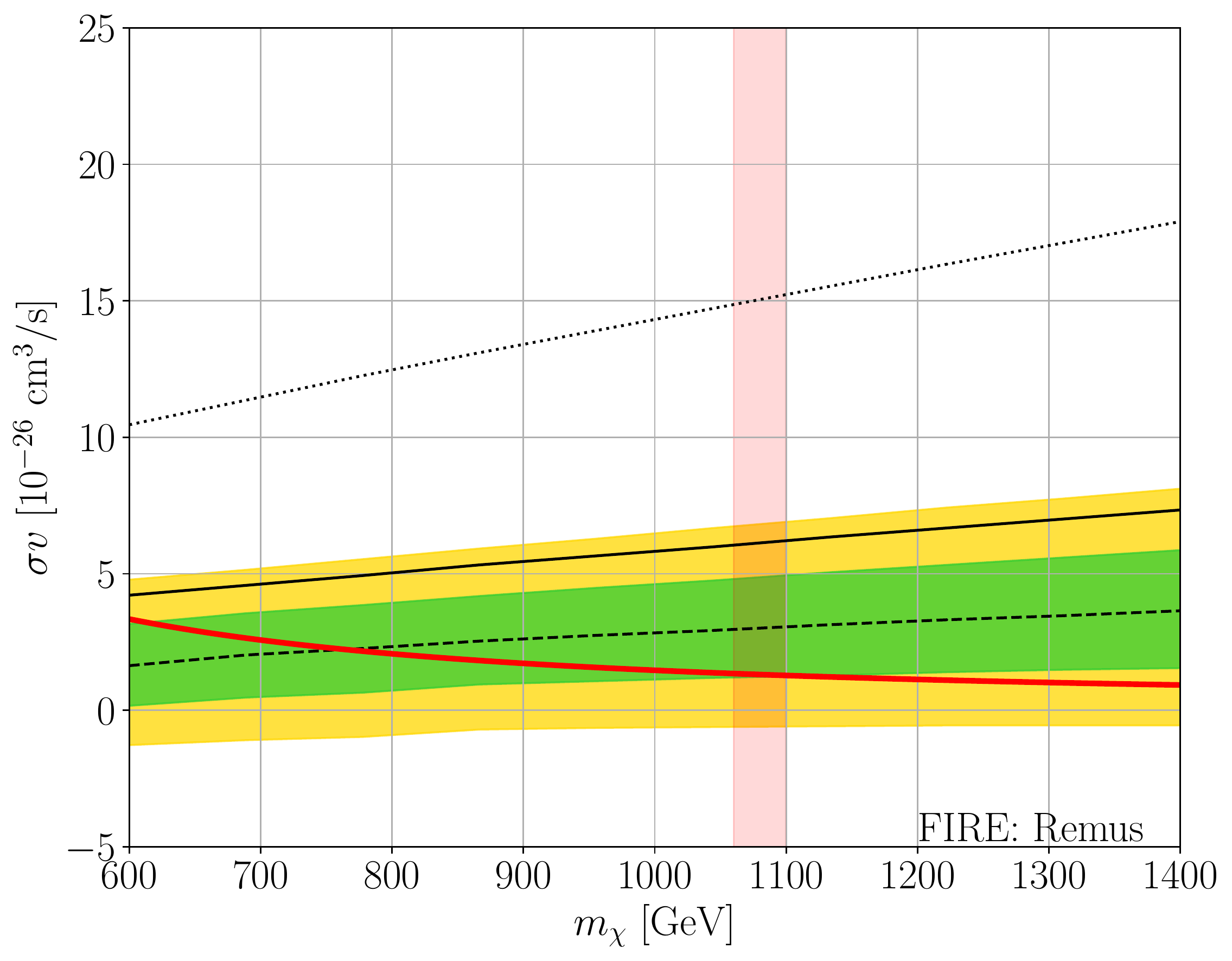}
		\includegraphics[width=0.49\textwidth]{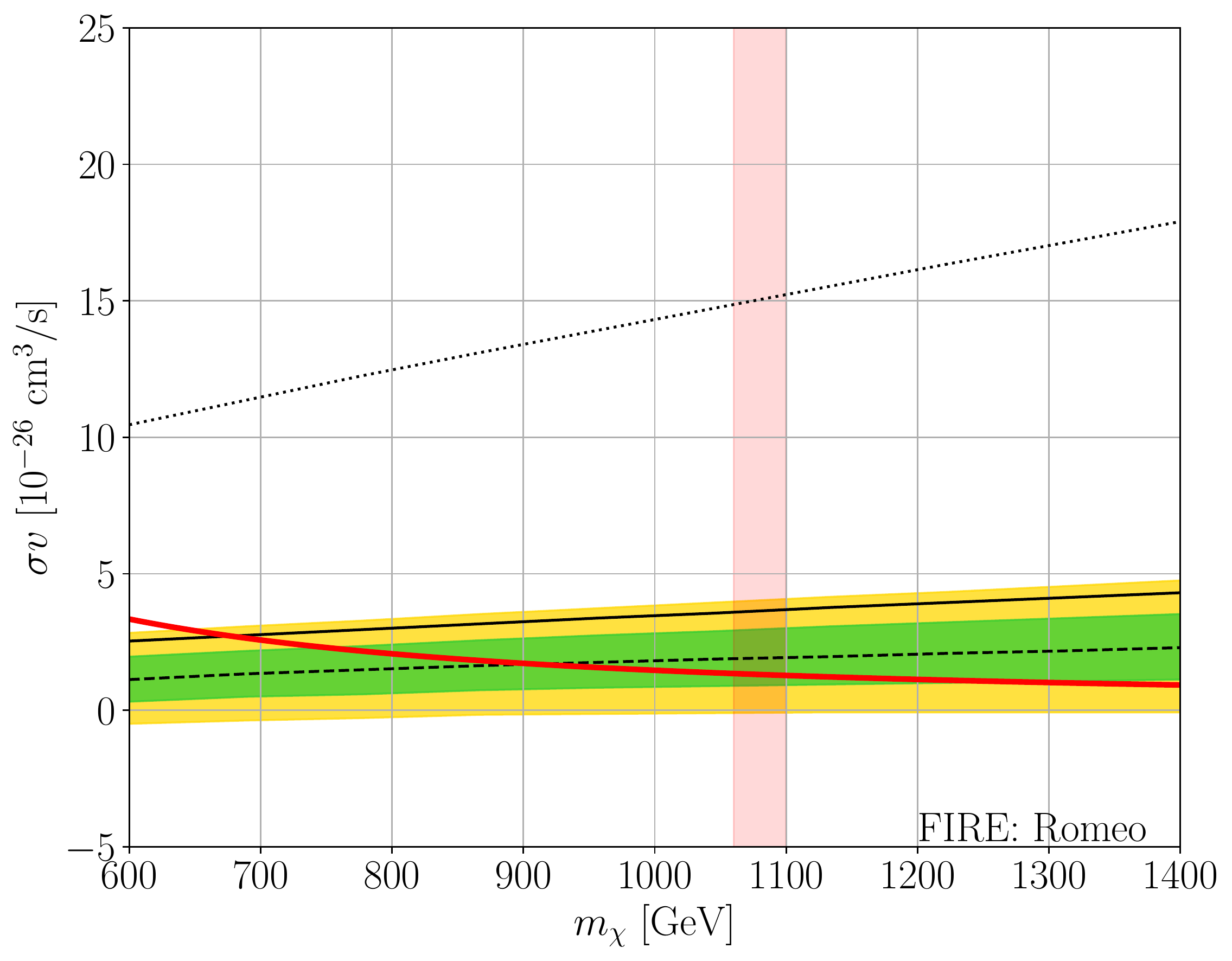}
		\includegraphics[width=0.49\textwidth]{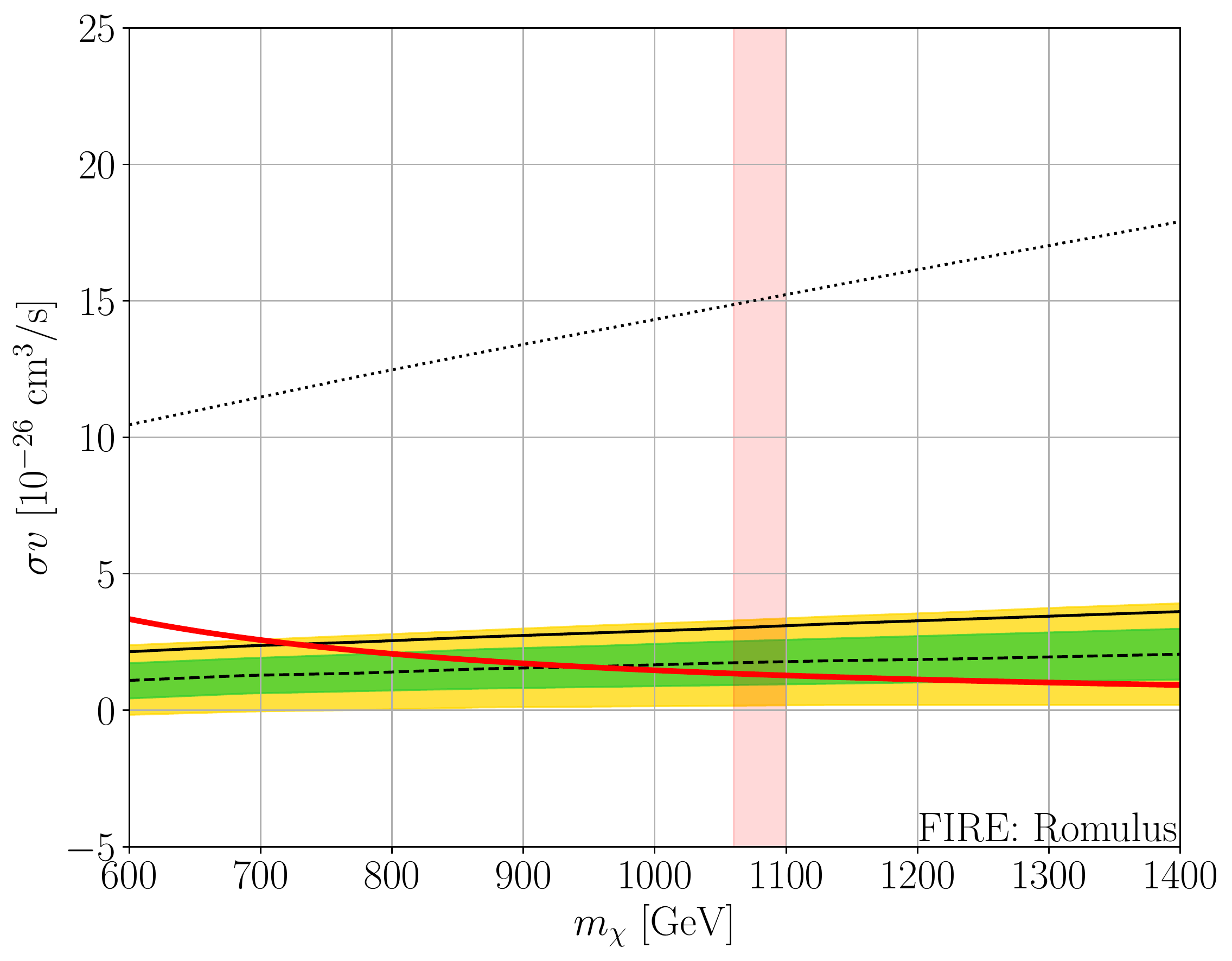}
		\includegraphics[width=0.49\textwidth]{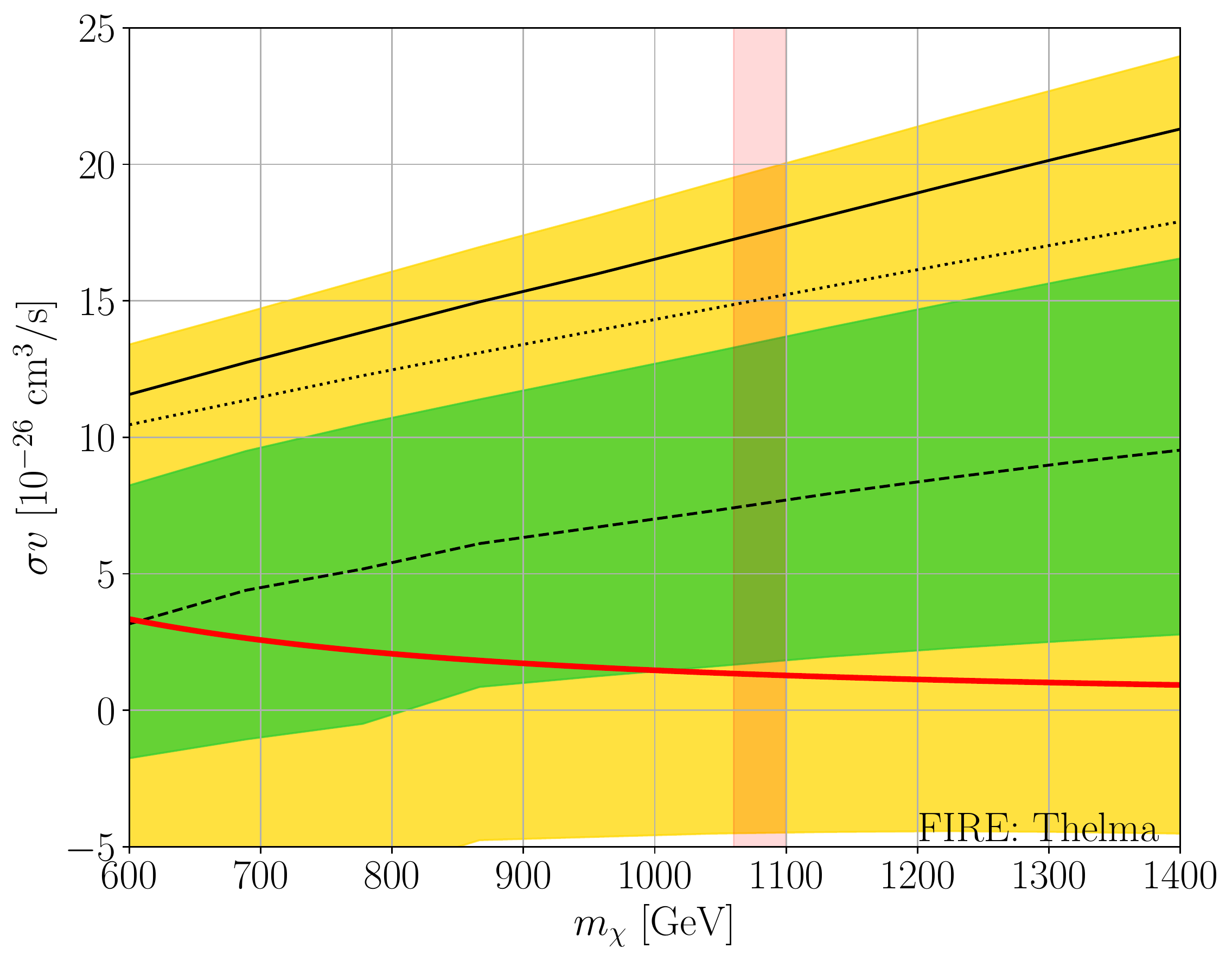}
		\end{center}
	\caption{ As in Fig.~\ref{fig:results} but for the first 6 FIRE-2 Milky Way analogue galaxy $J$-factors~\cite{2022MNRAS.513...55M}.  We compare the 95\% upper limits to that found using the NFW profile; in 11 of 12 cases the FIRE-2 galaxies lead to stronger upper limits relative to that from the NFW profile.  }
	\label{fig:FIRE_UL_1}
\end{figure*}

\begin{figure*}[!htb]
	\begin{center}
		\includegraphics[width=0.49\textwidth]{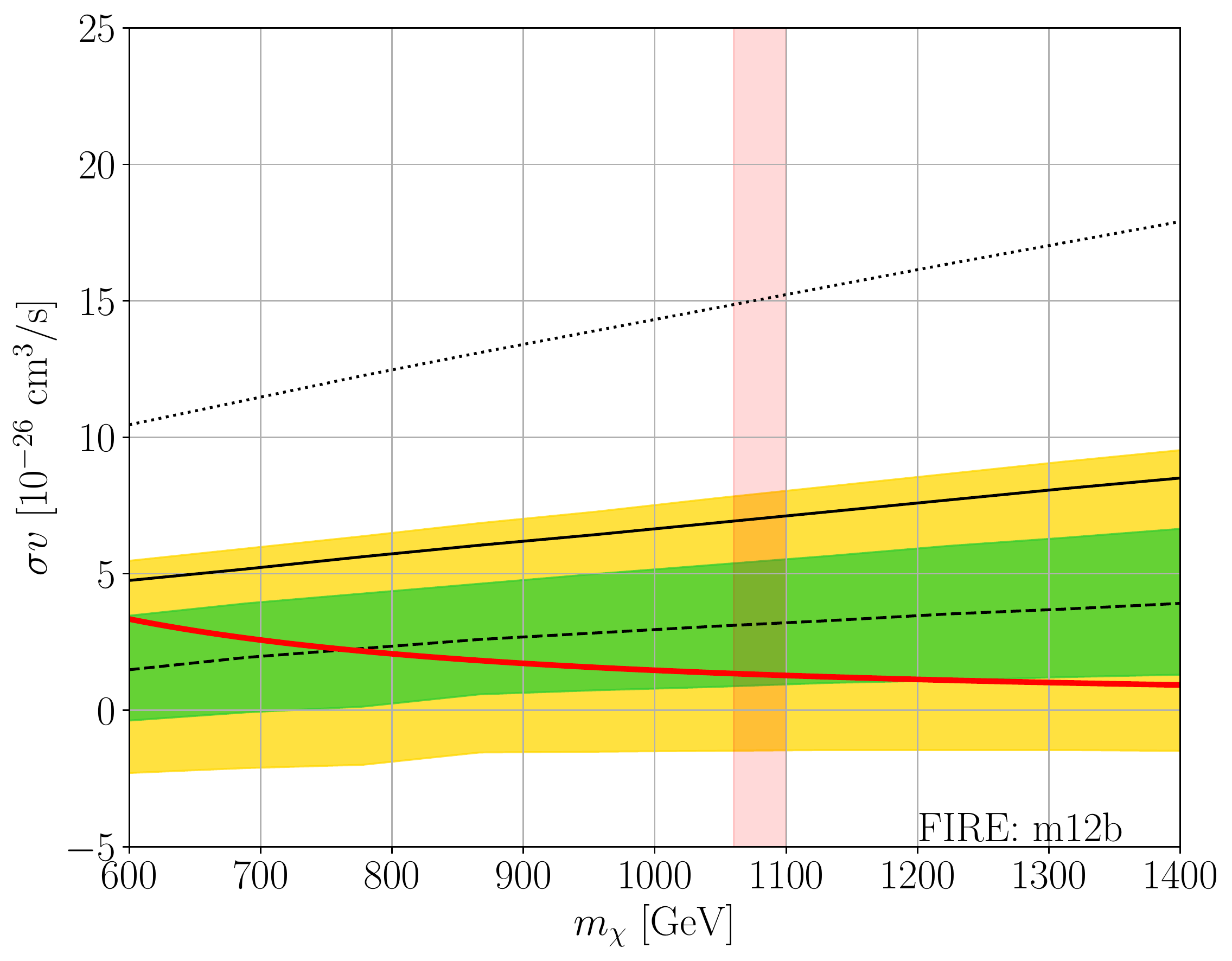}
		\includegraphics[width=0.49\textwidth]{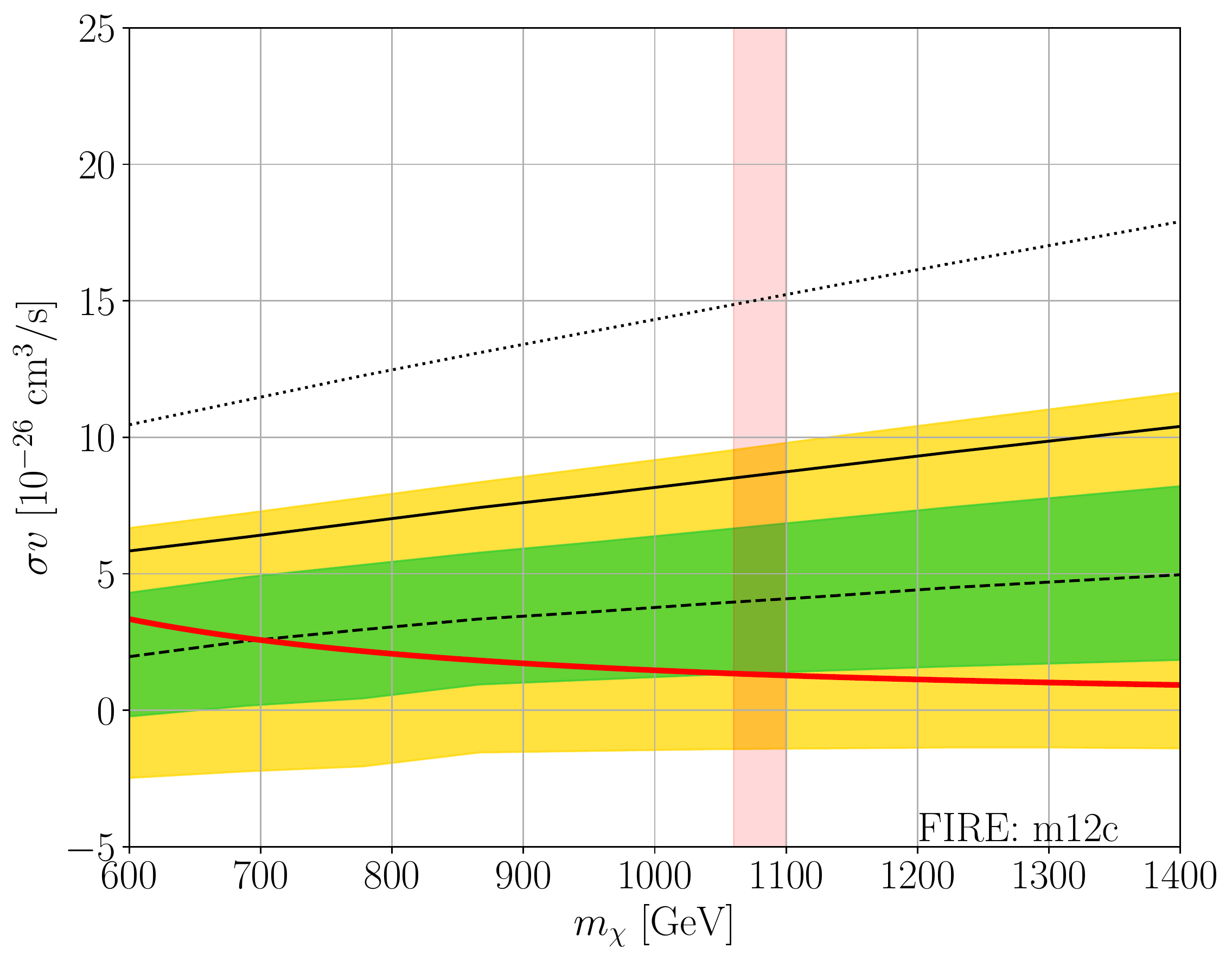}
		\includegraphics[width=0.49\textwidth]{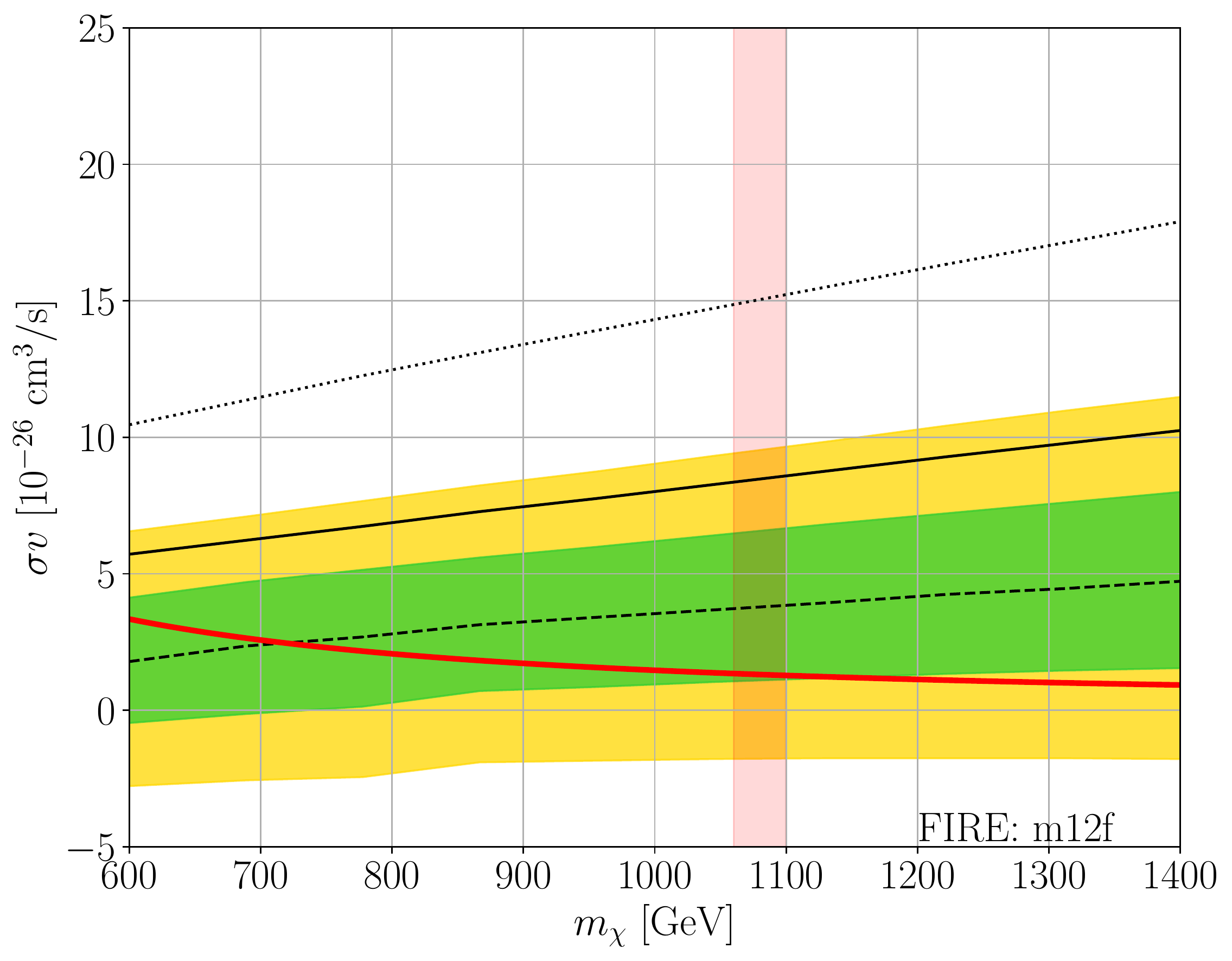}
		\includegraphics[width=0.49\textwidth]{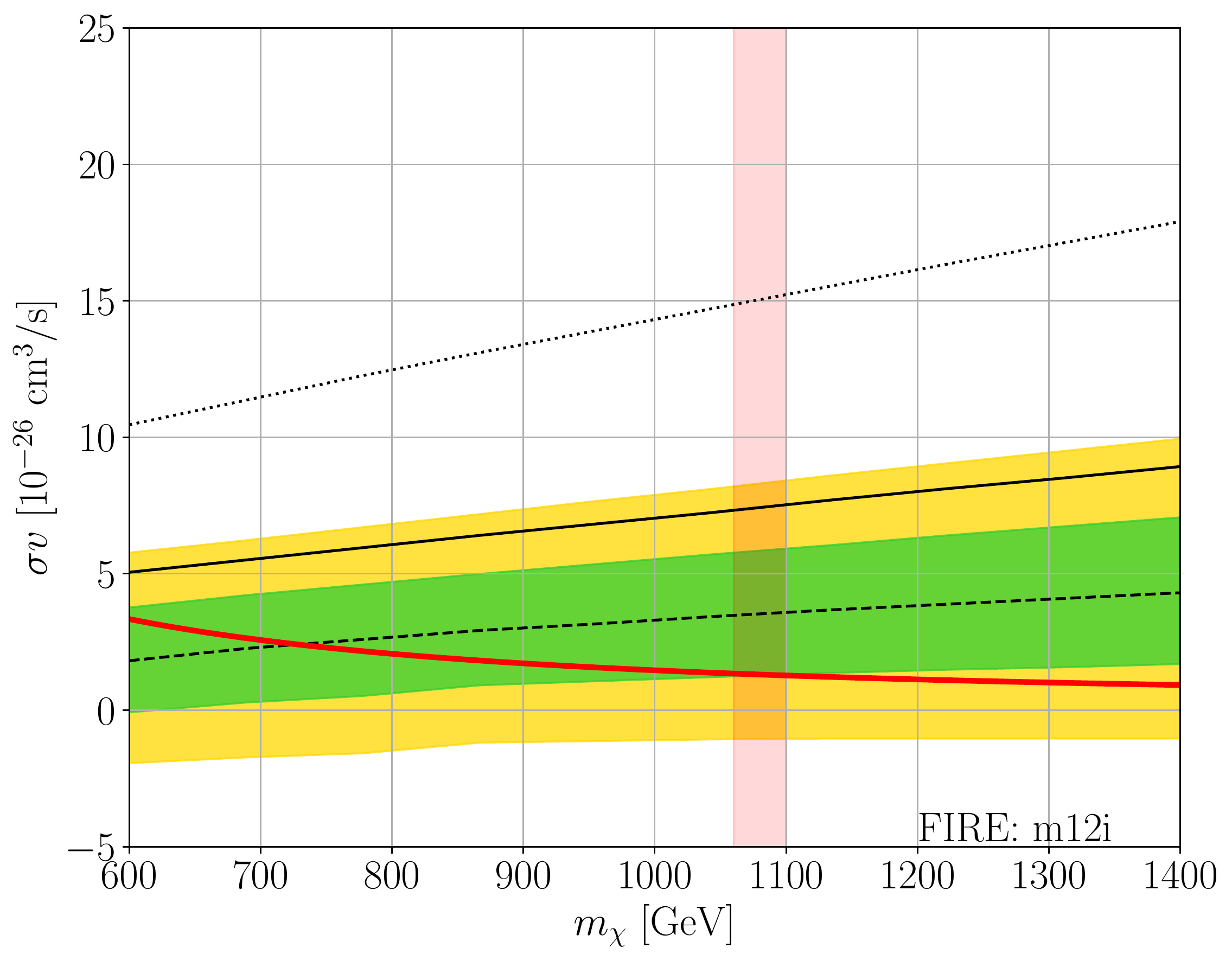}
		\includegraphics[width=0.49\textwidth]{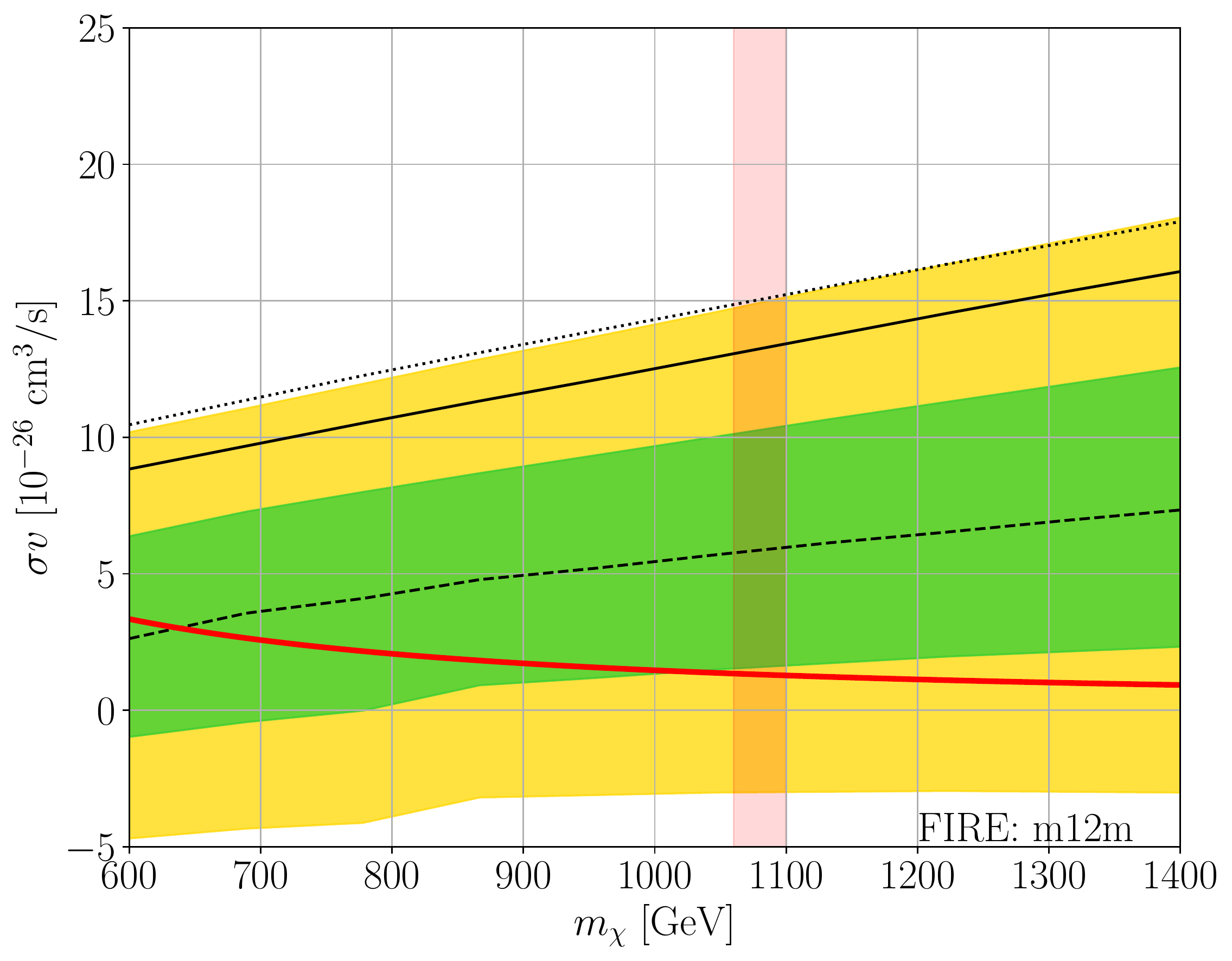}
		\includegraphics[width=0.49\textwidth]{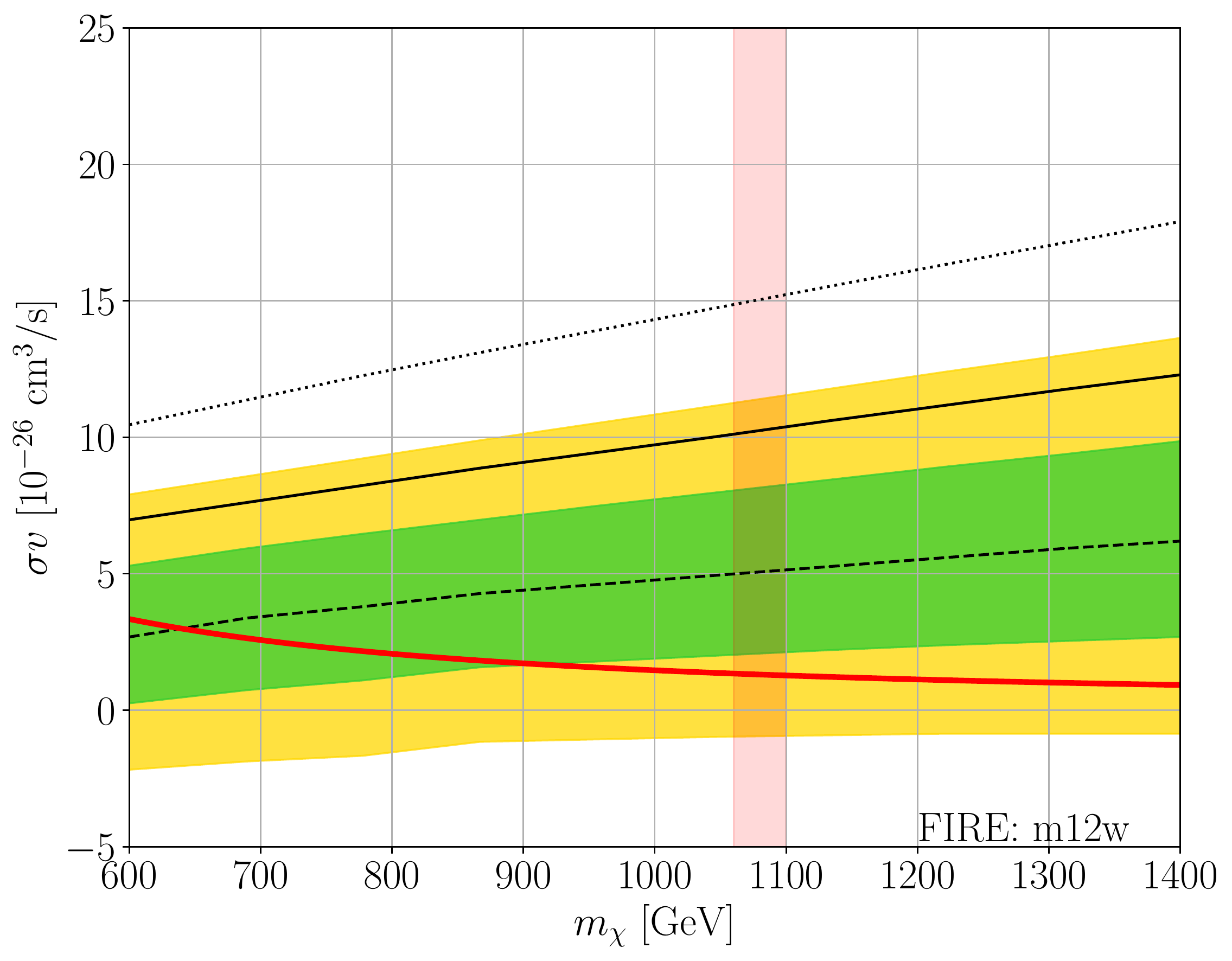}
		\end{center}
	\caption{As in Fig.~\ref{fig:FIRE_UL_1} for the latter 6 of 12 Milky Way analogue galaxies.}
	\label{fig:FIRE_UL_2}
\end{figure*}

In Figs.~\ref{fig:FIRE_UL_1} and~\ref{fig:FIRE_UL_2} we show the results of our analysis, as in Fig.~\ref{fig:results}, interpreted in the context of higgsino DM for all 12 FIRE-2 galaxies.  
Note that we also compare the 95\% upper limits to that found using the NFW DM profile. In 11 of 12 FIRE-2 galaxies we find a stronger upper limit than we do with the NFW profile, suggesting that adiabatic contraction with baryons likely enhances the DM signal in Milky Way type galaxies relative to the NFW expectation.  

\begin{figure*}[!htb]
	\begin{center}
		\includegraphics[width=0.55\textwidth]{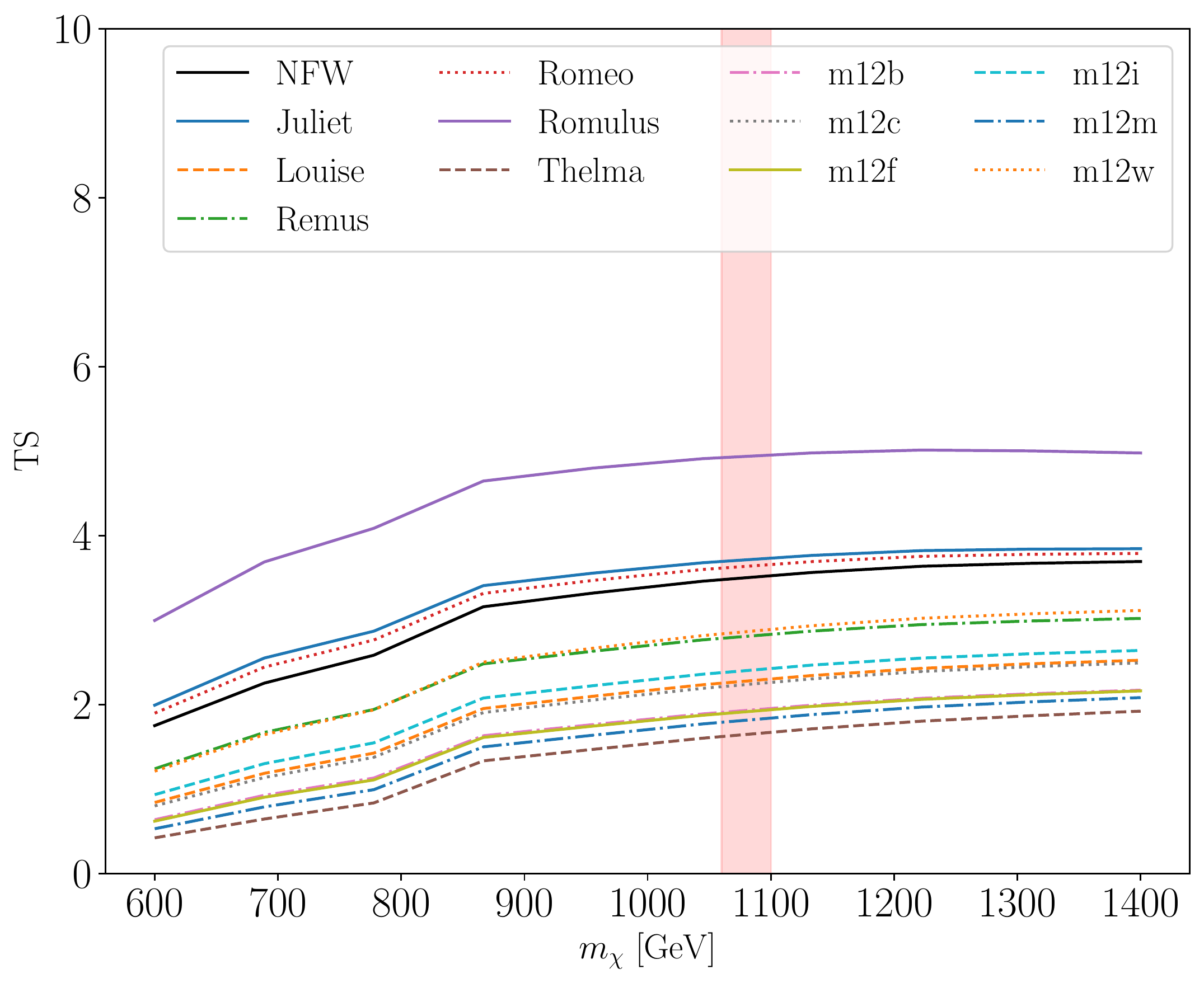}
		\end{center}
	\caption{The discovery TS in favor of the signal model over the null hypothesis. The Z-score for discovery is given approximately by the square root of the TS for a one-sided test.  We illustrate the TS as a function of $m_\chi$ for analyses using the NFW DM profile and the 12 FIRE-2 $J$-factor profiles.}
	\label{fig:TS}
\end{figure*}

In Fig.~\ref{fig:TS} we show the discovery TS as a function of the higgsino mass $m_\chi$ for analyses using the NFW DM profile and the 12 FIRE-2 $J$-factor profiles. 
The evidence in favor of the signal hypothesis is $\sim$1-2$\sigma$ in significance at all masses. Thus, we conclude that the data shows no evidence for higgsino DM.  The TS is slightly larger than expected under the null hypothesis, which could indicate the first hint of a DM signal, but this modest TS could also arise from one or a combination of mismodeling and random statistical chance.

\begin{figure*}[!htb]
	\begin{center}
		\includegraphics[width=0.49\textwidth]{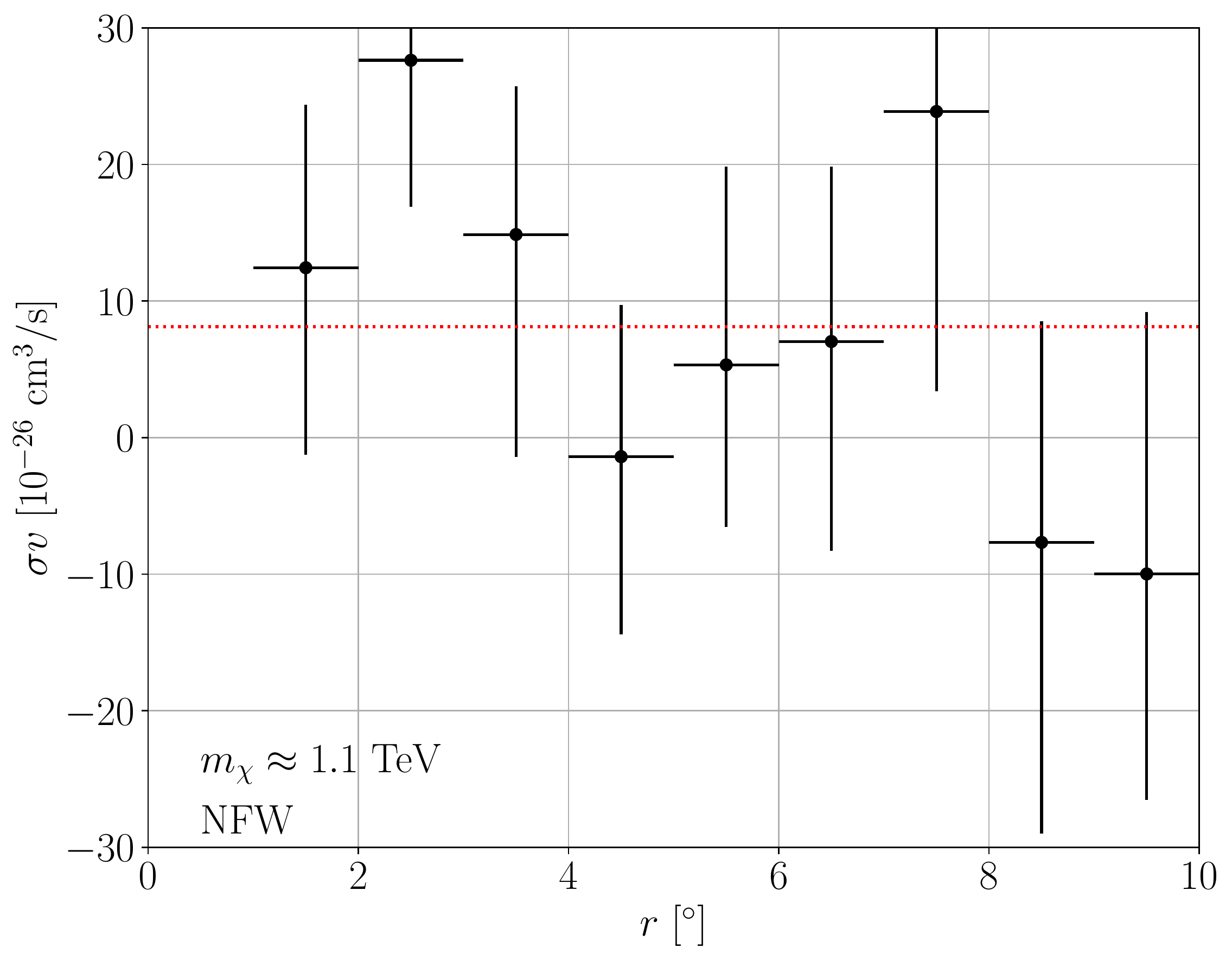}
		\includegraphics[width=0.49\textwidth]{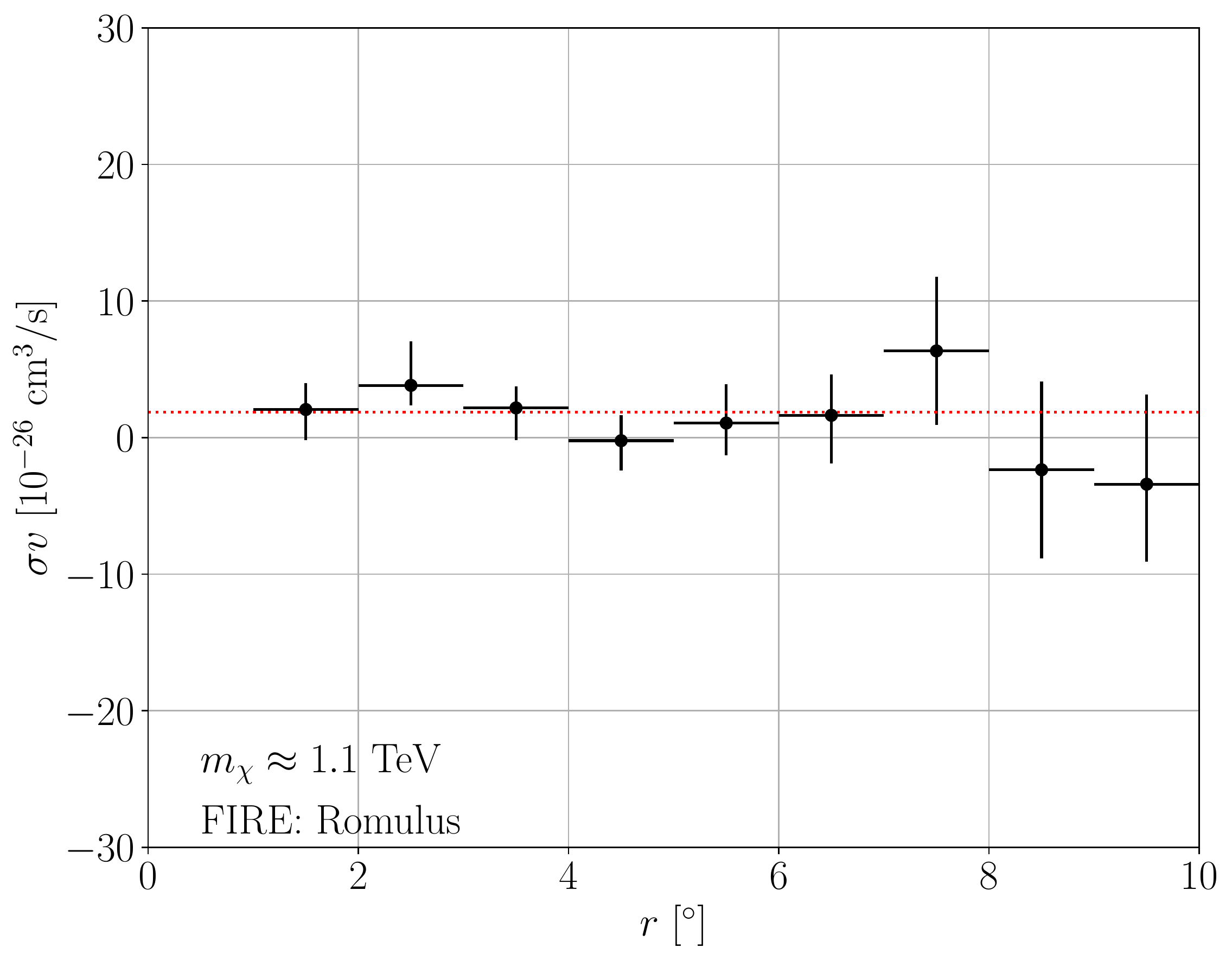}
		\end{center}
	\caption{The best fit annihilation cross-sections found in our analyses in each of the annuli independently, as interpreted using the NFW DM profile (left panel) and the \texttt{Romulus} profile (right panel), with $m_\chi = 1.1$ TeV.}
	\label{fig:bf_radius}
\end{figure*}

In Fig.~\ref{fig:bf_radius} we show the best-fit cross-section and 1$\sigma$ confidence interval in each radial bin independently, interpreted using the NFW DM profile (left panel) and the \texttt{Romulus} profile (right panel), with the higgsino mass fixed to be $m_\chi = 1.1$ TeV. These figures show that the analyses in the independent annuli, which are joined together in our fiducial analysis, produce consistent results up to statistical fluctuations. 

\begin{figure*}[!htb]
	\begin{center}
		\includegraphics[width=0.49\textwidth]{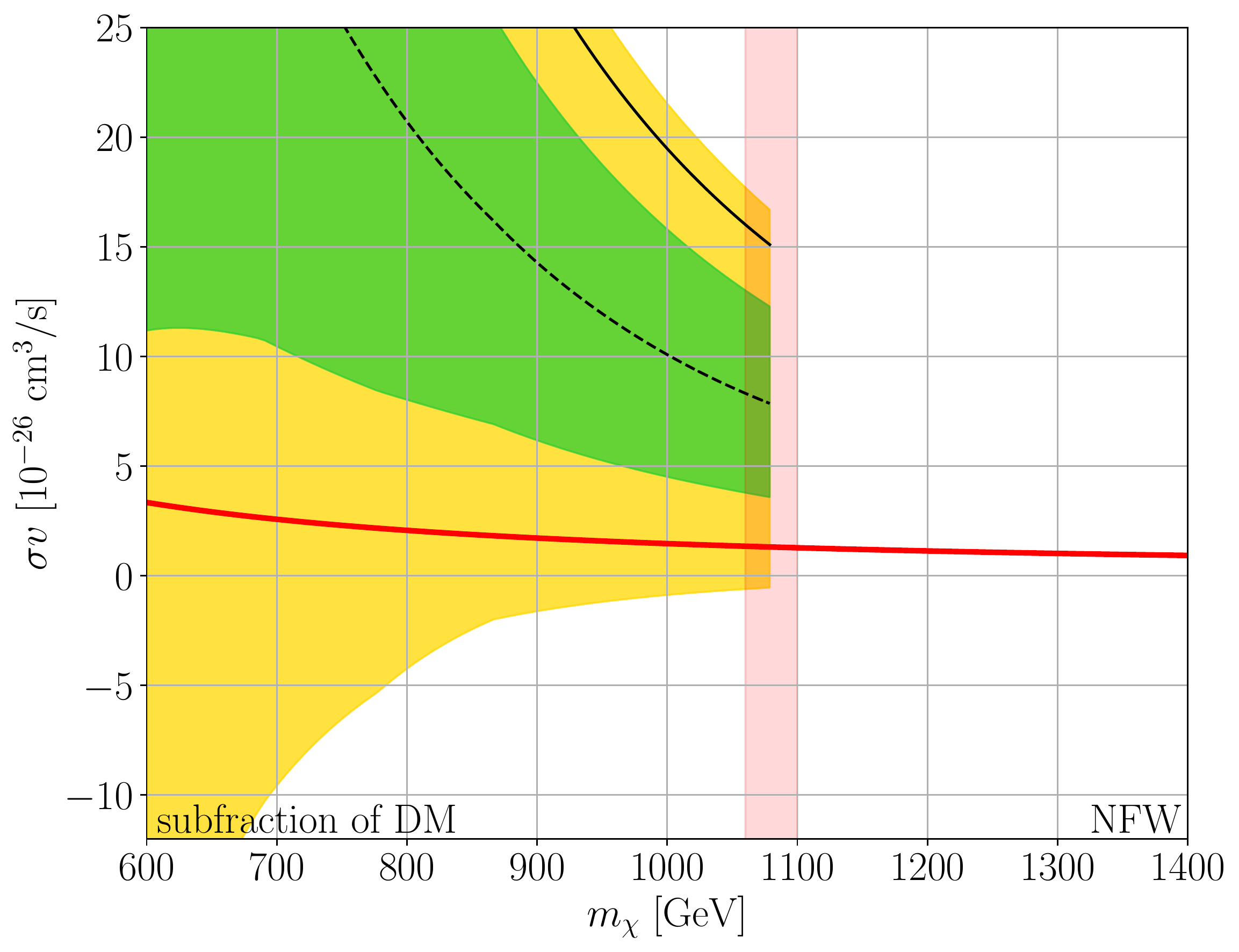}
		\includegraphics[width=0.49\textwidth]{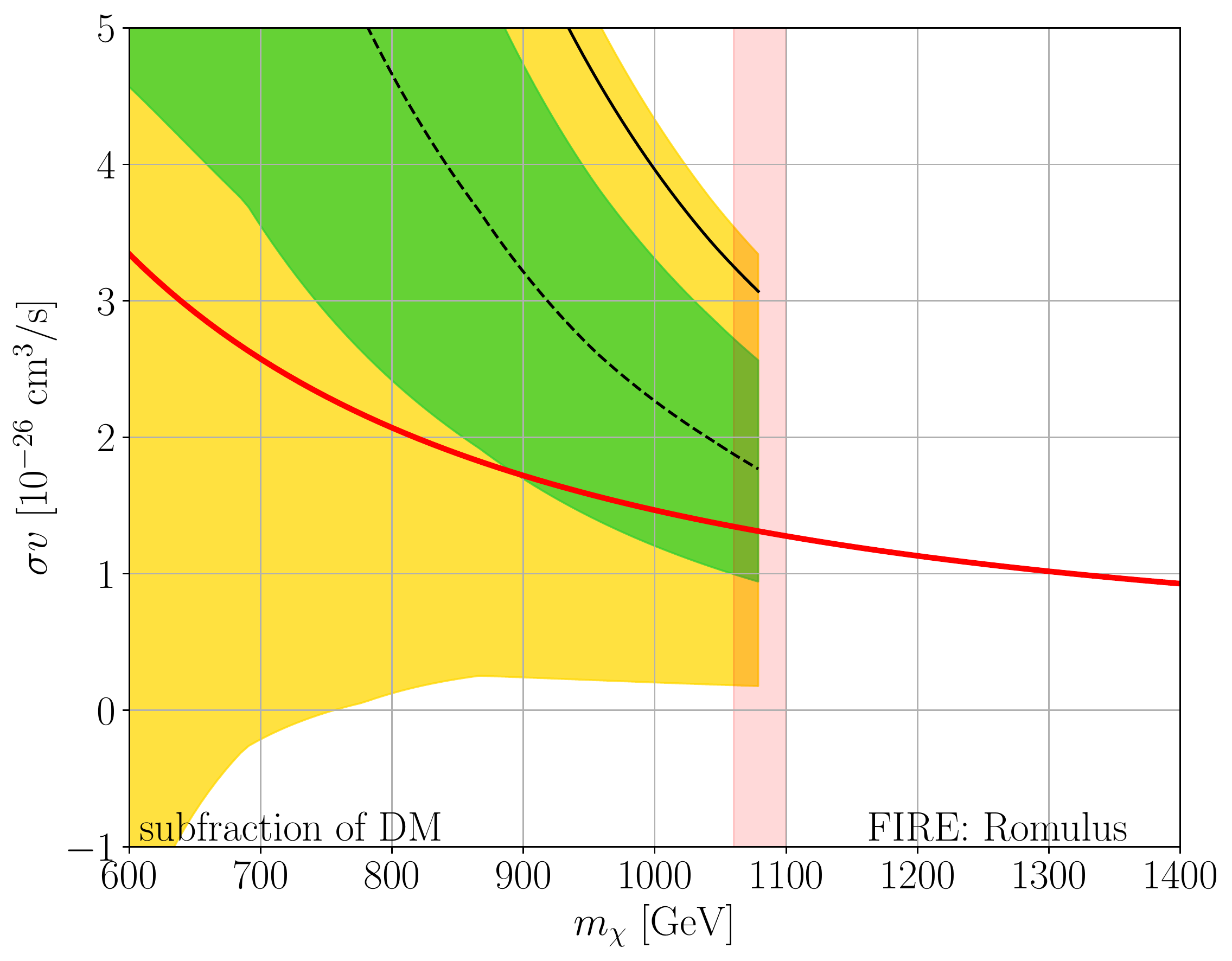}
		\end{center}
	\caption{As in Fig.~\ref{fig:results} but instead of assuming the higgsino is all of the DM regardless of $m_\chi$ we assume that the higgsino is a subfraction of the DM as computed under the assumption of a standard thermal cosmology and with a thermal mass of $m_\chi = 1.08$ TeV. We do not show our results for masses above $1.08$ TeV in this case since the DM abundance is overproduced in this region given our assumptions for this figure.}
	\label{fig:sf}
\end{figure*}

Throughout the main Letter we assume that regardless of the mass $m_\chi$, the higgsino makes up 100\% of the DM. An alternative possibility, however, is that for $m_\chi$ less than the thermal mass the higgsino is a subfraction of the total DM, with the fraction scaling like $(m_\chi/1.08 \, \, {\rm TeV})^2$ for a thermal mass of $1.08$ TeV and the standard thermal cosmology.  In this case, the DM annihilation signal is reduced by a factor $(m_\chi/1.08 \, \, {\rm TeV})^4$ for $m_\chi < 1.08$ TeV. In Fig.~\ref{fig:sf} we illustrate our results assuming that the higgsino is a subfraction of the DM for a thermal mass of $1.08$ TeV. Note that we do not show masses above $1.08$ TeV since such masses would overproduce the DM in this framework.


\FloatBarrier

\section{Systematic analysis variations}
\label{sec:analysis}

\begin{figure*}[!htb]
	\begin{center}
		\includegraphics[width=0.49\textwidth]{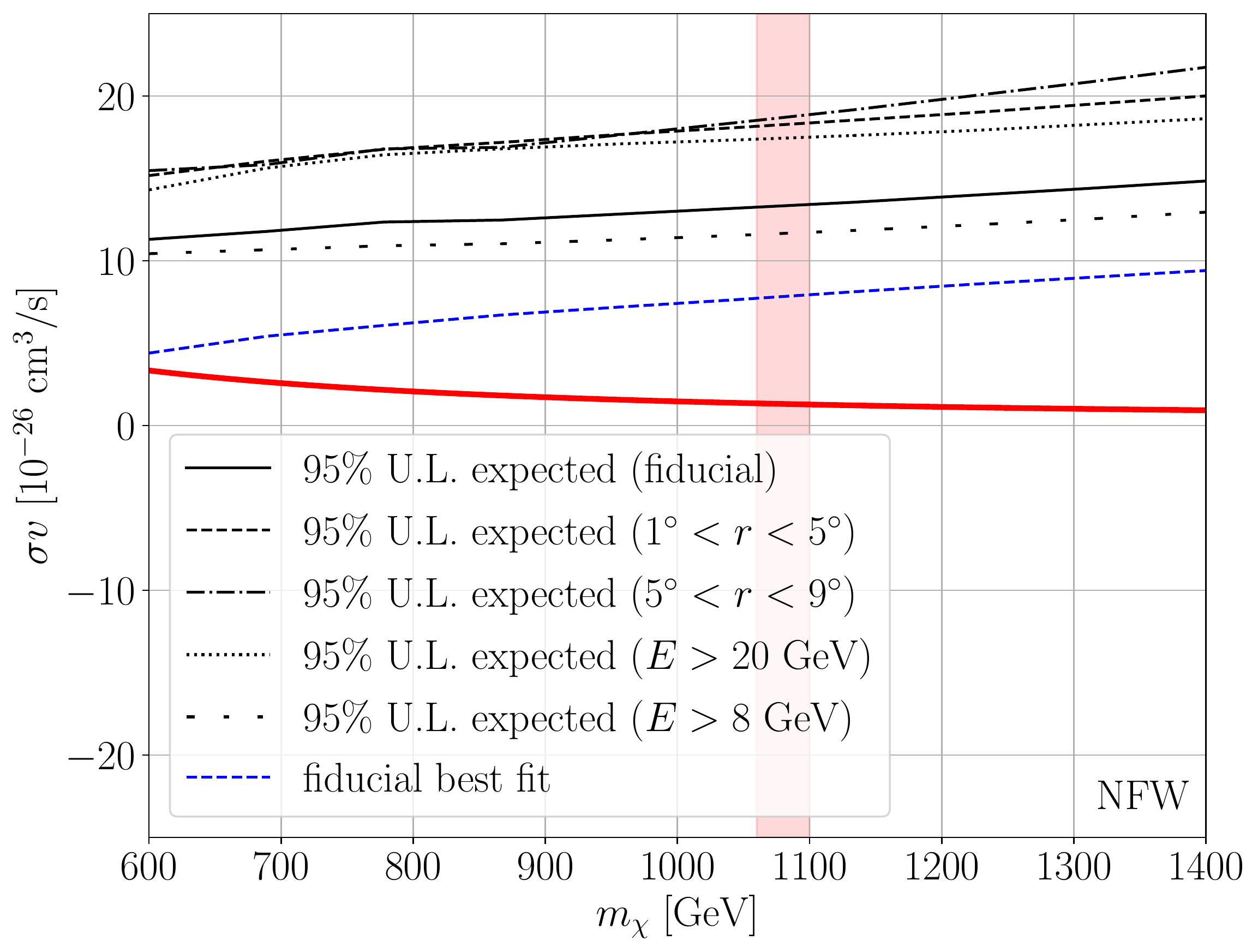}
		\includegraphics[width=0.49\textwidth]{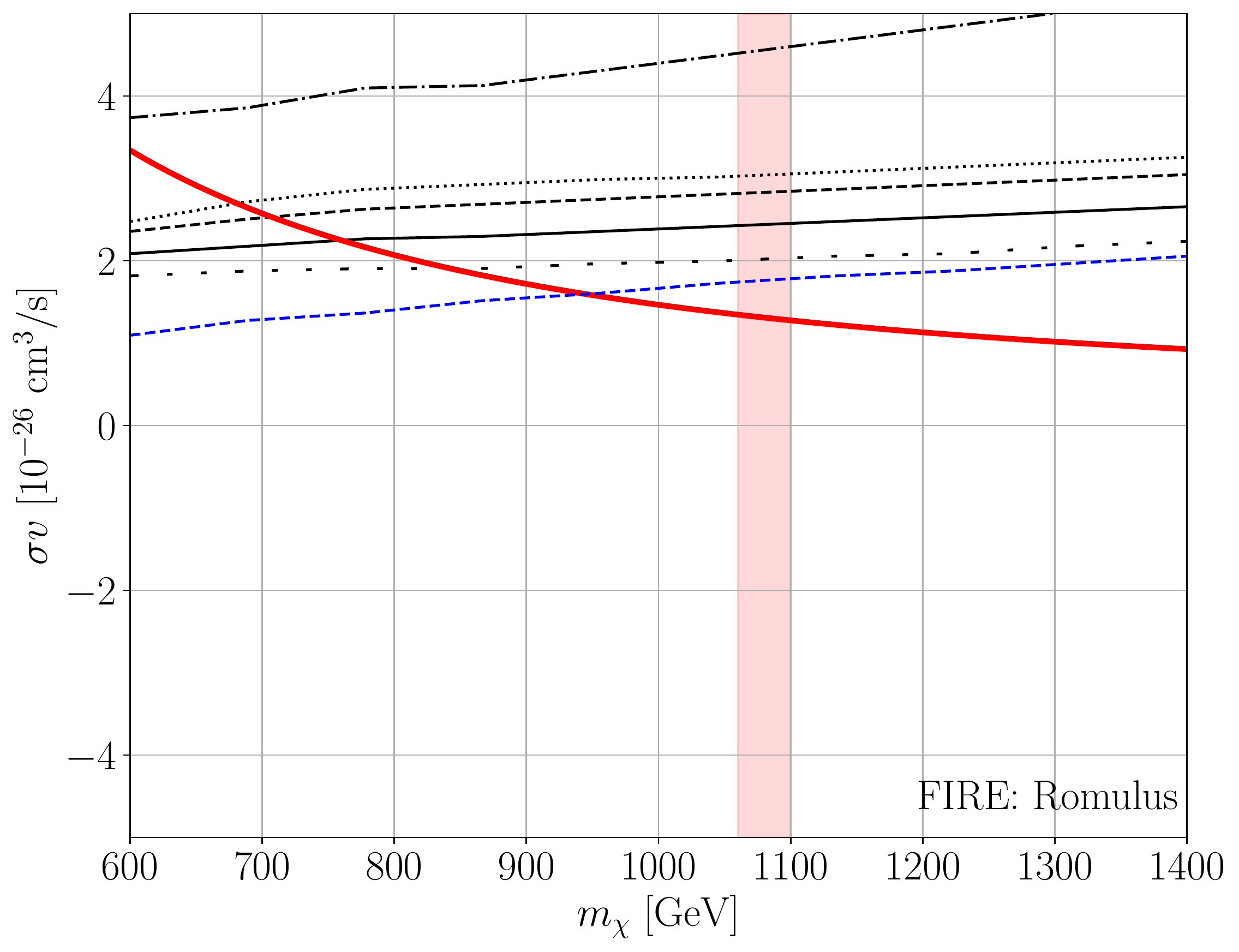}
		\end{center}
	\caption{The best fit annihilation cross-sections for our fiducial analysis compared to the expected 95\% upper limit for our fiducial analysis and some of the analysis variations considered in Secs.~\ref{sec:reduced} and~\ref{sec:energy}. 
	}
	\label{fig:asimov}
\end{figure*}

In this section we describe the results of systematic variations to our fiducial analysis.  Before describing the variations in detail, we refer to Fig.~\ref{fig:asimov}, which summarizes the expected changes to sensitivity for a subset of the systematic analysis variations described below in Sec.~\ref{sec:reduced} and Sec.~\ref{sec:energy}. In Sec.~\ref{sec:reduced} we consider changing our analysis ROI, while in Sec.~\ref{sec:energy} we change the analysis energy range.  In Fig.~\ref{fig:asimov} we show the expected 95\% upper limit in our fiducial analysis assuming the NFW DM profile (left panel) and \texttt{Romulus} profile (right panel).  These expected upper limits are computed using the Asimov procedure~\cite{Cowan:2010js}, whereby we assume the data is given by the best-fit null hypothesis model.  We also illustrate our fiducial best-fit cross-sections from the analyses on the actual data.

We compare our fiducial analysis expectations to those from the Asimov procedure as applied to our analysis variations in Sec.~\ref{sec:reduced} and Sec.~\ref{sec:energy}.  In particular, we show the effects of reducing the ROI and changing the minimum analysis energy, as indicated.  We illustrate these changes through the expected 95\% upper limit, as the changes to the other quantities of interest may be derived from this information.  All of our analysis variations have comparable sensitivity, making the cross-checks described in Sec.~\ref{sec:reduced} and Sec.~\ref{sec:energy} powerful probes of possible mismodeling. 

\subsection{Reduced ROI}
\label{sec:reduced}

\begin{figure*}[!htb]
	\begin{center}
		\includegraphics[width=0.49\textwidth]{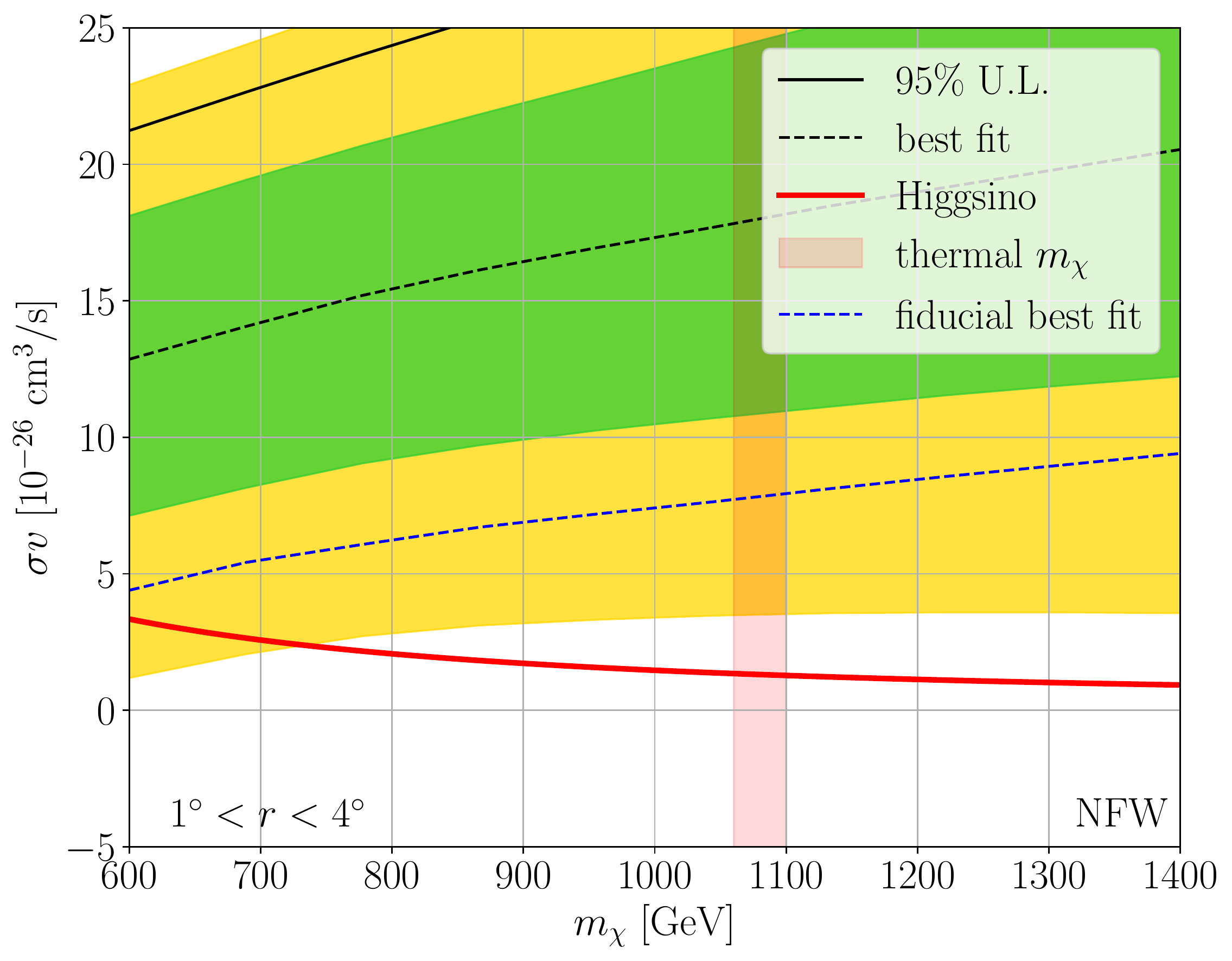}
		\includegraphics[width=0.49\textwidth]{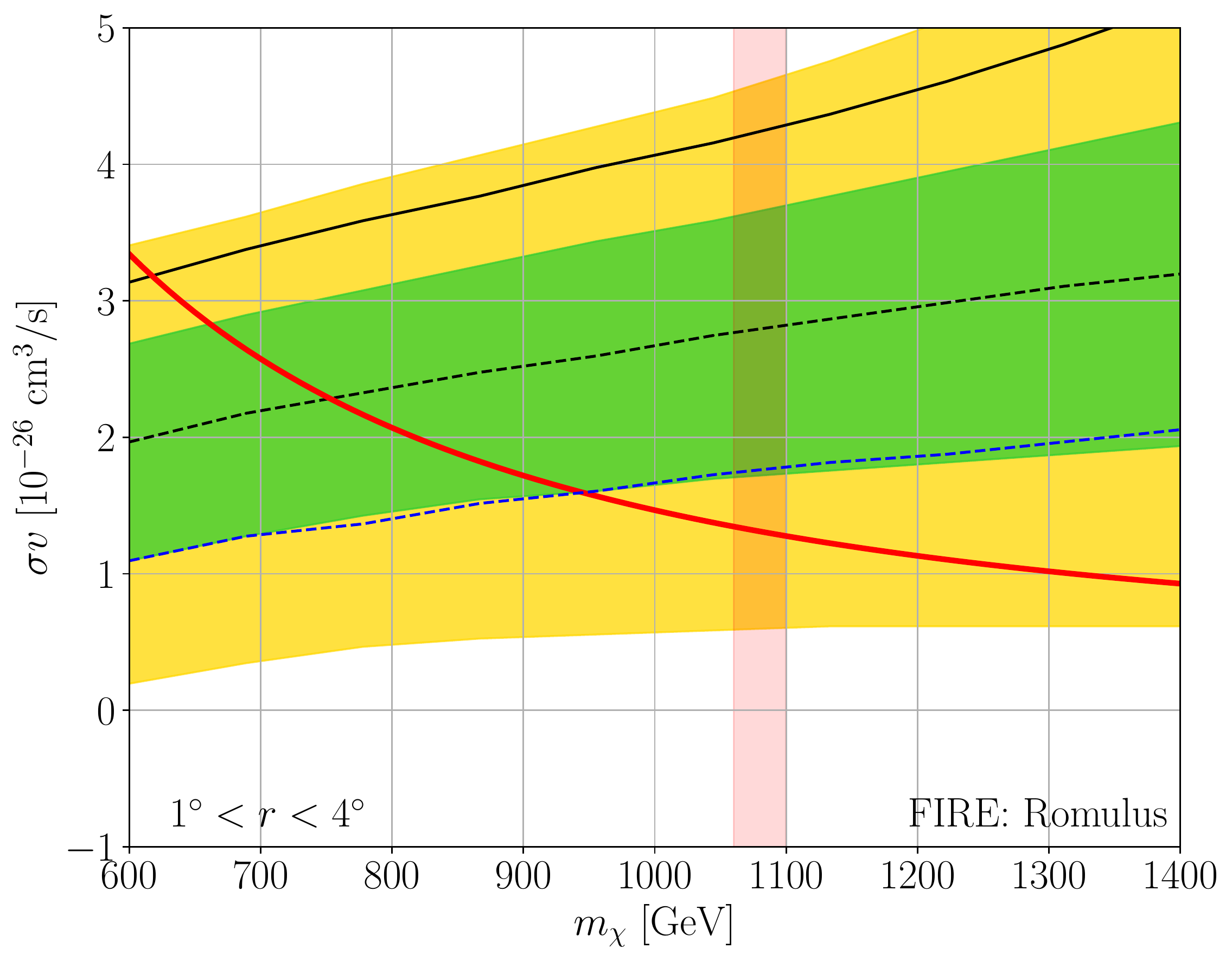}
		\end{center}
	\caption{As in Fig.~\ref{fig:results} but only including the first four rings and also including the best-fit cross-sections for our fiducial analyses.}
	\label{fig:UL_r2}
\end{figure*}

In Fig.~\ref{fig:UL_r2} we show the effect of only including the inner four rings ($1^\circ < r < 4^\circ$) in our analysis for the NFW profile (left) and \texttt{Romulus} profile (right). The results are consistent with those found in Fig.~\ref{fig:results}, as would be expected considering Fig.~\ref{fig:bf_radius}, though the best-fit cross-sections are slightly larger than in the analyses that include all annuli.

\begin{figure*}[!htb]
	\begin{center}
		\includegraphics[width=0.49\textwidth]{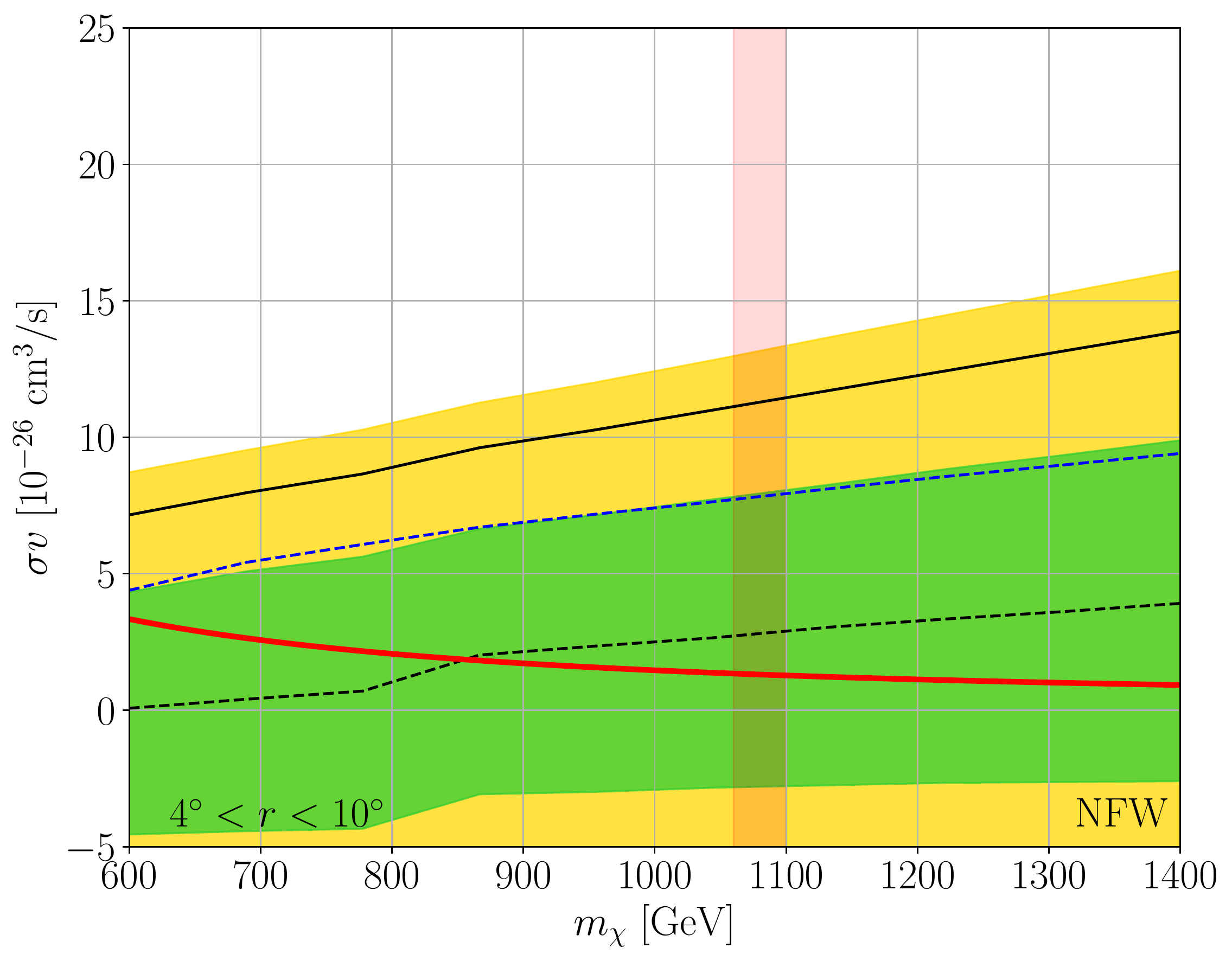}
		\includegraphics[width=0.49\textwidth]{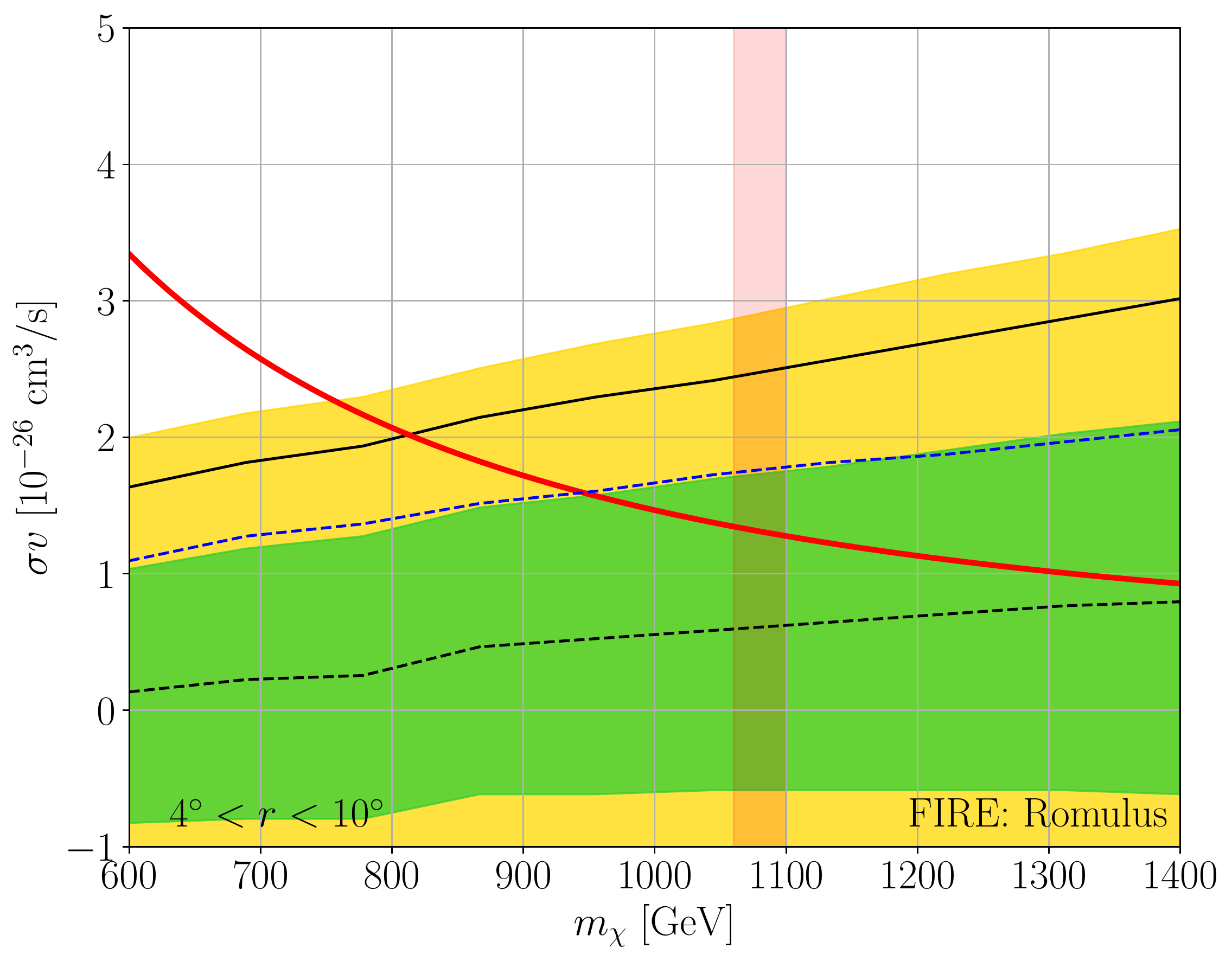}
		\end{center}
	\caption{As in Fig.~\ref{fig:results} and Fig.~\ref{fig:UL_r2} but only including the outer five rings.}
	\label{fig:UL_r4}
\end{figure*}

In Fig.~\ref{fig:UL_r4}, on the other hand, we remove the inner four rings and only keep the outer five rings, such that $4^\circ < r < 10^\circ$.  The results are consistent with those in Fig.~\ref{fig:results} and the inner Galaxy results in Fig.~\ref{fig:UL_r2}, with the best-fit cross-sections being slightly smaller.

\begin{figure*}[!htb]
	\begin{center}
		\includegraphics[width=0.49\textwidth]{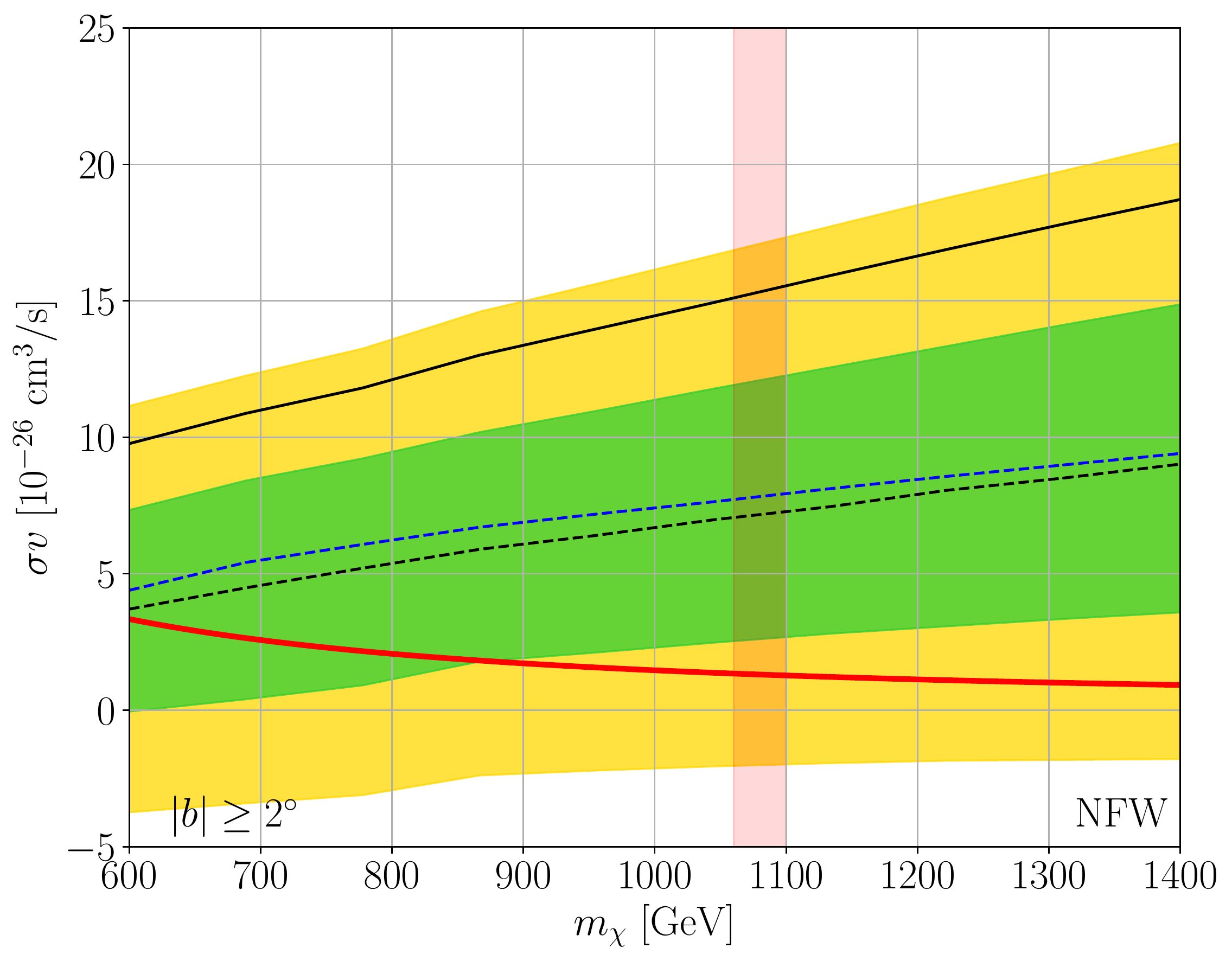}
		\includegraphics[width=0.49\textwidth]{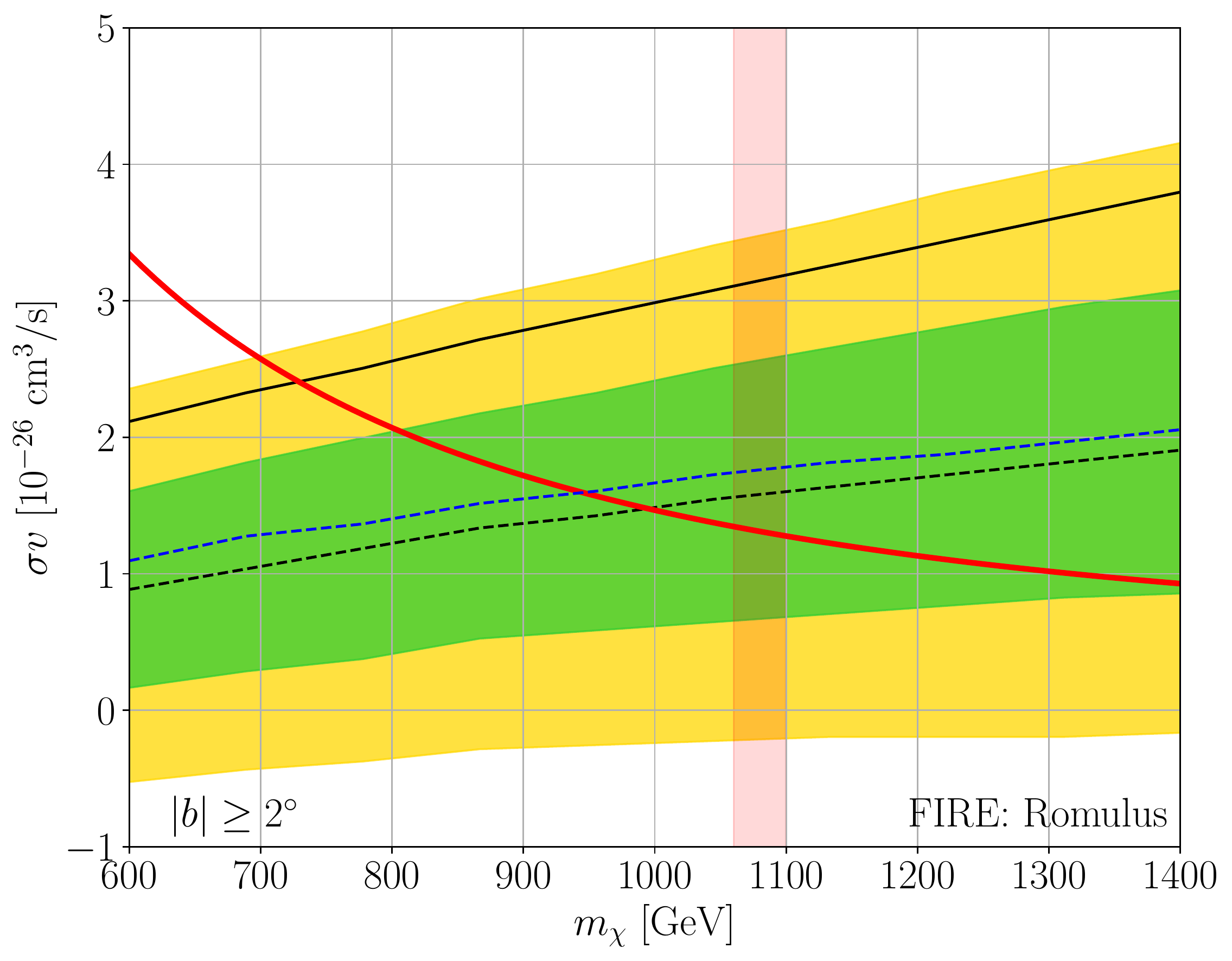}
		\end{center}
	\caption{As in Fig.~\ref{fig:results} and Fig.~\ref{fig:UL_r2} but with the Galactic plane masked at $|b| \geq 2^\circ$.  The innermost ring is not included since it is fully masked. }
	\label{fig:2p0}
\end{figure*}

In Fig.~\ref{fig:2p0} we increase the plane mask to only include $|b| \geq 2^\circ$.  As the result of the increased plane mask the innermost ring is completely masked, and so we only include annuli 2 through 9.  This analysis is meant to address possible systematic effects from mismodeling near the Galactic plane, though the results are consistent with those of the fiducial analysis.

\subsection{Energy range variations}
\label{sec:energy}

\begin{figure*}[!htb]
	\begin{center}
	\includegraphics[width=0.49\textwidth]{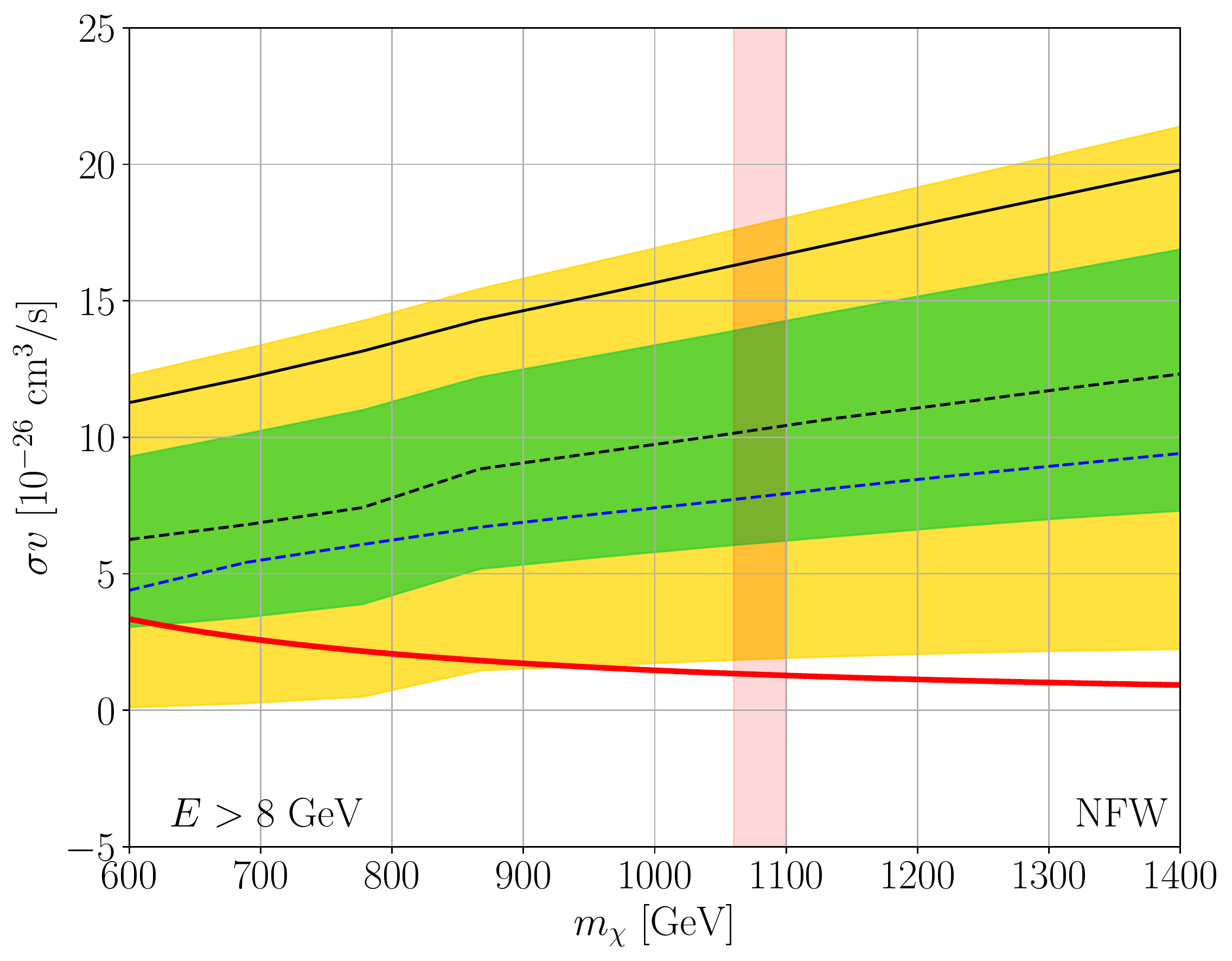}
		\includegraphics[width=0.49\textwidth]{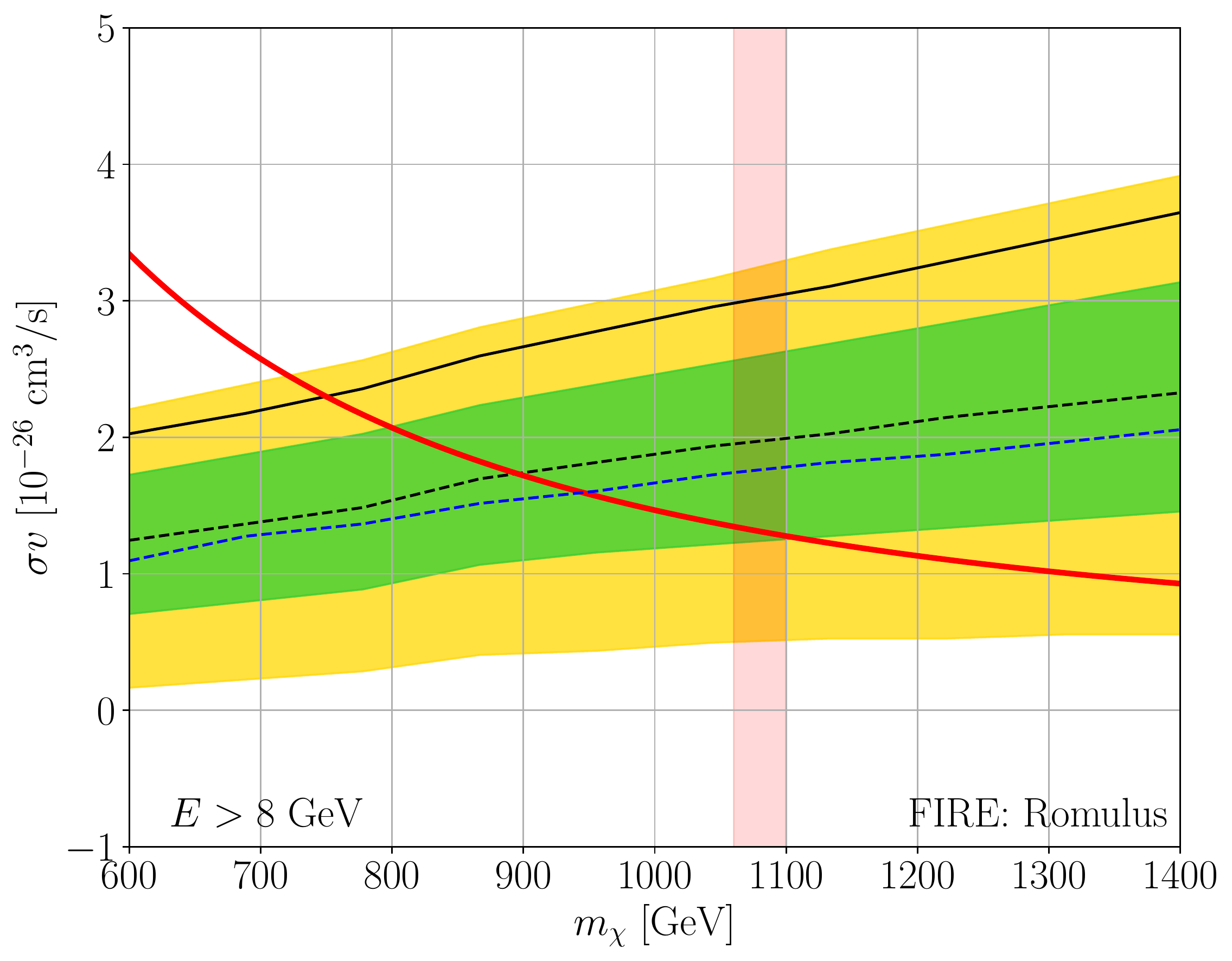}
		\includegraphics[width=0.49\textwidth]{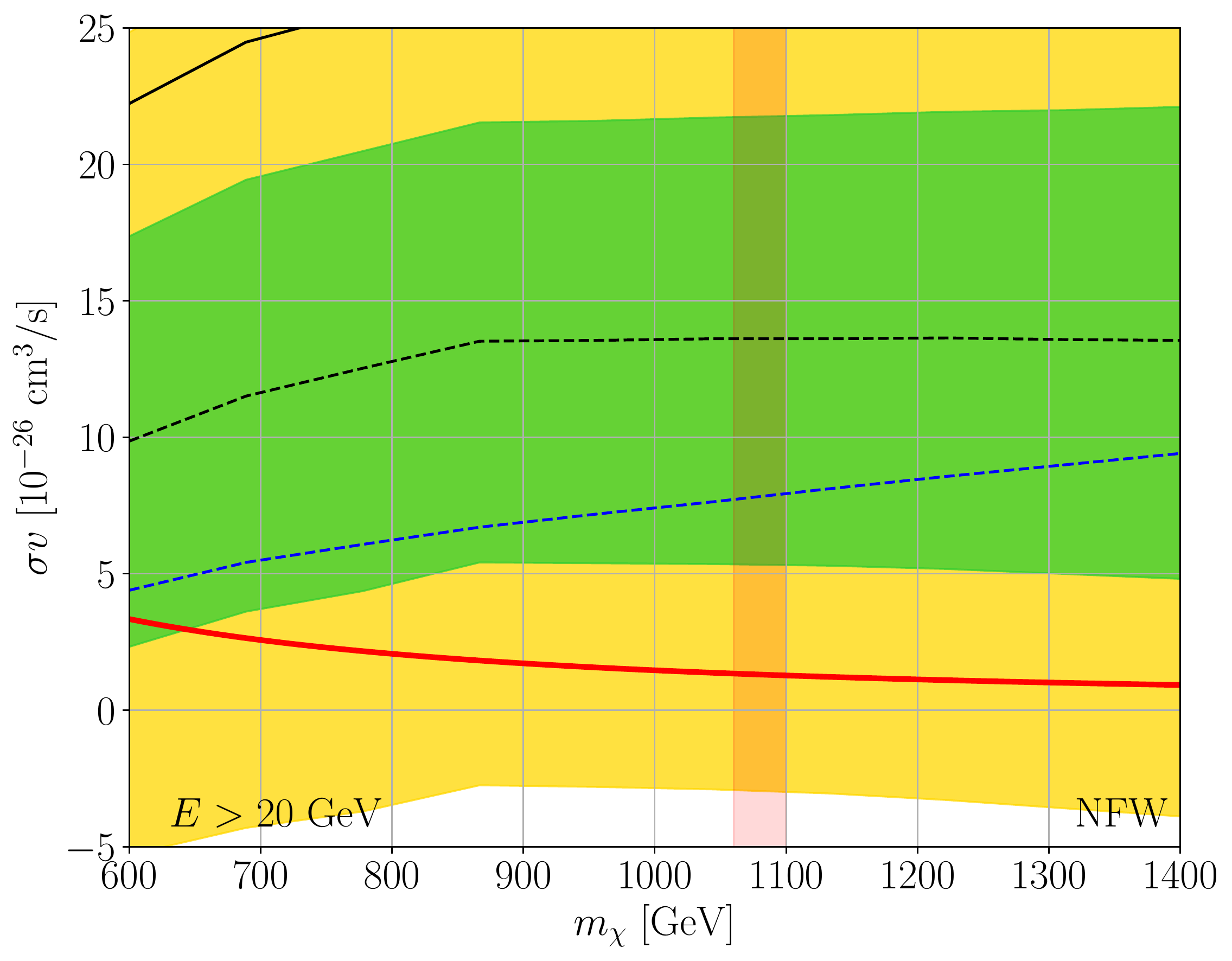}
		\includegraphics[width=0.49\textwidth]{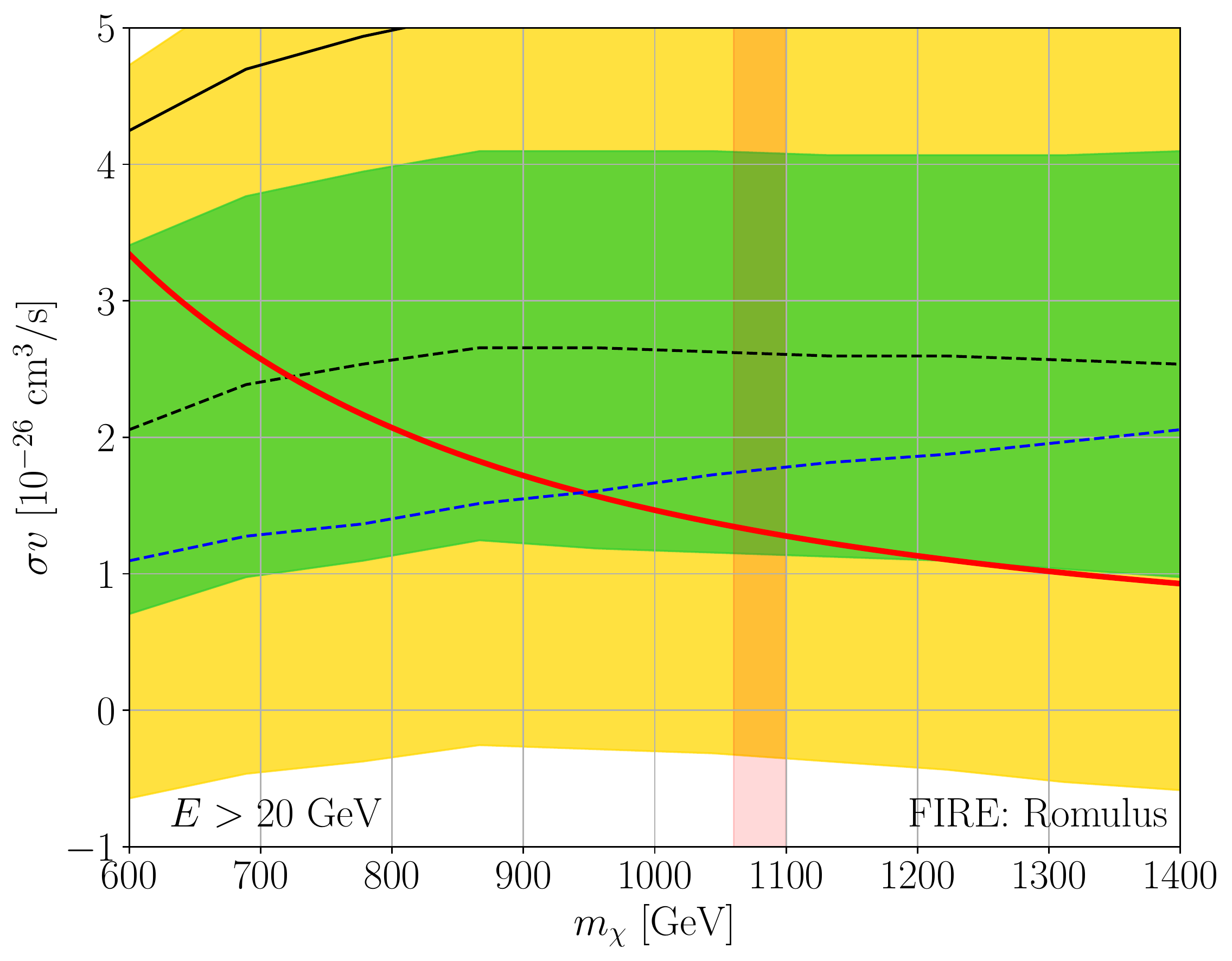}
		
		\includegraphics[width=0.49\textwidth]{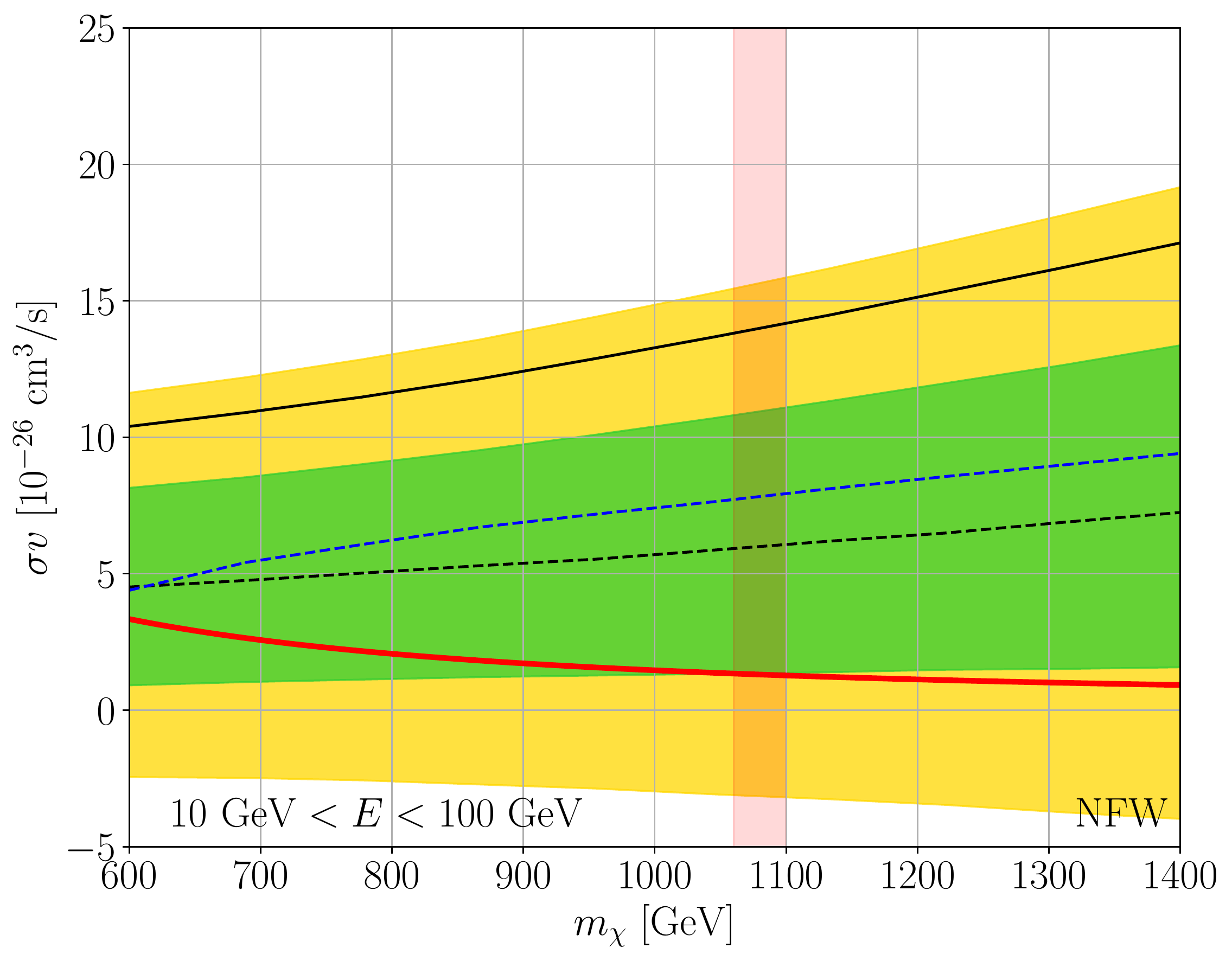}
		\includegraphics[width=0.49\textwidth]{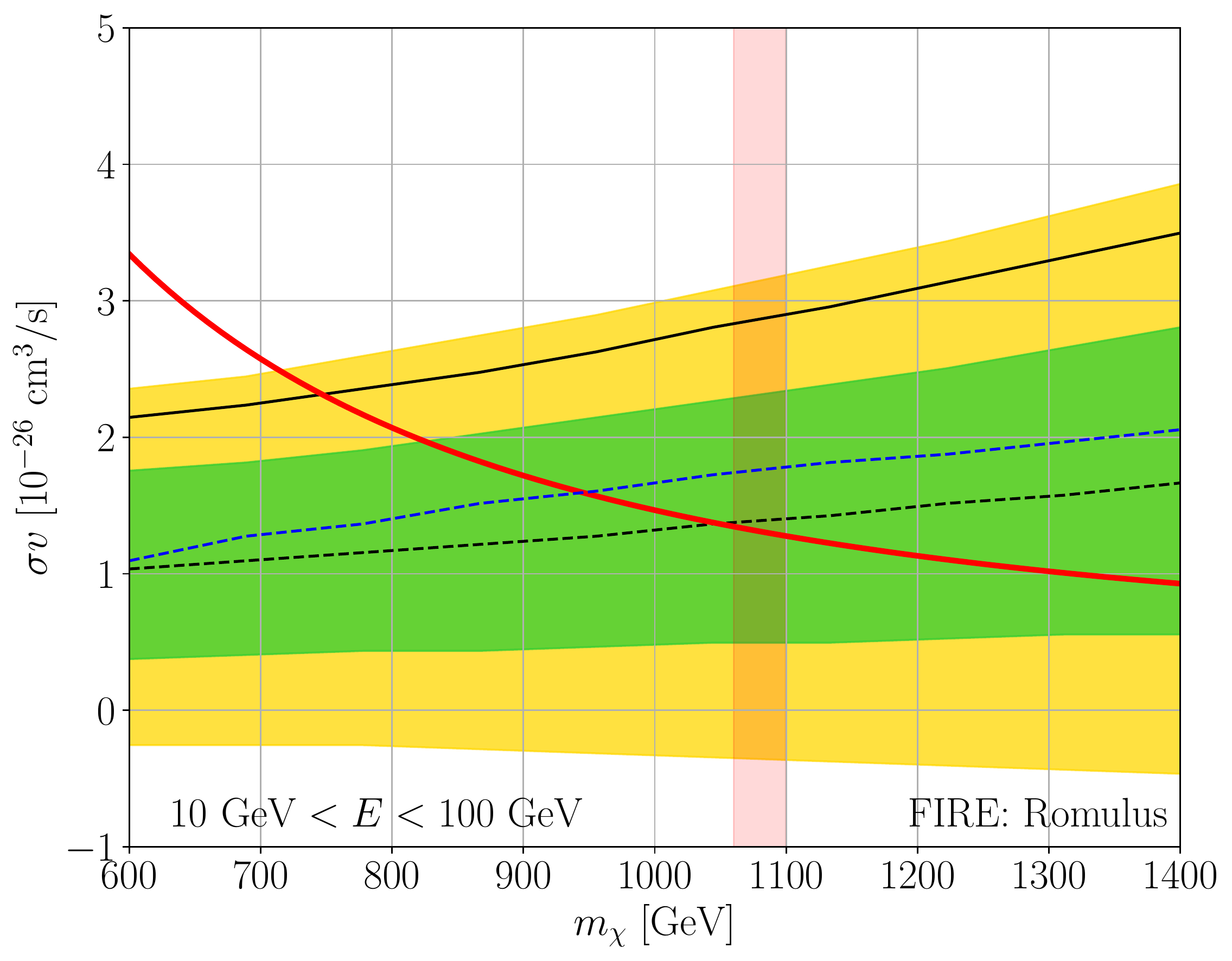}
		\end{center}
	\caption{As in Fig.~\ref{fig:results} and Fig.~\ref{fig:UL_r2} but adding an additional energy bin (top panel), removing three low-energy bins (middle panel), and removing high-energy bins (lower panel), with photon energy ranges as indicated. }
	\label{fig:systematic_energy}
\end{figure*}

We now revert to our fiducial analysis ROI but we change the number of energy bins included in the spectral fits. In particular, we consider adding in one additional (lower) energy bin, so that photons with $E \gtrsim 8$ GeV are included, and then removing 3 energy bins so that only photons with $E \gtrsim 20$ GeV are incorporated.  Then, we remove high-energy photons so that the energy constraint is $10 \, \, {\rm GeV} < E < 100 \, \, {\rm GeV}$.  The results of these analysis variations are illustrated in Fig.~\ref{fig:systematic_energy}. Including lower-energy photons produces consistent results to our fiducial analysis but slightly increases the discovery TS, while removing the first few energy bins ($E > 20$ GeV) lowers the TS but still produces a consistent best-fit and upper limit.  Removing the upper energy bins ($E < 100$ GeV) also leads to consistent results, as illustrated in the bottom panel. 

\begin{figure*}[!htb]
	\begin{center}
	\includegraphics[width=0.49\textwidth]{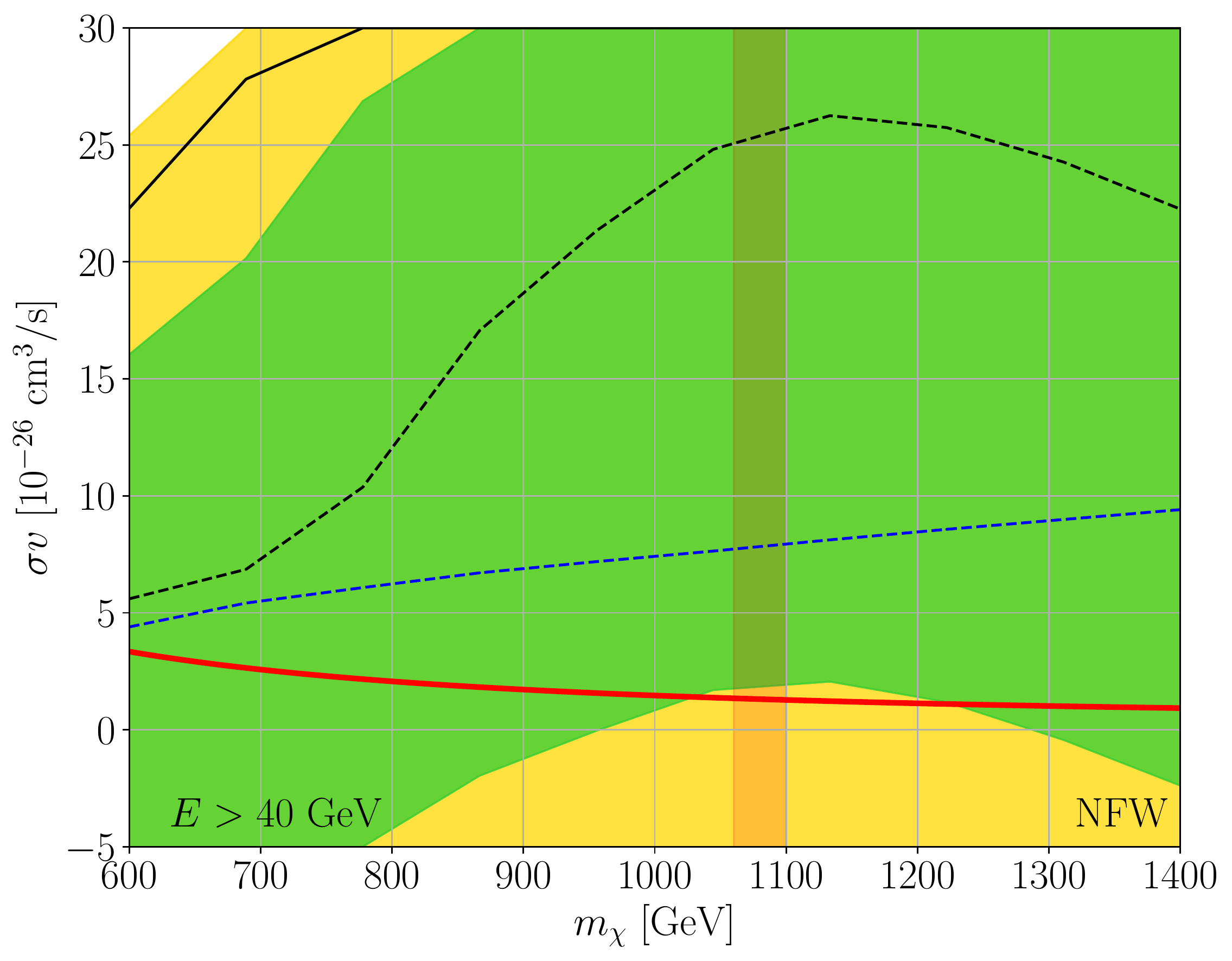}
		\includegraphics[width=0.49\textwidth]{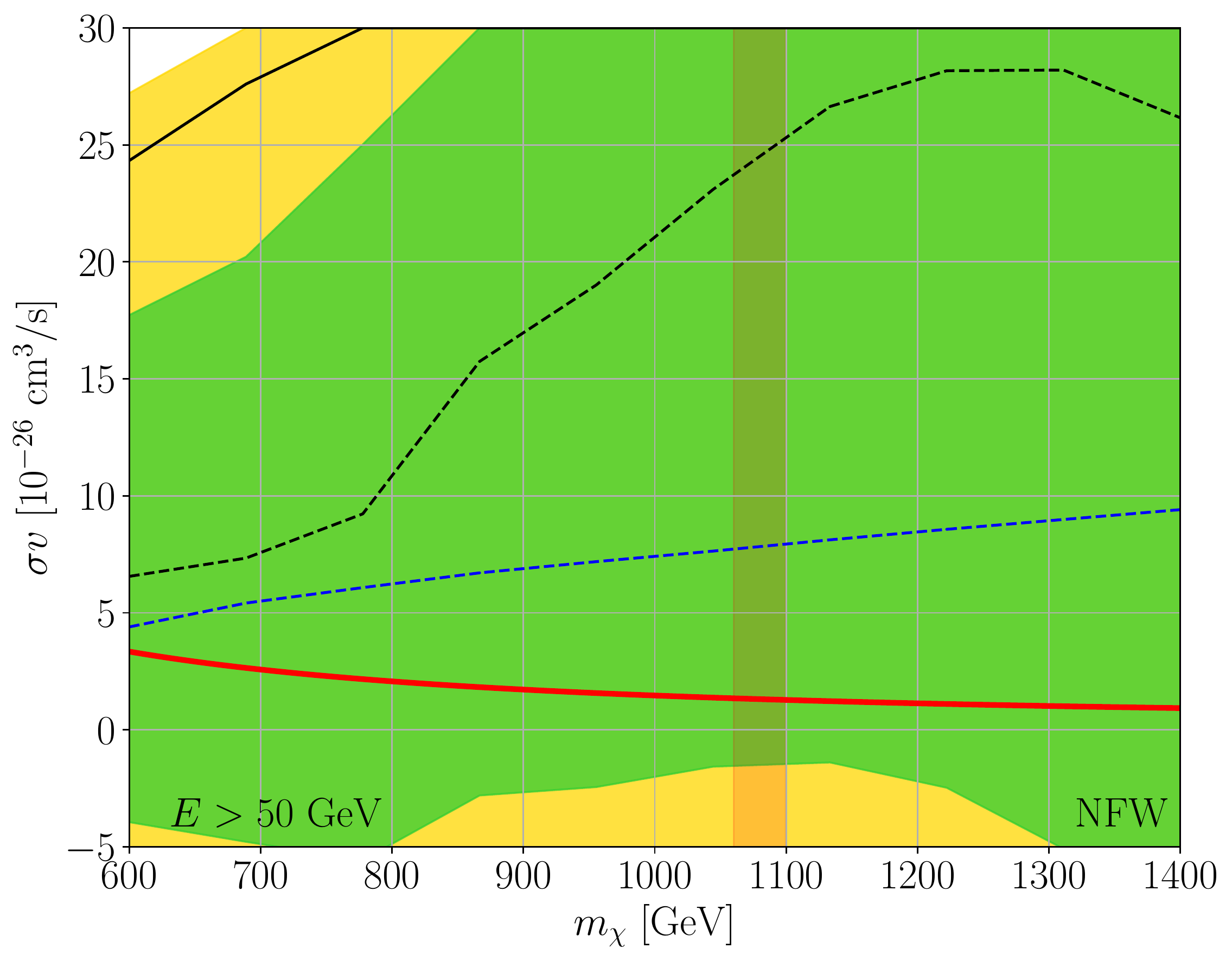}
		\end{center}
	\caption{As in Fig.~\ref{fig:systematic_energy} but for increasingly high lower energy thresholds, as indicated.  For illustrative purposes we show the results assuming the NFW DM profile.  }
	\label{fig:ultra_high}
\end{figure*}

As the lower energy cut-off is increased, we rapidly lose sensitivity to a putative higgsino signal. As an illustration, in Fig.~\ref{fig:ultra_high} we show the results of analyses where we restrict $E > 40$ GeV and $E > 50$ GeV. Interestingly, there is still a $\sim$1$\sigma$ preference for a higgsino, though the uncertainties are significantly increased compared to in our fiducial analyses.

\subsection{Removing the PS and isotropic spectral templates}

\begin{figure*}[!htb]
	\begin{center}
		\includegraphics[width=0.49\textwidth]{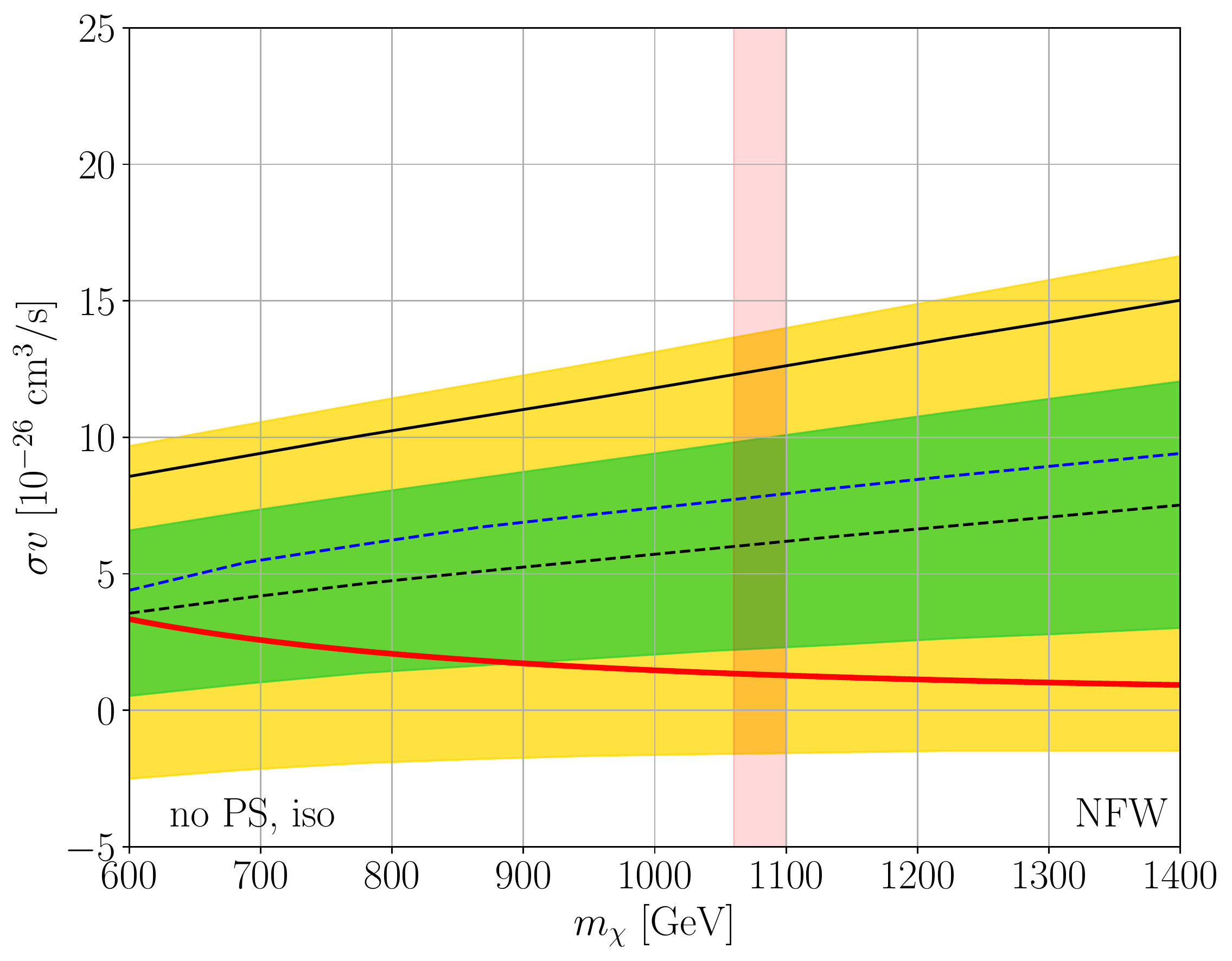}
		\includegraphics[width=0.49\textwidth]{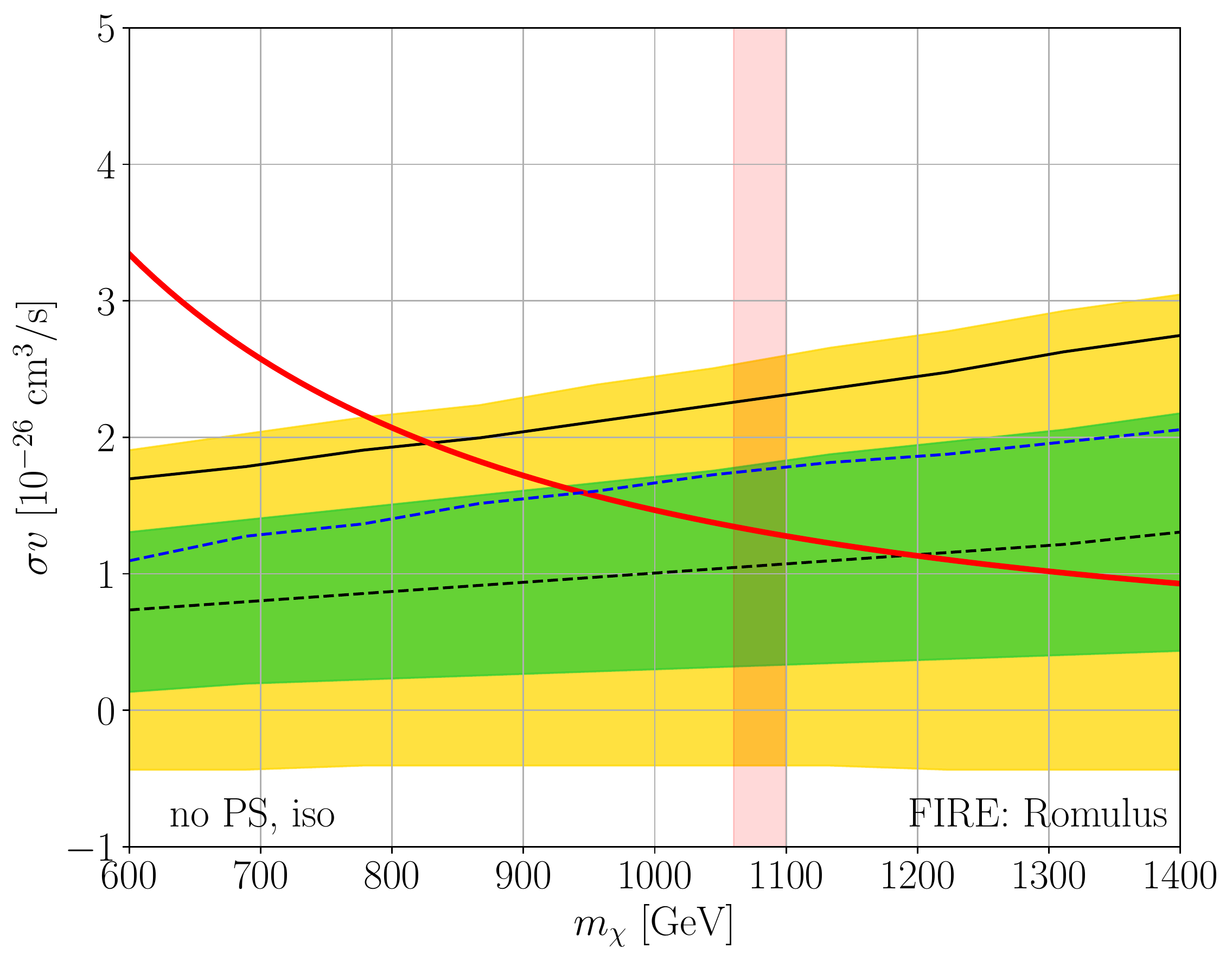}
		\end{center}
	\caption{As in Fig.~\ref{fig:results} and Fig.~\ref{fig:UL_r2} but without the PS and isotropic emission spectral templates.}
	\label{fig:UL_float_PS}
\end{figure*}

In our fiducial analysis we include Galactic emission, PS, and isotropic spectral templates. However, as shown in Fig.~\ref{fig:model}, the PS and isotropic templates are subdominant compared to the Galactic emission in our energy range and ROI.  As a cross-check of our analysis, we consider the effects of removing the PS and isotropic templates completely from our analysis. The results of this analysis are shown in Fig.~\ref{fig:UL_float_PS}.  Comparing these results to Fig.~\ref{fig:results}, the differences are minor and within statistical uncertainties. 

\subsection{Data-driven background model}
\label{sec:dd}

In our fiducial analysis we use the \texttt{p8r3} Galactic emission model in addition to the PS and isotropic models to describe the null hypothesis.  In this section we instead adopt a data-driven approach and determine the background model from the ROI described by $10^\circ < r < 15^\circ$, with PSs masked (using our fiducial mask) and the Galactic plane masked such that $|b| \geq 1^\circ$. In the background ROI we determine the flux spectrum $dF/dE$ numerically, in units of ${\rm cts}/{\rm cm}^2/{\rm s}/{\rm sr}/{\rm GeV}$, and we assume the background emission within our signal ROI takes on the same spectral shape. However, we still assign a nuisance parameters that re-scale the overall background flux $dF/dE$ in each annulus independently.  Crucially, we do not make use of the \texttt{p8r3} Galactic emission model in this analysis, and we make no assumptions about the spatial dependence of the background model.  We also do not include separate PS and isotropic spectral emission templates, since these should be captured by the data-driven template.

\begin{figure*}[!htb]
	\begin{center}
		\includegraphics[width=0.49\textwidth]{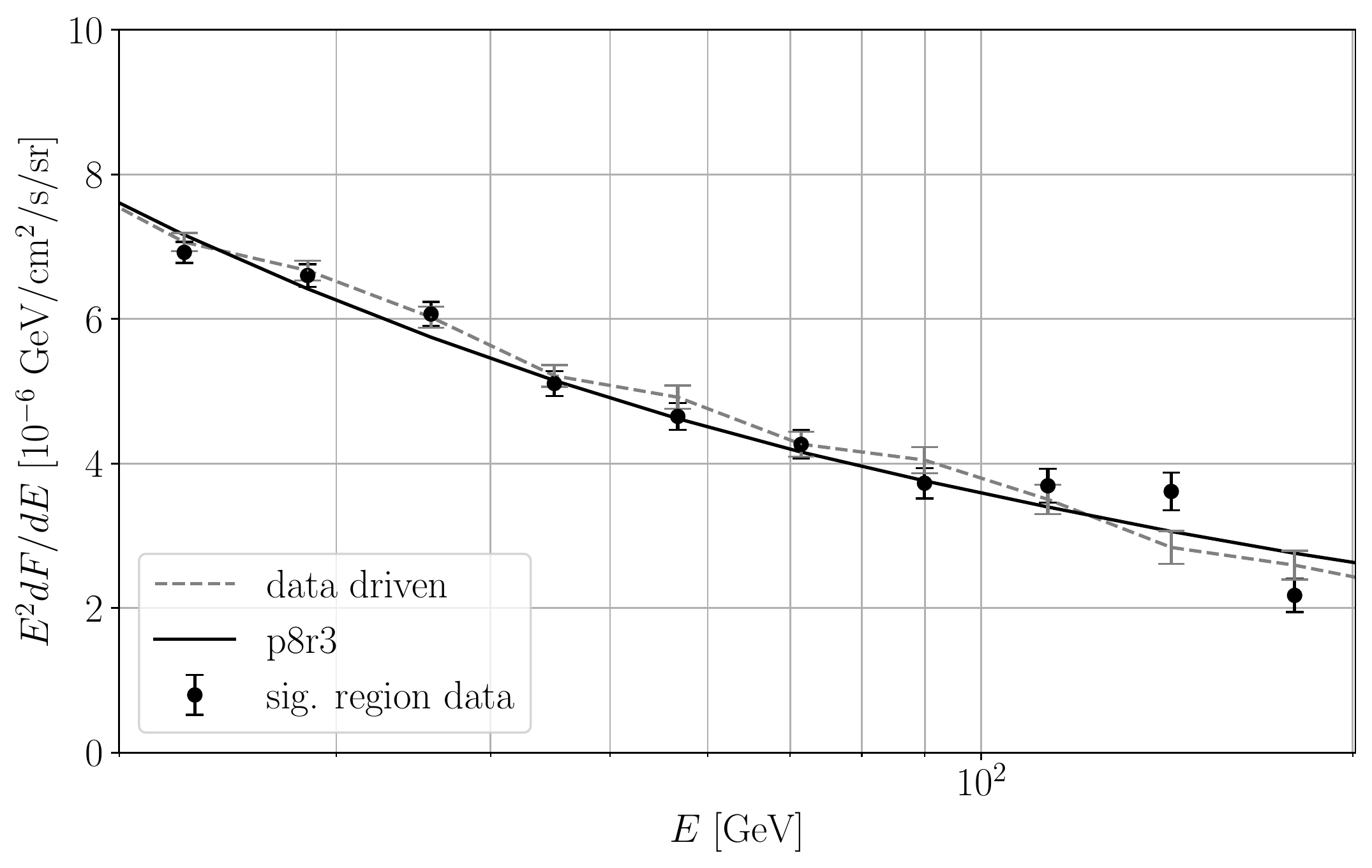}
		\end{center}
	\caption{The best-fit spectra for our fiducial analysis and the data-driven background model analysis, which is described in Sec.~\ref{sec:dd}.  Note that we restrict the energy range to $20 \, \, {\rm GeV} < E < 200$ GeV for the data-driven background model analysis. The spectra are shown summed over all 9 annuli, over the full ROI, though the analysis uses the joint likelihood constructed by analyzing the annuli independently.  We indicate statistical uncertainties in our determination of the background model, which are subdominant compared to the statistical uncertainties in our summed signal region by a factor $\sim$0.8 across the energy range.  }
	\label{fig:dd_spectrum}
\end{figure*}

Given that PSs and Galactic emission are more important at low energies, we restrict this analysis to $E > 20$ GeV. We also restrict $E < 200$ GeV because of the large statistical uncertainties in determining the background-region model at high energies. The best-fit data-driven model is shown relative to the signal region data (summed over all 9 annuli) in Fig.~\ref{fig:dd_spectrum}.  In that figure we also compare the spectrum to our best-fit fiducial model (labeled \texttt{p8r3}) which is also determined in our fiducial energy range. Note that while in this figure we show the summed spectrum over the full ROI the analysis is performed through a joint likelihood over the independent annuli, as in our fiducial analysis.  

We do not account for statistical uncertainties in our determination of the background-region spectrum in our analysis, since the background region is larger than the signal region.  The statistical uncertainties on the background-region spectrum are illustrated in Fig.~\ref{fig:dd_spectrum} relative to the summed signal region statistical uncertainties.  As a further cross-check, we consider only including the inner four annuli ($1^\circ < r < 5^\circ$), such that the signal region is much smaller than the background region and such that the two regions are separated by $5^\circ$. Including all annuli out to $10^\circ$, the best-fit cross-section for the NFW (\texttt{Romulus}) profile at the thermal mass is $10 \pm 9 \times 10^{-26}$ cm$^3/$s ($2.2 \pm 1.5 \times 10^{-26}$ cm$^3/$s), while restricting to the inner four annuli the best fit changes to $18 \pm 12 \times 10^{-26}$ cm$^3/$s ($3.0 \pm 1.8 \times 10^{-26}$ cm$^3/$s).  Thus, shrinking the signal region for the data-driven analysis leads to consistent results as in the full signal-region analysis.

\begin{figure*}[!htb]
	\begin{center}
		\includegraphics[width=0.49\textwidth]{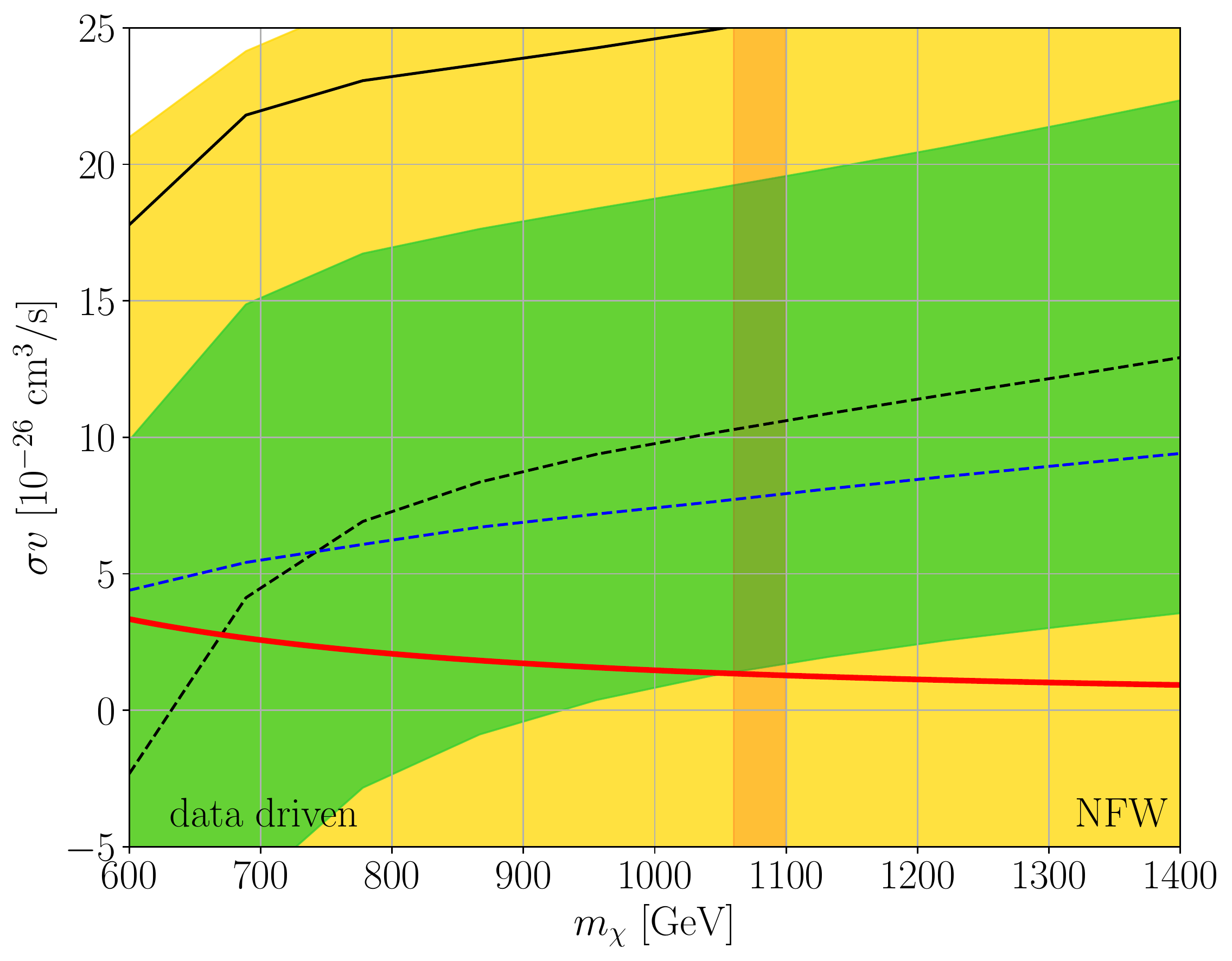}
		\includegraphics[width=0.49\textwidth]{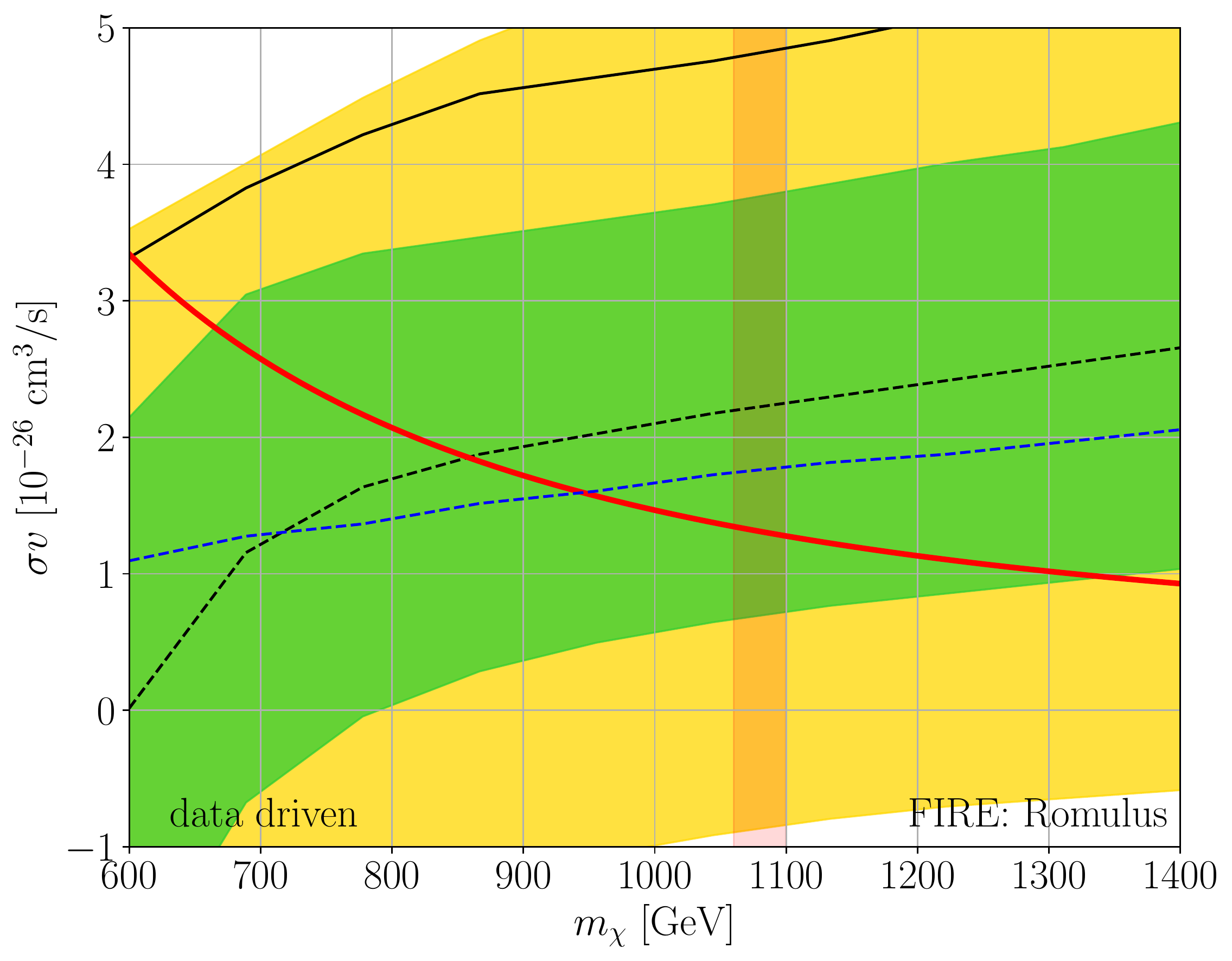}
		\end{center}
	\caption{As in Fig.~\ref{fig:results} and Fig.~\ref{fig:UL_r2} but with a data-driven background model that consists of the spectrum from $15^\circ > r > 10^\circ$ with $|b| < 1^\circ$ masked in addition to PSs masked.  We restrict the energy range to $20 \, \, {\rm GeV} < E < 200$ GeV for this analysis. }
	\label{fig:dd}
\end{figure*}

In Fig.~\ref{fig:dd} we show the results of the analysis searching for the higgsino signal (including all signal-region annuli) with the data-driven background model. The results are consistent with those in Fig.~\ref{fig:results}, though the search is somewhat less constraining because of the reduced energy range.

\subsection{Including all data quartiles}
\label{sec:all_data}

In this section we increase the data volume relative to our fiducial analysis. In the main Letter we use the top 3 quartiles of events as ranked by PSF.  Here, we include all 4 quartiles of events. Naively including more events should increase the sensitivity to a higgsino signal (with {\it e.g.} detection significances increasing, on average, by around 15\% in the event of a signal), however this is less straightforward in our analysis framework since including the last quartile of events increases the size of the PSF mask.

\begin{figure*}[!htb]
	\begin{center}
		\includegraphics[width=0.49\textwidth]{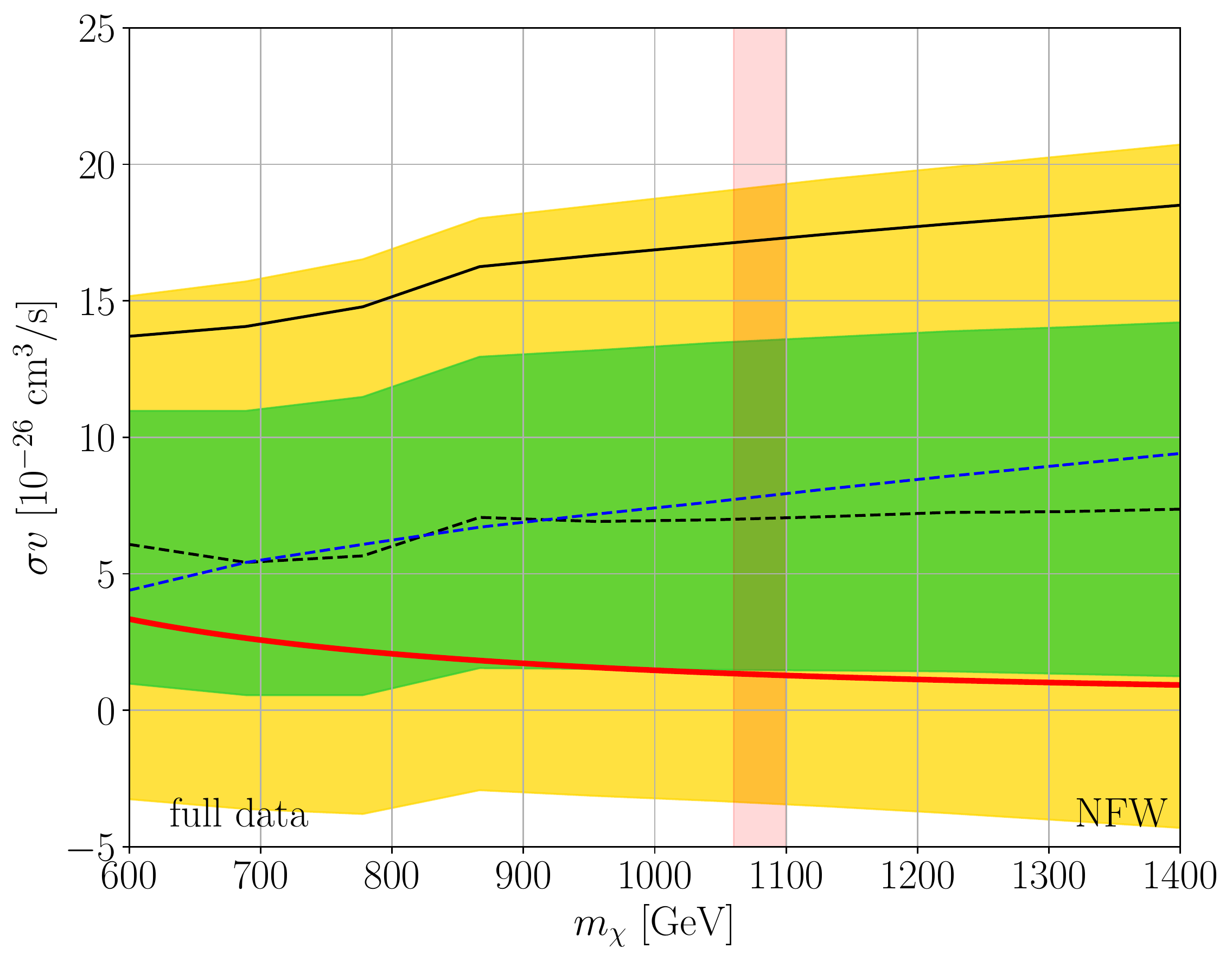}
		\includegraphics[width=0.49\textwidth]{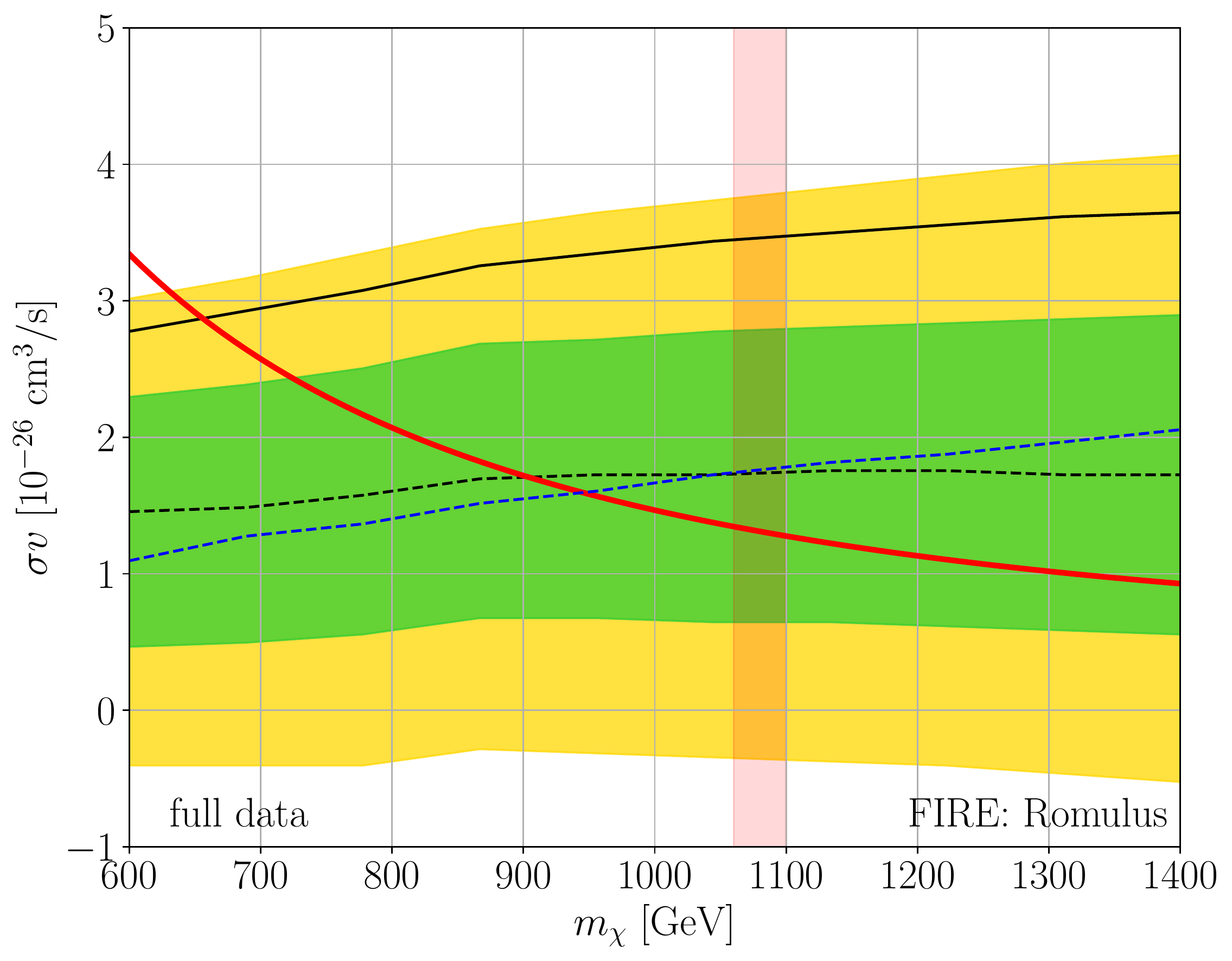}
		\end{center}
	\caption{As in Fig.~\ref{fig:results} and Fig.~\ref{fig:UL_r2} but including all four quartiles of data as ranked by PSF instead of only the top 3, as in our fiducial analysis. }
	\label{fig:full}
\end{figure*}

In Fig.~\ref{fig:full} we show the results of the higgsino search using the full data set, as presented in Fig.~\ref{fig:results} for our fiducial analysis.  The results including the last quartile of data are consistent with those in our fiducial analysis. 

\subsection{Including an inverse Compton template}
\label{sec:IC}

\begin{figure*}[!htb]
	\begin{center}
		\includegraphics[width=0.49\textwidth]{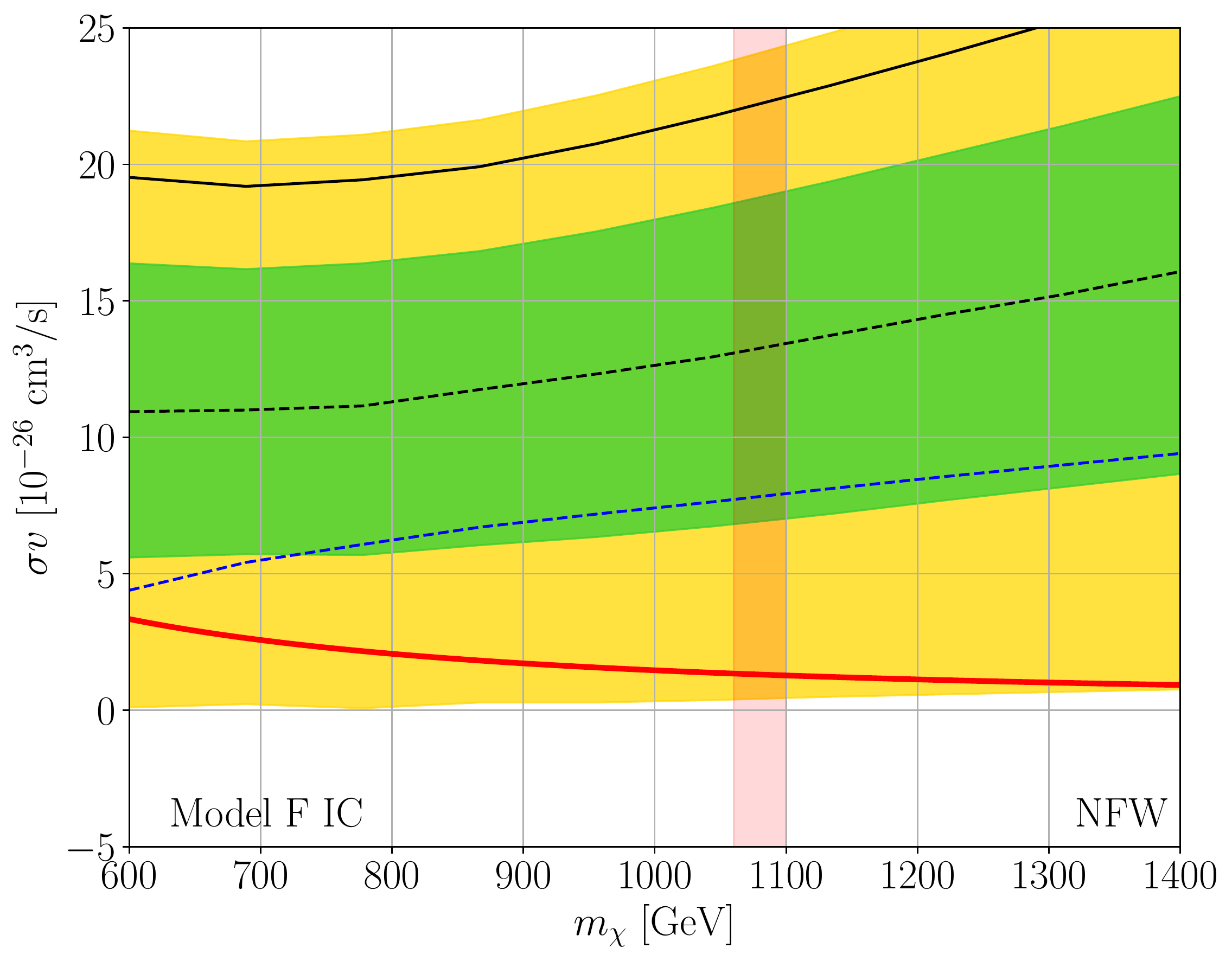}
		\includegraphics[width=0.49\textwidth]{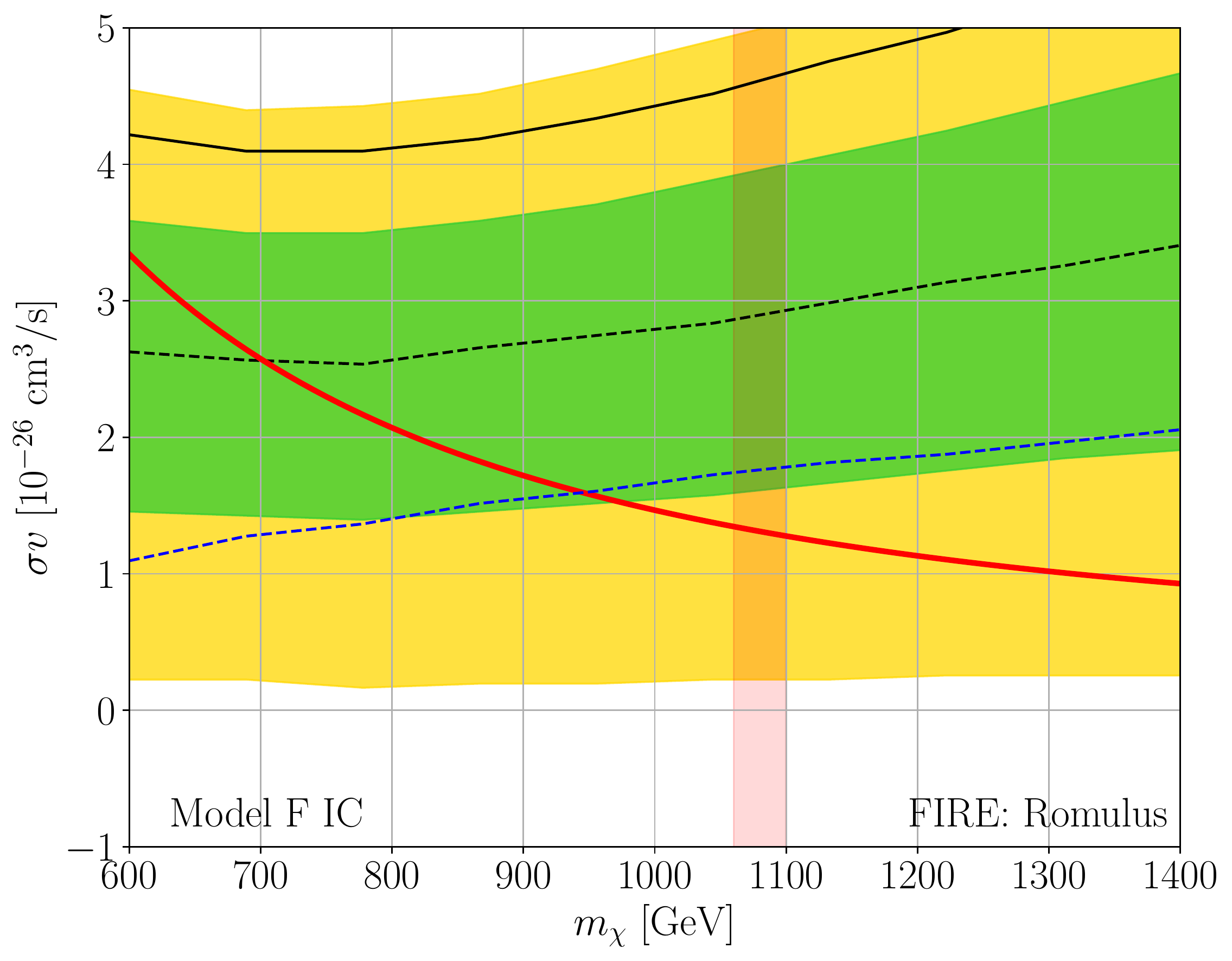}
		\end{center}
	\caption{As in Fig.~\ref{fig:results} and Fig.~\ref{fig:UL_r2} but including an additional IC template that may take on a positive or negative normalization. }
	\label{fig:IC}
\end{figure*}

In our fiducial analysis we use the \texttt{p8r3} Galactic emission model, which is constructed from multiple components.  Broadly speaking, \texttt{p8r3} includes gas-correlation emission templates in addition to IC templates that correlate with the radiation field and the cosmic ray distribution.  Since these two sources of emission have relatively uncorrelated systematics, it makes sense to try to assign them individual nuisance parameters in the analysis. We implement this by using the \texttt{p8r3} emission template in addition to an extra IC spectral template, which is allowed to have an unconstrained positive or negative normalization.  We use the IC template from \texttt{Model F} of~\cite{Calore:2014xka}, reprocessed for our data set.  The IC template is illustrated in Fig.~\ref{fig:model}.  The results of including this template are shown in Fig.~\ref{fig:IC}.  The best-fit cross-section is consistent with our fiducial result, though slightly larger.

\subsection{Spatial template fit analysis}

We now present the results of the spatial template fit analysis, which is described in Sec.~\ref{sec:methods} and which makes use of the likelihood in~\eqref{eq:template}.  As a reminder, we include three spatial templates in each energy bin over our ROI, with independent nuisance parameters in each energy bin that rescale the overall template normalizations. The templates are Galactic emission (\texttt{p8r3}), PS emission from the 4FGL that extends beyond our PS mask, and DM annihilation as given by the $J$-factor map.  Our analysis ROI is a slight variation of that found by joining all of our annuli, as described below. We consider all energy bins above 10 GeV and below the DM mass, as in our fiducial spectral analysis. 

\begin{figure*}[!htb]
	\begin{center}
	\includegraphics[width=0.49\textwidth]{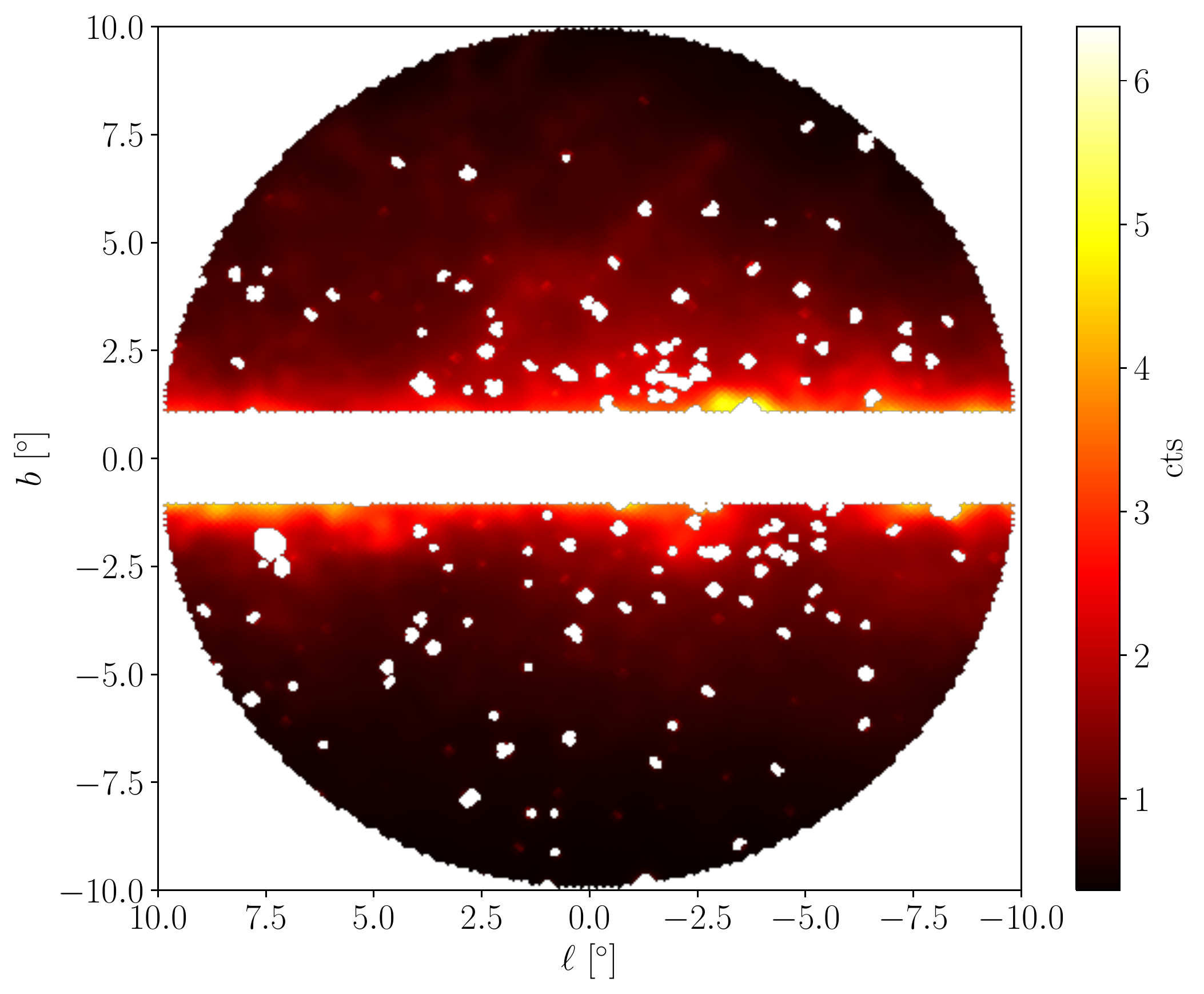}
	\includegraphics[width=0.49\textwidth]{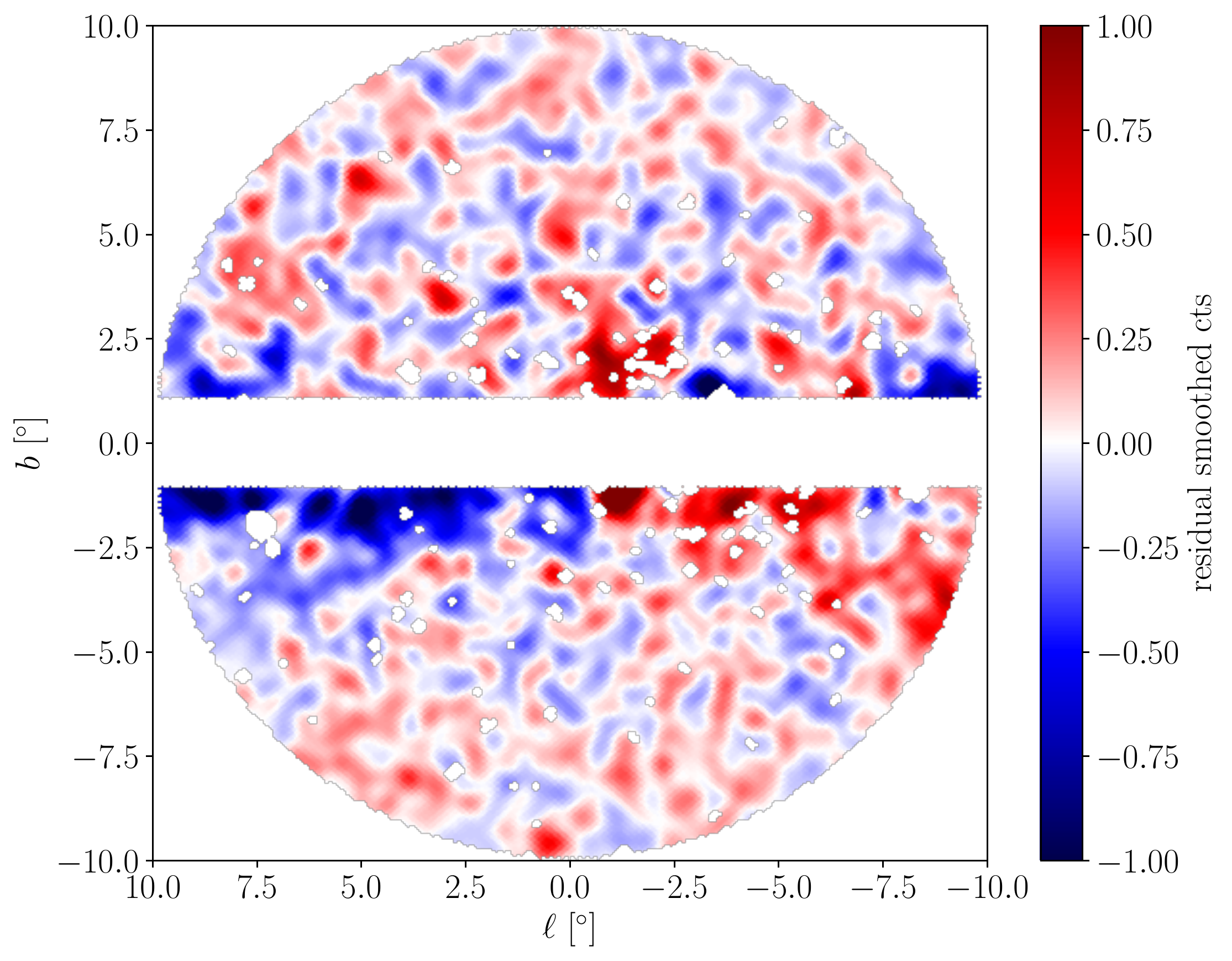}
	\includegraphics[width=0.49\textwidth]{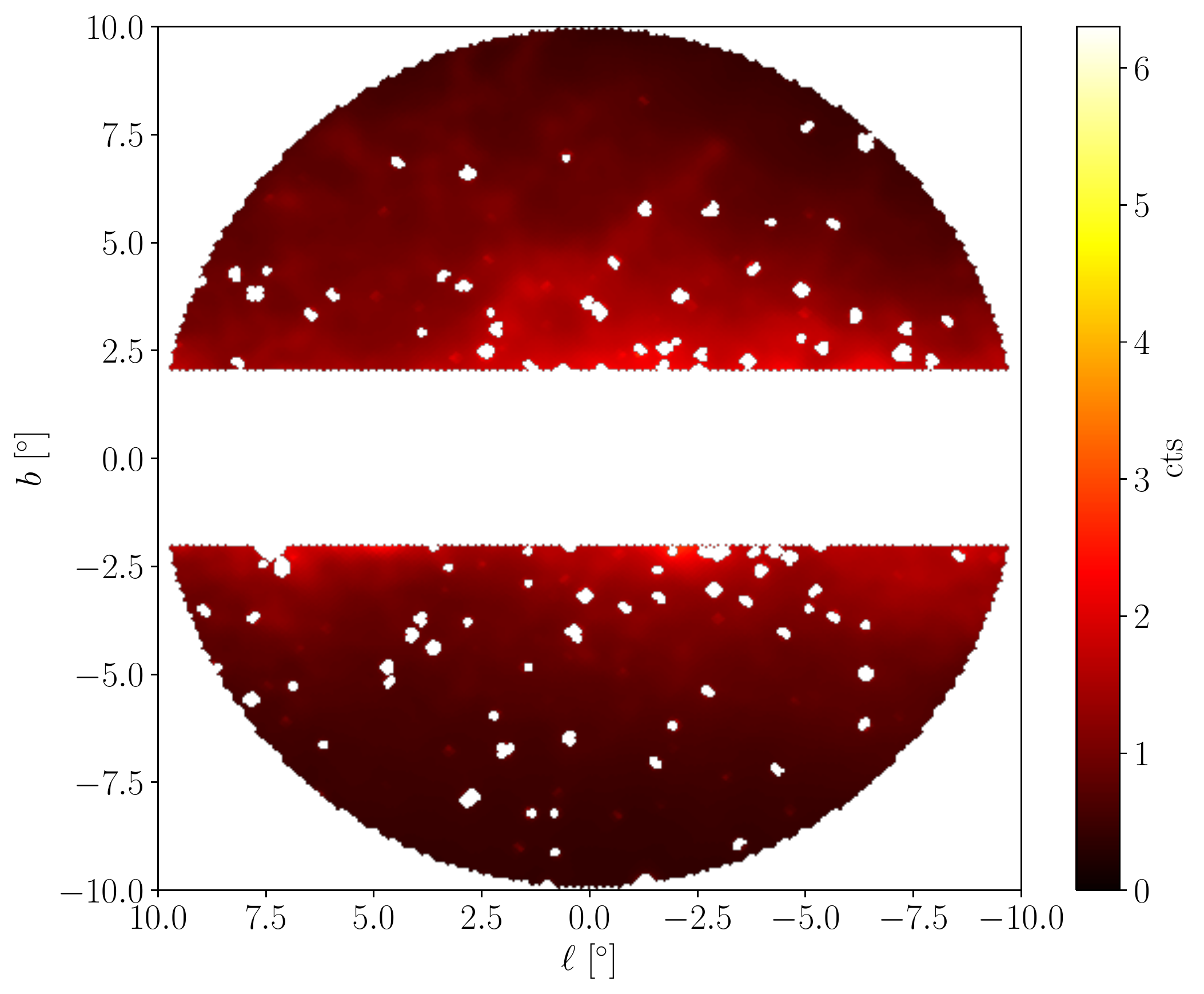}
	\includegraphics[width=0.49\textwidth]{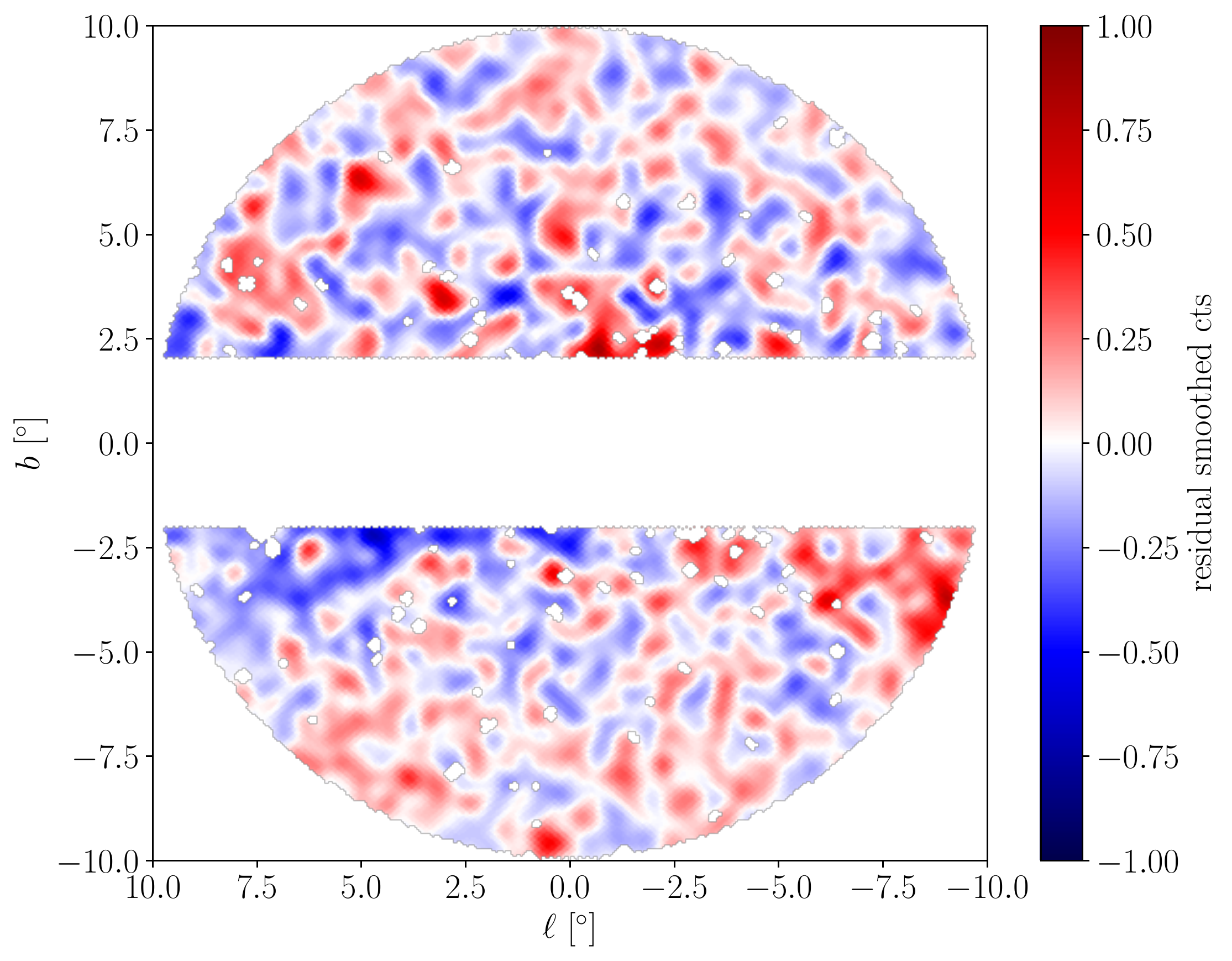}
		\end{center}
	\caption{As in Fig.~\ref{fig:data} but for the spatial template analysis, showing the best-fit null hypothesis models (left panels) and residuals (right panels) for the analyses masking the plane at 1.5$^\circ$ (top panels) and 2.0$^\circ$ (bottom panels).  The residuals are smoothed at $0.5^\circ$ for visual clarity.  The results of the analyses in the individual energy bins (for $10 \, \, {\rm GeV} < E < 1.1$ TeV) are summed together for illustrative purposes.  Mismodeling is visible near the Galactic plane, particularly in the northern hemisphere in the $|b| \geq 1.5^\circ$ analysis.  }
	\label{fig:spatial_maps}
\end{figure*}

The spatial template analysis is more susceptible to mismodeling since we now have to describe the data in each spatial pixel in addition to each energy bin. The Galactic emission templates are notoriously poor descriptions of gamma-ray emission in the inner Galaxy (see, {\it e.g.},~\cite{Buschmann:2020adf}).  Indeed, we find (by visual inspection) that the analysis in our fiducial ROI produces a poor fit to the data, with large residuals and regions of over-subtraction near the Galactic plane in particular. We partially mitigate the mismodeling by increasing the size of our Galactic plane mask, increasing it to $|b| \geq 1.5^\circ$ and to $|b| \geq 2^\circ$. Of course, in increasing the plane mask we also mask more region near the GC, which decreases our sensitivity to a putative DM signal.  In Fig.~\ref{fig:spatial_maps} we show the best-fit null-hypothesis models (left panels) and smoothed residuals (right panels) from analyses with $|b| \geq 1.5^\circ$ (top panels) and $|b| \geq 2^\circ$ (bottom panels). While we perform the analyses in each energy bin independently, for these figures we sum the results over all energy bins (taking $m_\chi = 1.1$ TeV so that we only include photons with $E \lesssim 1.1$ TeV). 

In the top right panel of Fig.~\ref{fig:spatial_maps} large regions of over-subtraction are visible by eye, particularly in the northern hemisphere near the Galactic plane. The over-subtraction is better though still present in the $|b| \geq 2^\circ$ analysis. 

\begin{figure*}[!htb]
	\begin{center}
	\includegraphics[width=0.49\textwidth]{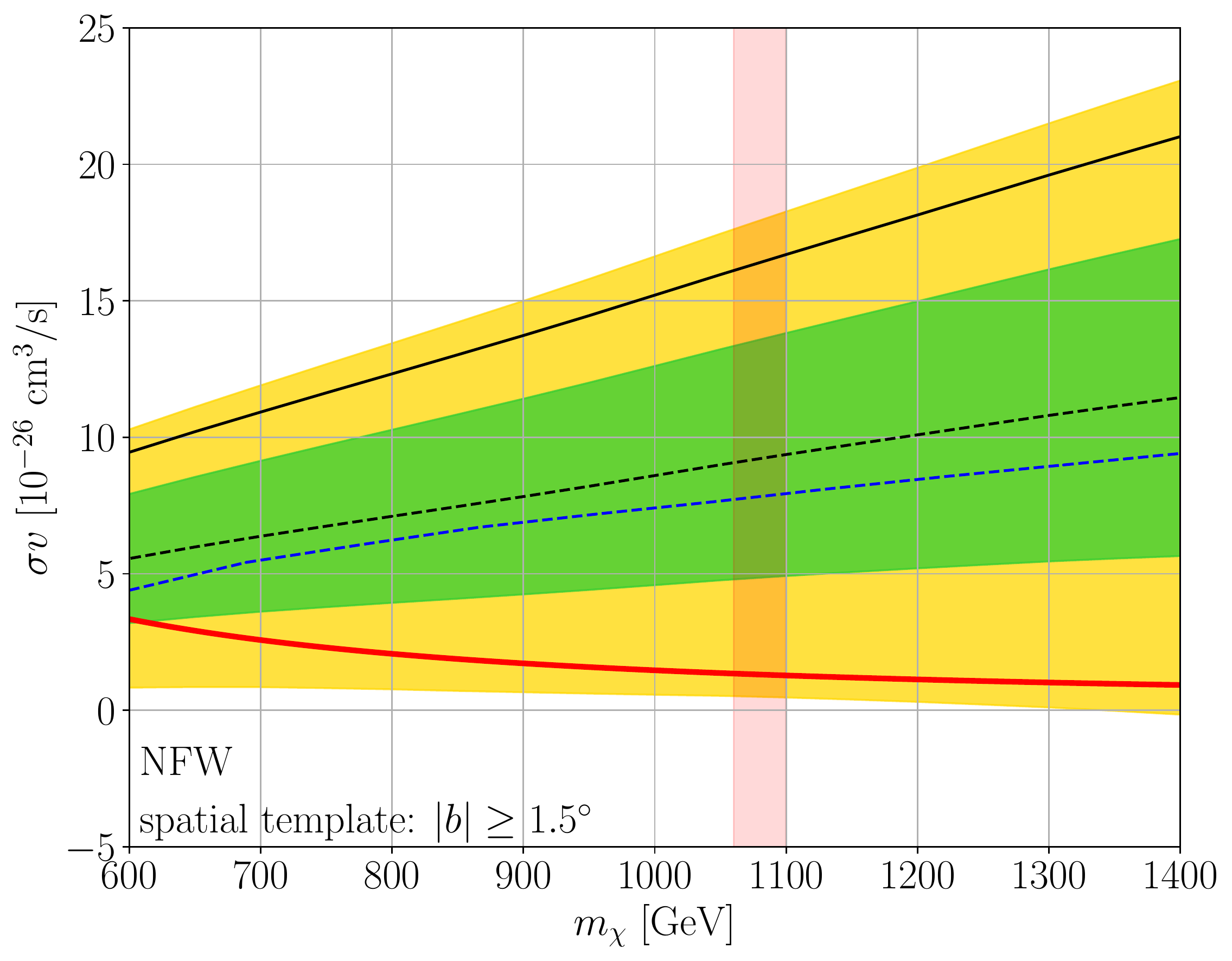}
	\includegraphics[width=0.49\textwidth]{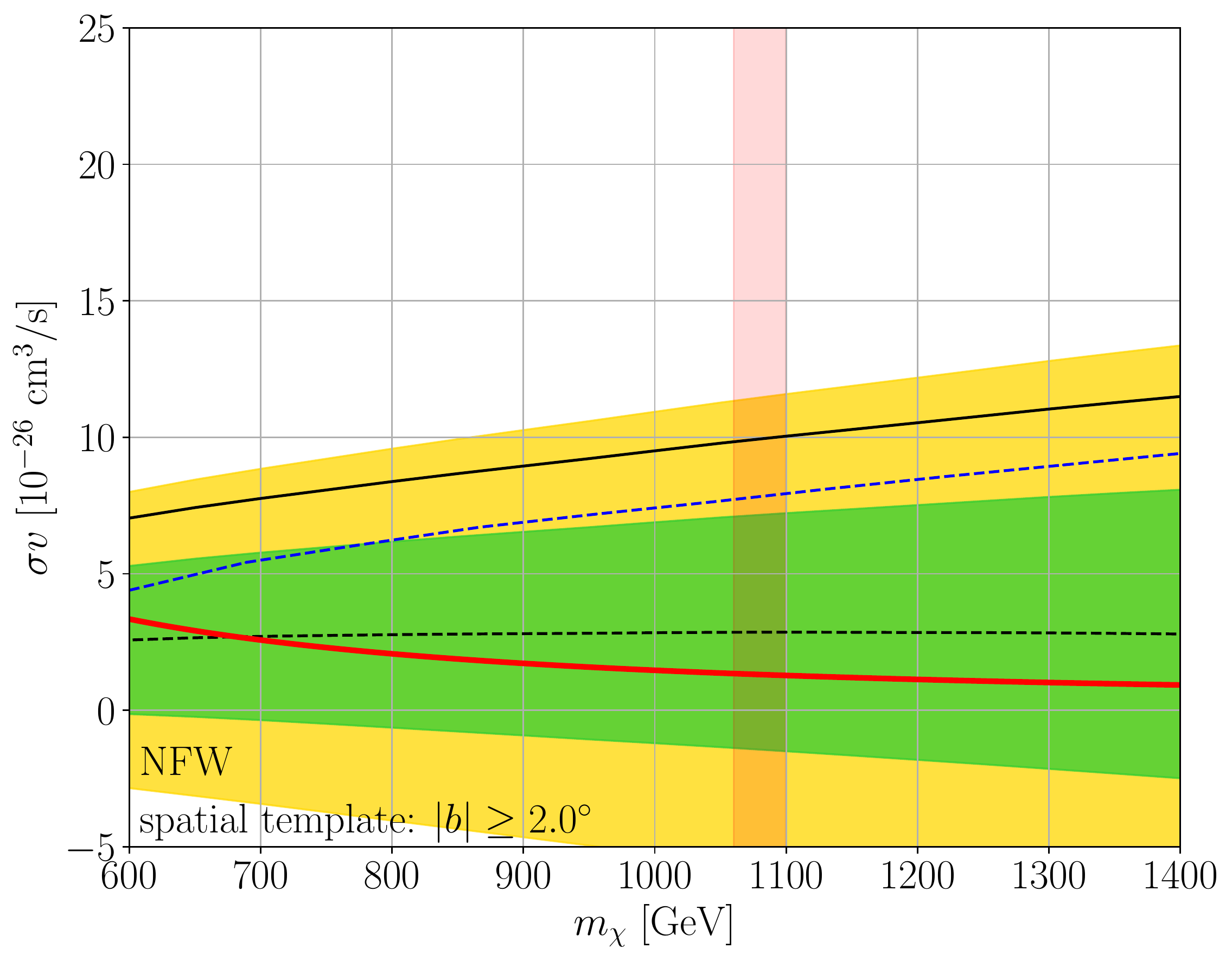}
		\end{center}
	\caption{As in Fig.~\ref{fig:results} and Fig.~\ref{fig:UL_r2} but for the spatial template analyses using the NFW DM profile  with $|b| \geq 1.5^\circ$ and $|b| \geq 2^\circ$, as indicated.
	}
	\label{fig:spatial}
\end{figure*}

The results of the spatial template analysis (to be compared to Fig.~\ref{fig:results}) are illustrated in Fig.~\ref{fig:spatial}. Interestingly, the results of this analysis are broadly consistent with those found in the fiducial spectral analysis, with comparable upper limits and consistent (positive) best-fit values, though the detection significance in slightly reduced in the spatial analyses. The results of the $|b| \geq 1.5^\circ$ and $|b| \geq 2^\circ$ analyses are consistent, though as expected the $|b| \geq 1.5^\circ$ analysis is slightly more sensitive.  Note that we only present results using the NFW DM profile since the FIRE-2 profiles are not at high enough spatial resolution to be used in the template analysis without further smoothing.

\FloatBarrier

\end{document}